\documentclass[aps,pra,amsmath,amssymb,superscriptaddress,10pt]{revtex4-1}

\usepackage[usenames,dvipsnames]{xcolor}
\usepackage{graphicx}
\usepackage{etoolbox}
\usepackage{braket}
\usepackage{subcaption}

\newcommand{\versionsupp}{supp_v3_mjs}

\newcommand{\refsupp}[1]{section~\ref{#1} of the supplementary material}
\newcommand{\refmain}[1]{section~\ref{#1} of the main paper}
\newcommand{\eqrefmain}[1]{Eq.~\eqref{#1} of the main paper}
\newcommand{\refsec}[1]{Section~\ref{#1}}

\newcommand{\expect}[1]{{\left \langle #1 \right \rangle}}
\newcommand{\etal}{\emph{et al.\/}}

\newcommand{\dd}{\mathrm{d}}
\newcommand{\ds}{\mathrm{d}s}
\newcommand{\dx}{\mathrm{d}x}
\newcommand{\dy}{\mathrm{d}y}
\newcommand{\dk}{\mathrm{d}k}
\newcommand{\dz}{\mathrm{d}z}
\newcommand{\dkp}{\mathrm{d}k^\prime }
\newcommand{\qp}{q^\prime }
\newcommand{\kp}{k^\prime }
\newcommand{\alphap}{\alpha^\prime }
\newcommand{\vecdR}{\dd\mathbf{R}}
\newcommand{\dR}{\dd R}

\newcommand{\dxdy}{\dx\dy}
\newcommand{\dq  }{\mathrm{d}q}
\newcommand{\dqp}{\mathrm{d}q^\prime}

\newcommand{\vd}{\mathbf{d}}
\newcommand{\ve}{\mathbf{e}}
\newcommand{\vf}{\mathbf{f}}
\newcommand{\vh}{\mathbf{h}}
\newcommand{\vu}{\mathbf{u}}
\newcommand{\vecr}{\mathbf{r}}
\newcommand{\vecrp}{\mathbf{r^\prime}}

\newcommand{\vE}{\mathbf{E}}
\newcommand{\vF}{\mathbf{F}}
\newcommand{\vH}{\mathbf{H}}
\newcommand{\vecR}{\mathbf{R}}
\newcommand{\vS}{\mathbf{S}}
\newcommand{\vU}{\mathbf{U}}

\newcommand{\gammap}{\gamma{^\prime}}
\newcommand{\sP}{\text{\tiny P}}
\newcommand{\sS}{\text{\tiny S}}
\newcommand{\so}{\text{\tiny o}}

\newcommand{\rp}{{r^\prime}}
\newcommand{\epso}{\epsilon_0}
\newcommand{\muo}{\mu_0}

\newcommand{\vpi}{{\boldsymbol \pi}}
\newcommand{\vD}{\mathbf{D}}
\newcommand{\vB}{\mathbf{B}}

\newcommand{\unitn}{\hat{\mathbf{n}}}
\newcommand{\unitu}{\hat{\mathbf{u}}}
\newcommand{\unitx}{\hat{\mathbf{x}}}
\newcommand{\unity}{\hat{\mathbf{y}}}
\newcommand{\unitz}{\hat{\mathbf{z}}}
\newcommand{\cc}{\text{c.c.}}
\newcommand{\hc}{\text{h.c.}}

\newcommand{\calL}{{\cal L}}
\newcommand{\calM}{{\cal M}}
\newcommand{\me}{\mathrm{e}}

\newcommand{\drrr}{\mathrm{d}\vecr}
\newcommand{\drrrp}{\mathrm{d}\vecrp}
\newcommand{\parparr}[2]{\frac{\partial #1}{\partial r^{#2}}}
\newcommand{\falq}{f_{\alpha q}}
\newcommand{\ualq}{u_{\alpha q}}
\newcommand{\vfalq}{\vf_{\alpha q}}

\newcommand{\stiff}[1]{c^{#1}}
\newcommand{\stiffxy}[1]{c^{#1}(x,y)}
\newcommand{\sqrtrhoxy}{\sqrt{\rho(x,y)}}
\newcommand{\sqrtrho}{\sqrt{\rho}}

\newcommand{\powdensA}{{\cal P}^\text{A}}
\newcommand{\powA}{P^\text{A}}
\newcommand{\pmodeA}{p^\text{A}}
\newcommand{\powEM}{P^\text{EM}}
\newcommand{\pmodeEM}{p^\text{EM}}
\newcommand{\falqonsrho}{\frac{\falq^k}{\sqrtrho}}
\newcommand{\eiqz}{\me^{iqz}}

\newcommand{\hamilA}{H^\text{A}}
\newcommand{\hamilEM}{H^\text{EM}}
\newcommand{\ePE}{\mathrm{ePE}}

\newcommand{\dRc}{\dd R_c}
\newcommand{\barGam}{\bar{\Gamma}}
\newcommand{\Omalq}{{\Omega_{\alpha q}}}
\newcommand{\Omsqalq}{{\Omega^2_{\alpha q}}}

\newcommand{\Real}{\mathrm{Re}}

\providebool{includesupp}
\booltrue{includesupp}

\ifbool{includesupp}{}{%
\usepackage{xr}
\externaldocument{\versionsupp}
}

\begin{document}

\title{A Hamiltonian treatment of stimulated Brillouin scattering in nanoscale integrated waveguides}

\author{J. E. Sipe }
\email[]{sipe@physics.utoronto.ca}

\affiliation{Department of Physics and Institute for Optical Sciences, University of Toronto, Toronto, Ontario M5S 1A7, Canada}
\affiliation{Macquarie University Quantum Science and Technology Centre (QSciTech), 
Department of Physics \& Astronomy, Macquarie University, NSW 2109, Australia}

\author{M. J. Steel}
\affiliation{Macquarie University Quantum Science and Technology Centre (QSciTech), 
Department of Physics \& Astronomy, Macquarie University, NSW 2109, Australia}
\affiliation{ Centre for Ultrahigh bandwidth Devices for Optical Systems (CUDOS)
and MQ Photonics Research Centre,
Department of Physics \& Astronomy, Macquarie University, NSW 2109, Australia}

\date{\today} 
\begin{abstract} 
We present a multimode Hamiltonian formulation
for the problem of opto-acoustic interactions in optical waveguides.
We establish a Hamiltonian representation of the acoustic field and then
introduce a full system with a simple opto-acoustic coupling that includes both
photoelastic/electrostrictive and radiation pressure/moving boundary effects.
The Heisenberg equations of motion are used to obtain coupled mode equations
for quantized envelope operators for the optical and acoustic fields. We show
that the coupling coefficients obtained coincide with those established
earlier, but our formalism provides a much simpler demonstration of the
connection between radiation pressure and moving boundary effects than in
previous work~[C. Wolff \etal, Physical Review A \textbf{92}, 013836 (2015)].
\end{abstract}

\maketitle

\section{Introduction} 
Almost a century after it was first
proposed~\cite{Brillouin1922,Mandelstam1926} and fifty years since the
invention of the laser allowed its first observation~\cite{Chiao1964}, 
the phenomenon of stimulated Brillouin scattering (SBS) may only now be
entering its golden age.  At its simplest, SBS refers to the stimulated
interaction between a pair of coherent optical waves and a resonant hypersonic
acoustic wave.  
SBS has traditionally been  encountered as the 
scattering of an optical pump beam into a backward traveling Stokes 
beam of slightly lower frequency by an acoustic wave
oscillating at the optical beat frequency.
The acoustic wave is generated by the process of
electrostriction~\cite{BookBoydNonlinearOptics,Shen1965}, and both the Stokes and
acoustic wave grow by a process of positive feedback. 
In optical fiber, this process can be highly efficient and is
often described as the ``strongest'' fiber nonlinearity.
This can be problematic, as SBS prevents the propagation
of high power narrow-bandwidth pumps. Nevertheless, 
SBS in fibers has long provided a mechanism for  producing narrow 
linewidth lasers and amplifiers~\cite{Kobyakov2010,Debut2000,Abedin2012},
filters and other spectral components 
for microwave photonics~\cite{Loayssa2006,Vidal2007,Zhang2012,Marpaung2013}, 
as well as various sensors~\cite{Horiguchi1995,Bao2009}.

Like most optical nonlinearities, the emergence of sub-wavelength scale
waveguides with strong confinement and tunable dispersion has greatly increased 
the efficiency, utility and reach of stimulated Brillouin processes.  
This includes the generation of frequency combs by cascaded Brillouin generation in
 microstructured small-core fibers in both backward~\cite{Dainese2006} and 
forward~\cite{Kang2009} configurations, slow and fast light effects~\cite{Thevenaz2008},
and novel microstructured fiber lasers~\cite{Tow2012,Kabakova2014}.

Motivated by such studies, the development of on-chip SBS in highly-nonlinear
\emph{integrated} waveguides has recently been pursued
aggressively~\cite{Kabakova2015}.  Attaining efficient on-chip SBS is complicated
by the requirement of simultaneous confinement of both the optical and acoustic fields.  
This is non-trivial because optically dense materials suitable for optical
waveguide cores are commonly mechanically stiff and therefore susceptible to leakage of
the acoustic wave into softer substrates; the silicon on silica system is an important example.  
Consequently, on-chip SBS was first achieved~\cite{Pant2011} 
in rib waveguides made from nonlinear chalcogenide glasses, which combine high refractive index and
nonlinearity with relative mechanical softness.  Subsequently, SBS in silicon waveguides has been
observed in Si/SiN membranes~\cite{Shin2013} and elevated rails~\cite{VanLaer2015,Casas-Bedoya2015}, 
which both exploit physical isolation of the waveguide to minimize acoustic losses.
Considerable development will be needed to reach designs suitable for mass-fabrication,
but a practical platform for on-chip SBS would enable numerous applications \cite{Eggleton2013}
in microwave photonics~\cite{Vidal2007,Chin2010,Li2013,Marpaung2013,Morrison2014,Pant2014}, 
sensing, isolators~\cite{Huang2011,Poulton2012} and chip-based lasers~\cite{Kabakova2013a,Hu2014}.

A key driver for developing SBS in sub-micron waveguides was the realization by
Rakich~\etal~\cite{Rakich2010,Rakich2012} that at small scales, there are new
contributions to SBS associated with radiation pressure of light on the
waveguide boundaries, and the back-action of ``moving boundaries'' on the optical
field~\cite{Laer2014}. Depending on the particular waveguide configuration and combination
of optical modes, these contributions can either reinforce or counteract the more familiar
bulk contributions from electrostriction and photoelasticity~\cite{Rakich2010}.
As well, waveguides in which the acoustic fields are strongly confined
can enhance the scattering efficiency of near-stationary quasi-transverse
acoustic waves, a requirement for efficient forward SBS where the pump and Stokes
wave co-propagate~\cite{Kang2009,Rakich2012}.

Following this realization there was some variation in the literature as to
how best to incorporate the new effects into a coupled mode theory self-consistently.
In conventional SBS, electrostriction (the mechanical stress induced by the optical field) is
accompanied by the complementary process of photoelasticity (the change
in the dielectric response induced by the acoustic strain)~\cite{BookBoydNonlinearOptics}.  The two processes are
captured by identical coupling terms in the coupled mode theory, as
required by the Manley-Rowe relations~\cite{Wolff2015,VanLaer2015}.
One should expect the same symmetry between the effect of radiation pressure 
on the waveguide boundaries driving the acoustic field, and the reverse
effects of moving boundaries on the optical field.
However, the initial formulations in terms of optical forces and the Maxwell stress
tensor led to some confusion about whether certain additional coupling terms arise 
or whether they are essentially ``double-counting''.  
These include concepts of an electrostrictive ``boundary pressure''~\cite{Shin2013}
and bulk contributions to the radiation pressure term~\cite{Rakich2010,Rakich2012}.  

Recently, our group provided a new derivation of the coupled mode equations~\cite{Wolff2015} 
that avoids the formalism of optical forces.  Instead we 
used thermodynamic arguments to unambiguously identify the correct coupling term
describing both radiation pressure and moving boundaries effects.
We found that this term indeed matches expressions suggested
in several papers~\cite{Rakich2010,Qiu2013} and the matter seems to be resolved.  
It turns out, for instance, that the appearance of an electrostrictive boundary
pressure depends on whether electrostriction is viewed in terms of a stress
or a force density~\cite{Wolff2015}.  Nevertheless, the argument establishing the correct form for the moving boundary coupling 
was quite involved, and a simpler, more direct derivation would be desirable.

In this work, we provide such a derivation.  Rather than the standard
approach of applying slowly-varying envelope approximations to the wave equation,
we extend a quantized multimode Hamiltonian formalism of integrated
optical waveguides~\cite{Sipe2004} to include opto-acoustic interactions.  The photoelastic
and radiation pressure couplings are introduced through a single interaction energy term in 
the Hamiltonian, and fully quantum equations of motion are obtained from the Heisenberg
equations. In the classical limit, the coupled mode equations of Wolff~\etal~\cite{Wolff2015} 
emerge naturally and simply, with no ambiguity about double-counting.

There are number of other advantages to our approach.
The fundamental quantum process
underlying SBS---the stimulated decay of a pump photon into a lower energy Stokes 
photon and an acoustic phonon---is manifestly visible in the interaction term.
Further, obtaining quantum equations that respect the appropriate operator commutation relations
provides an important starting point for the investigation of effects at the 
boundary of quantum and classical opto-acoustics.  This is likely to become more important
as the distinction between cavity optomechanics and guided wave opto-acoustics becomes
increasingly blurred~\cite{VanLaer2015a,Li2012}, and phonon confinement 
strategies are improved. The propagation of guided wave acoustic fields with
strongly modified phonon density of states is likely not far off, which raises the 
prospect of Brillouin interactions in the quantum regime.
Finally, the Hamiltonian formalism has proved very powerful for the description of other integrated 
quantum nonlinear processes such as
spontaneous four wave mixing~\cite{Helt2010,Helt2012} and spontaneous
parametric downconversion~\cite{Yang2008}.  

We should note that quantum or analytical dynamics approaches to guided wave
opto-acoustics themselves have some pedigree. 
Hamiltonian approaches to SBS date back to the first rigorous treatment 
by Shen and Bloembergen~\cite{Shen1965} who gave an analysis for plane waves in the semi-classical
limit.
Drummond and Corney incorporate Raman
gain into their quantized theory of nonlinear fiber propagation~\cite{Drummond2001}.
In that case, the Raman response, which depends on the detailed glass composition and network,
 is introduced through a phenomenological measured response function. In contrast, for 
the Brillouin couplings we consider here, the phonon response is entirely determined by the bulk  
elastic properties and waveguide geometry, and so can be calculated
exactly using acoustic mode solvers.
Finally, van Laer \etal~\cite{VanLaer2015a} have recently discussed 
the connections between quantum optomechanics for single or few resonator 
systems and classical SBS~\cite{VanLaer2015a}.  They identify an elegant connection between
the opto-acoustic coupling in waveguide SBS and the corresponding coupling in 
quantum optomechanical systems. The latter is treated with 
a single mode Hamiltonian approach which is appropriate for
the optomechanics of a resonator consisting of a single cavity,
but limits its application to longer structures with
continuous phonon spectra.  In contrast, for the waveguide problem that is our focus, 
a full multi-mode treatment is appropriate in order to treat arbitrary input optical fields.

The paper is structured as follows.
In \refsec{sec:hamilacoustics} we construct a Hamiltonian description of acoustics, including useful
expressions for group velocity and power flow.
In \refsec{sec:hamilem} we review some necessary results from guided wave electromagnetic quantization.
In \refsec{sec:hamilall}, which is the core of the paper, we provide the full optoacoustic Hamiltonian,
find expressions for the coupling terms, and show directly how the symmetry between radiation pressure and moving boundary effects emerges.
In \refsec{sec:quantumcme} we derive quantum coupled mode equations for the system, make connections to prior expressions for the coupling strength,
and recover the classical coupled mode equations for SBS.
Finally, in \refsec{sec:discussion} we discuss directions for future examination
including the important issue of phonon dissipation.
A comprehensive supplementary materials document provides detailed derivations of many of the results.

\section{Hamiltonian formulation of guided wave acoustics }
\label{sec:hamilacoustics} 
The classical theory of guided elastic waves is of
course very mature and can be formulated in many guises. Auld~\cite{Auld1990} provides
an excellent introduction for readers with an optics background. Since our goal
is a Hamiltonian operator representing the complete opto-acoustic system we
begin by re-framing guided acoustic wave propagation in a quantum Hamiltonian
picture; we have not found such a formulation in the literature.

For purely classical applications, one might build a Hamiltonian 
from which the dynamics are determined by Hamilton's equations.
For generality, we construct the theory in a quantum form, using
canonical quantization with commutators that follow from
the standard association with the Poisson brackets of the classical 
formulation:
\begin{equation}
\frac{1}{i\hbar }\left[ \;, \;\right] \Leftrightarrow \left\{ \;, \;\right\} .
\end{equation}

\subsection{Hamiltonian operator} \label{sec:hamilelastic}

To identify a classical theory suitable
for canonical quantization we should begin with canonical variables, which will
become the canonical operators in the quantum theory, and a classical
Hamiltonian that both yields the standard equations of motion in the
form of the elastic wave equation, and is numerically equal to the classical
energy of the system.
To that end we introduce vector field
variables $\vu(\vecr)$  describing the displacement 
and $\vpi(\vecr) $ as their conjugate momenta;
these will become operators in the quantum theory, although we will
not explicitly include ``hats'' in our notation.

\begin{figure}
\centering
\includegraphics[width=8cm]{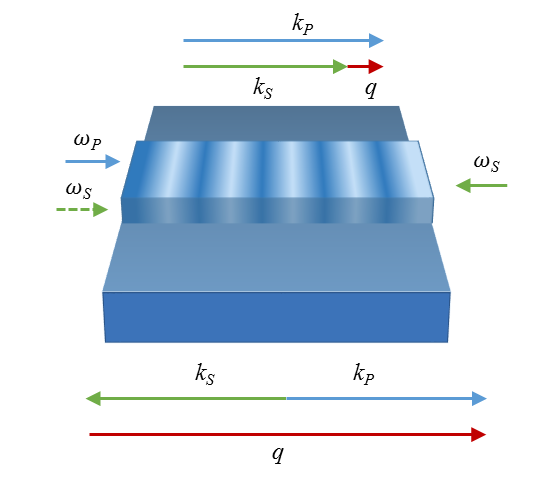}
\label{fig:wguide}
\caption{ Schematic geometry of the interacting waves in integrated
SBS.  Arrows at the top denote wavenumbers for forward SBS; arrows  
at the bottom denote wavenumbers for backward SBS.} 
\end{figure}

Following standard quantum mechanics, we naturally choose the commutation relations 
\begin{equation}
\left[ u^{n}(\vecr),\pi^{m}(\vecrp)\right] =i\hbar
\delta^{nm}\delta (\vecr-\vecrp),  \label{canonical2}
\end{equation}
where superscripts stand for Cartesian components.  The obvious acoustic Hamiltonian operator is 
\begin{equation}
\hamilA=\int \frac{\pi^i (\vecr)\pi^i (\vecr)}{2\rho (\vecr)}\drrr 
+\frac{1}{2}\int S^{ij}(\vecr)c^{ijkl}(\vecr)S^{kl}(
\vecr) \, \drrr , \label{Hamiltonian}
\end{equation}
where the integration is over all space.
Here $\rho (\vecr)$ and $c^{ijkl}( \vecr)$ are position-dependent c-number quantities
representing the density and stiffness tensor respectively, while
\begin{equation} \label{eq:straindef}
S^{ij}(\vecr)=\frac{1}{2}\left( \frac{\partial u^i (\vecr)}{
\partial r^j }+\frac{\partial u^j (\vecr)}{\partial r^i }\right) ,
\end{equation}
is the strain tensor operator~\cite{Auld1990}, and again we emphasise that
unless stated otherwise, $\vu(\vecr)$ and $\vpi(\vecr)$ are to be read as operators. 
In~\eqref{Hamiltonian} and throughout, repeated indices are to be
summed over the Cartesian coordinates $x,y,z$.
We will formally
assume these quantities are continuous functions of position, although they may change
drastically as one moves from solid to air, for example. The limit to
these functions changing discontinuously can be taken at the end of the
calculation when matrix elements and the like are evaluated, but in
all dynamical equations and derivations $\rho (\vecr)$ and $ c^{ijkl}(\vecr)$ 
should be taken as continuous functions. 

Let us comment on the precise meaning of the operators $\vu$ and $\vpi$.
In elastic theory, the displacement field applies to ``volume elements'' 
much larger than the atomic scale, but much smaller than
the characteristic wavelength of any acoustic excitation. 
Therefore, the displacement and momentum operators do not correspond 
directly to any individual physical oscillator
but describe collective excitations of the mesoscopic bulk medium. 
Nevertheless the low-energy phonon excitations that emerge from the theory 
are very real and their quantization is physical.
For example, completeness relations are correct up to an appropriate wavevector
cutoff, far above the typical wavevector of any Brillouin-induced excitation.

The strain tensor~\eqref{eq:straindef} is obviously symmetric in its two indices, 
and since the stiffness tensor appears with two strain tensors in (\ref{Hamiltonian}), 
$c^{ijkl}(\vecr)$ can be taken to
satisfy $c^{ijkl}(\vecr)=c^{ijlk}(\vecr),$ as well as $c^{ijkl}(
\vecr)=c^{klij}(\vecr)$.
In all then~\cite{LandauLifshitzElasticity} ,
\begin{equation}\label{tensorsymmetry}
c^{ijkl}(\vecr)=c^{jikl}(\vecr)=c^{ijlk}(\vecr), %
\end{equation}
so we can write (\ref{Hamiltonian}) as simply 
\begin{equation}
\hamilA=\int \frac{\pi^i (\vecr)\pi^i (\vecr)}{ 2\rho (\vecr)}\, \drrr
+\frac{1}{2}\int \frac{ \partial u^i (\vecr)}{\partial r^j} 
 c^{ijlm}(\vecr)\frac{\partial u^l (\vecr)}{
\partial r^m}\, \drrr .
\end{equation}
Evaluating the Heisenberg equations of motion
(see supplementary material section~\ref{supp:heisenbergacoustic})
\begin{subequations}\label{eq:heisacoustic}
\begin{eqnarray}
i\hbar \frac{\partial }{\partial t}u^{n}(\vecr,t) &=&\left[ u^{n}(
\vecr),\hamilA\right] , \\
i\hbar \frac{\partial }{\partial t}\pi^{n}(\vecr,t) &=&\left[ \pi^{n}( \vecr),\hamilA\right] ,
\end{eqnarray}
\end{subequations}
yields 
\begin{subequations}\label{eq:eomacoustic}
\begin{eqnarray}
\frac{\partial }{\partial t}u^{n}(\vecr,t) &=&\frac{\pi^{n}(\vecr ,t)}{\rho (\vecr)},  \label{solution} \\
\frac{\partial }{\partial t}\pi^{n}(\vecr,t) &=&\frac{\partial }{ \partial r^j }
      \bigg( c^{njkl}(\vecr)S^{kl}(\vecr)\bigg) . 
\end{eqnarray}
\end{subequations}
Classically these equations indicate, as expected, that the conjugate momentum
is the product of the mass density  and the velocity field $\dot{u}^n$, 
and that the momentum evolves according to Newton's laws.

From~\eqref{eq:eomacoustic},
we recover the standard acoustic wave equation~\cite{Auld1990}:
\begin{equation}
\rho (\vecr)\frac{\partial^{2}u^{n}(\vecr,t)}{\partial t^{2}}=
\frac{\partial }{\partial r^j }\bigg( c^{njkl}(\vecr)S^{kl}(\vecr)\bigg) .  \label{dynamics}
\end{equation}
Thus we have 
the desired equations of motion in both the classical and quantum
descriptions.  Moreover, substituting the relation (\ref{solution})
into (\ref{Hamiltonian}) and taking the classical limit by dropping the operator 
nature of $\vu$ and $S^{ij}$, we recover the classical Hamiltonian
\begin{eqnarray*}
\hamilA &\rightarrow &\frac{1}{2}\int \rho (\vecr)\left( \frac{\partial \vu(
\vecr,t)}{\partial t}\right)^{2}\drrr
+\frac{1}{2}\int S^{ij}(\vecr)c^{ijkl}(\vecr)S^{kl}(\vecr) \, \drrr \\
&=&T+V.
\end{eqnarray*}
This is clearly numerically equal to 
the sum of kinetic and potential energy~\cite{Auld1990}, and so (\ref{Hamiltonian}) meets
all the criteria to be
an appropriate Hamiltonian for quantization.
Note that at this point, no acoustic dissipation (viscosity) has been included.
Direct incorporation of losses is less straightforward in Hamiltonian approaches than in Lagrangian
approaches, in which a Rayleigh dissipation function can be introduced.
The phonon loss is important to the physics of SBS, but can be incorporated later perturbatively.
We discuss this in \refsec{sec:discussion}.

\subsection{Modes and new fields}

We now seek to reduce the Hamiltonian to standard harmonic oscillator form;
the low energy states will describe the quasi-particle or phonon excitations
of the medium.
To diagonalise the acoustic Hamiltonian 
we look for classical solutions of (\ref{dynamics}) of the form 
\begin{equation}
\vu(\vecr,t)= \vU_{\Lambda }(\vecr)\me^{-i\Omega_{\Lambda }t}+\cc,
\end{equation}
where $\Lambda$ is a mode index.
This implies
\begin{equation}
-\frac{
\partial }{\partial r^j }\left( c^{njkl}(\vecr)\frac{\partial
U_{\Lambda }^k (\mathbf{r)}}{\partial r^l }\right)   \label{EV}
=
\rho (\vecr)\Omega_{\Lambda }^{2}U_{\Lambda }^{n}(\vecr),
\end{equation}
where again we have used the fact that $c^{njkl}(\vecr)$ is symmetric
in its last two indices. 
It is convenient however, 
if the linear operator generating mode functions
is Hermitian, which the form in Eq.~\eqref{EV} is not. 
To obtain a Hermitian system, we introduce new fields
\begin{eqnarray} \label{newfields} 
\tilde{\vu}(\vecr)&=&\sqrt{\rho(\vecr)}\, \vu(\vecr) \\
\tilde{\vpi}(\vecr) &=&\frac{\vpi (\vecr)}{\sqrt{\rho(\vecr)}},
\nonumber
\end{eqnarray}
which clearly preserve the commutation relations. That is,
\begin{equation}
\left[ \tilde{u}^{n}(\vecr),\tilde{\pi}^{m}(\vecrp)\right]
=i\hbar \delta^{nm}\delta (\vecr-\vecrp). 
 \label{canonicaltilde}
\end{equation}
We now look for mode solutions in terms of the new fields in the form 
\begin{subequations}\label{eq:newmodesoln}
\begin{eqnarray} 
\tilde{\vu}(\vecr,t)&=&\mathbf{\tilde{U}}_{\Lambda }(\vecr)\me^{-i\Omega_{\Lambda }t}+\cc , \\
\tilde{\vpi}(\vecr,t)&=&\mathbf{\tilde{\Pi}}_{\Lambda }(\vecr)\me^{-i\Omega_{\Lambda }t}+\cc 
\end{eqnarray}
\end{subequations}
We introduce the operator $\mathcal{M}^{nk}$ which acts on a general vector function  
$\mathbf{C}(\vecr)$ as 
\begin{equation} \label{Moperator} 
\mathcal{M}^{nk}(\vecr)C^k (\vecr)
=-\frac{1}{\sqrt{\rho(\vecr)}} \frac{\partial }{\partial r^j }
\left( c^{njkl}(\vecr)\frac{\partial }{\partial r^l } 
\left(\frac{C^k(\vecr)}{\sqrt{\rho(\vecr)}} \right) \right) .
\end{equation}
\refsec{app:MHermitian} of the supplementary material
shows that $\mathcal{M}^{nk}(\vecr)$ is indeed Hermitian.
Then, using integration by parts we can write the Hamiltonian \eqref{Hamiltonian} 
in terms of the new fields (\ref{newfields}) in the simple form
\begin{eqnarray} \label{Hwork} 
\hamilA &=&
\frac{1}{2}\int \tilde{\pi}^i (\vecr)\tilde{\pi}^i (\vecr) \, \drrr 
+\frac{1}{2}\int \tilde{u}^i (\vecr)\mathcal{M}^{ik}(\mathbf{
r})\tilde{u}^k (\vecr)\, \drrr  .
\end{eqnarray}
As the basis of our acoustic modal expansion, Eq.~\eqref{Moperator} plays an analogous role
to that which the vector Helmholtz equation (commonly known as the ``master equation'' in the photonic
crystal literature~\cite{Joannopoulos2010Book},) plays in the quantization of the 
electromagnetic field (see~\eqref{eq:vechelmholtz} below).
Its Hermitian form is useful both for developing the formalism but also for formulating
mode-solving algorithms based on energy functional minimization~\cite{Joannopoulos2010Book}.

Now if we construct eigenfunctions $\vF _{\Lambda }(\vecr)$ of $\mathcal{M}^{nk}$ such that
\begin{equation}\label{eq:mcaleig} 
\mathcal{M}^{nk}(\vecr)F_{\Lambda }^k (\vecr)=\Omega_{\Lambda }^{2}F_{\Lambda }^{n}(\vecr),
\end{equation}
it follows from \eqref{EV} and \eqref{newfields} that we can take
\begin{eqnarray}
\mathbf{\tilde{U}}_{\Lambda }(\vecr) &=&\vF _{\Lambda }(\vecr
),  \label{tildes} \\
\mathbf{\tilde{\Pi}}_{\Lambda }(\vecr) &=&-i\Omega_{\Lambda }\vF 
_{\Lambda }(\vecr) , \nonumber
\end{eqnarray}
where the second equation follows from substituting 
\eqref{eq:newmodesoln} in~\eqref{solution}.

From the Hermiticity of $\mathcal{M}^{nk}$,
the eigenfunctions $\vF _{\Lambda }(\vecr)$ with
different eigenvalues will automatically be orthogonal, and if there are
eigenfunctions with the same eigenvalue we can construct them to be orthogonal. 
Thus we can write 
\begin{equation} \label{vnorm}
\int \vF _{\Lambda }^*(\vecr)\cdot \vF _{\Lambda
^\prime }(\vecr)\, \drrr=\delta_{\Lambda \Lambda^\prime }.
\end{equation}
Here the integration is over a finite volume that, in the end, can be allowed to pass
to infinity if we wish to generate a continuum of eigenfunctions. 
We take the set of eigenfunctions with positive eigenvalues $\Omega
_{\Lambda }^{2}$ as complete, 
\begin{equation} \label{complete}
\sum_{\Lambda }F_{\Lambda }^{n}(\vecr)\left( F_{\Lambda }^{m}(\vecr
^\prime )\right)^*=\delta^{nm}\delta (\vecr-\vecrp),
\end{equation}
at least for the problems of interest. For each $\Omega_{\Lambda }^{2}$
we choose a positive $\Omega_{\Lambda }$ and take it as the frequency of
the eigenfunction.  Some additional properties of the mode
functions that are required for reduction of the acoustic
Hamiltonian to canonical harmonic oscillator form are developed in \refsupp{supp:partners}.

Following similar arguments to those used in quantization of the vacuum 
electromagnetic field~\cite{Grynberg2010book} (or the electromagnetic field in nondispersive 
media~\cite{Sipe2004}), one can show using~\eqref{complete} that introducing new 
operators $b_\Lambda$ and $b^\dagger_\Lambda$ with commutation relations
\begin{eqnarray}
\left[ b_{\Lambda },b_{\Lambda^\prime }\right] &=&0,  \label{acomm} \\
\left[ b_{\Lambda },b_{\Lambda^\prime }^\dagger \right] &=&\delta
_{\Lambda \Lambda^\prime },  \nonumber
\end{eqnarray}
we can preserve the commutation relations~\eqref{canonicaltilde},
by expanding the fields as 
\begin{subequations}
\label{eq:upiexpansion}
\begin{align} 
\mathbf{\tilde{u}(r}) &=\sum_{\Lambda }\sqrt{\frac{\hbar }{2\Omega_{\Lambda }}}b_{\Lambda }
\vF _{\Lambda }(\vecr)+\hc , \\
\tilde{\vpi}(\vecr) &=-i\sum_{\Lambda }\sqrt{\frac{\hbar \Omega_{\Lambda }}{2}}b_{\Lambda }
\vF _{\Lambda }(\vecr)+\hc  ,
\end{align}
\end{subequations}
where $\hc$ is the Hermitian conjugate.
Then, through a series of manipulations using properties of the $F_\Lambda(\vecr)$, 
the Hamiltonian can be reduced to the canonical harmonic oscillator form
\begin{eqnarray} \label{eq:hamil}
\hamilA &=&\sum_{\Lambda }\hbar \Omega_{\Lambda } b_{\Lambda }^{\dagger
}b_{\Lambda },
\end{eqnarray}
where we have dropped the zero point energy which has no dynamical effect.
Derivations of Eqs.~\eqref{eq:upiexpansion} and~\eqref{eq:hamil} are provided in
\refsupp{supp:acousticexpansion}.

Finally, returning to the physical modes by introducing
\begin{subequations} \label{eq:physmodes}
\begin{align}
\vU_{\Lambda }(\vecr)&=\frac{1}{\Omega_{\Lambda }\sqrt{\rho (
\vecr)}}\vF _{\Lambda }(\vecr) ,  \\
\mathbf{\Pi }_{\Lambda }(\vecr)&=-i\sqrt{\rho (\vecr)}\vF 
_{\Lambda }(\vecr), 
\end{align}
\end{subequations}
we can write the full displacement and momentum field operators as 
\begin{subequations}\label{finalwork2} 
\begin{eqnarray}
\vu(\vecr) &=&\sum_{\Lambda }\sqrt{\frac{\hbar \Omega_{\Lambda }}{2}}
b_{\Lambda }\vU_{\Lambda }(\vecr)+\hc,  \\ 
\vpi(\vecr) &=&\sum_{\Lambda }\sqrt{\frac{\hbar \Omega_{\Lambda }}{2}}
b_{\Lambda }\mathbf{\Pi }_{\Lambda }(\vecr)+\hc,  
\end{eqnarray}
\end{subequations}
with prefactors in the expansions reminiscent of simple harmonic oscillator
physics.

\subsection{Waveguide acoustics}
The results to this point apply to any acoustic structure. 
We now specialize to waveguides
running along the $z$ direction, and choose a
box of length $L$ in that direction, which includes all $x$ and $y$. 
Focusing on acoustic modes confined to the
waveguide, we label the modes by a wavenumber $q=2\pi n/L$, for integer $n$, 
and a band $\alpha $ which identifies the transverse spatial mode structure
in the $xy$ plane. Then the eigenfunctions of the previous section can be written
\begin{equation}
\vF _{\alpha q}(\vecr)=\frac{\mathbf{f}_{\alpha q}(x,y)\eiqz}{
\sqrt{L}},  \label{modeintro}
\end{equation}
with the normalization~\eqref{vnorm} guaranteed by requiring
\begin{equation}
\int \dx\dy\, \mathbf{f}_{\alpha q}^*(x,y)\cdot \mathbf{f}_{\alpha q}(x,y)=1,  \label{normalization}
\end{equation}
where the integration is over the whole $x$-$y$ plane.
The derived mode amplitudes~\eqref{eq:physmodes} are then of the form 
\begin{eqnarray*} \label{eq:modamp}
\vU_{\alpha q}(\vecr) &=&\frac{\vu_{\alpha q}(x,y)\eiqz
}{\sqrt{L}}, \\
\mathbf{\Pi }_{\alpha q}(\vecr) &=&\frac{\vpi_{\alpha q}(x,y)\eiqz}{\sqrt{L}},
\end{eqnarray*}
with 
\begin{subequations}\label{lrelate1}
\begin{align}
\vu_{\alpha q}(x,y) & =\frac{1}{\Omalq \sqrt{\rho (x,y)}} \, \vfalq(x,y),   \label{lrelate1a} \\
\vpi _{\alpha q}(x,y)& =-i\sqrt{\rho (x,y)}\, \vfalq(x,y).  %
\end{align}
\end{subequations}
Note that the normalization condition (\ref{normalization}) can be written, 
using (\ref{lrelate1a}), as 
\begin{equation}\label{eq:unorm}
\Omega^2_{\alpha q}\int \dx\dy\, \rho (x,y)\, \vu_{\alpha q}^*(x,y)\cdot \vu_{\alpha q}(x,y) =1  .
\end{equation}

If we now let $L\rightarrow \infty $, moving to a continuous distribution of modes, the
commutation relations, Hamiltonian and field operator expansions respectively become 
\begin{eqnarray*}
\left[ b_{\alpha q},b_{\alphap \qp}\right]  &=&0, \\
\left[ b_{\alpha q},b_{\alphap \qp}^\dagger \right] 
&=&\delta_{\alpha \alpha^\prime }\delta (q-\qp),
\end{eqnarray*}
\begin{equation}
\hamilA=\sum_{\alpha }\int \dq  \,\hbar \Omalq b_{\alpha q}^{\dag }b_{\alpha q},
\end{equation}
\begin{eqnarray}
\label{eq:uexp}
\vu(\vecr) &=&\sum_{\alpha }\int \frac{\dq}{\sqrt{2\pi}}  \sqrt{\frac{\hbar \Omalq }{
2 }}b_{\alpha q}(t) \vu_{\alpha q}(x,y)\, \eiqz+\hc,  \label{uandpi} \\
\vpi(\vecr) &=&\sum_{\alpha }\int \frac{\dq}{\sqrt{2\pi}}  \sqrt{\frac{\hbar \Omalq 
}{2 }}b_{\alpha q}(t) \vpi_{\alpha q}(x,y)\, \eiqz+\hc,  \nonumber
\end{eqnarray}
while \eqref{normalization}--\eqref{eq:unorm} remain unchanged.
Here the $q$ integrals are over the range of $q$ for which each mode $\alpha$ exists, taking
account of any modal cutoffs. 
Normally we work in the Heisenberg regime so the $b_{\alpha q}$ are time-dependent.
Note our convention that while $\vu$ and $\vpi$ are operators, the mode functions
$\vu_{\alpha q}$ and $\vpi_{\alpha q}$, which carry modal index subscripts, are c-number quantities.

\subsection{Envelope functions}

To make the connection to the more familiar waveguide representation of slowly-varying envelopes,
we now introduce envelope functions for each type of acoustic mode.
We assume the excitation is centered at some wavenumber $q_\so $ and factor out that
dependence to produce a function varying slowly in space. However we
retain the full time-dependence in the operators $b_{\alpha q}(t)$ writing
\begin{equation}
\phi_{\alpha }(z,t)=\int \frac{\dq  }{\sqrt{2\pi }}b_{\alpha
q}(t)\me^{i(q-q_\so )z} . \label{phononenvelope}
\end{equation}
If the integral in (\ref{phononenvelope}) is taken to range over all $q$ it
is easily checked that we obtain the canonical equal-time continuous commutators
\begin{equation}\label{eq:phicomm}
\left[ \phi_{\alpha }(z,t),\phi_{\alphap }^\dagger (z^\prime,t )
\right] =\delta_{\alpha \alpha^\prime }\delta (z-z^\prime ).
\end{equation}
In reality, the range of integration in~\eqref{phononenvelope} is restricted by modal cutoffs, 
which will temper the Dirac delta function in~\eqref{eq:phicomm}.
However, assuming spectrally narrow envelopes far from any cut-offs, Eq.~\eqref{eq:phicomm} should
normally be an excellent approximation. From (\ref{uandpi}) we see that if $\Omega
_{\alpha q}$, $\vu_{\alpha q}(x,y)$, and $\vpi_{\alpha q}(x,y)$ 
vary little over the range of significant $q$ we can approximate 
the field expansions in terms of envelope functions centered at $q=q_\so$:
\begin{eqnarray*}
\vu(\vecr) &\simeq &\sum_{\alpha q_\so } \me^{iq_\so z}
\sqrt{\frac{\hbar \Omega_{\alpha q_\so } }{2}}\vu_{\alpha q_\so }(x,y)
\int \frac{\dq  }{\sqrt{2\pi }} b_{\alpha q}(t)\me^{i(q-q_\so )z}+\hc, \\
 &= & \sum_{\alpha q_\so} \me^{iq_\so z}\sqrt{\frac{\hbar \Omega_{\alpha q_\so } }{2}}\vu_{\alpha q_\so }(x,y)\, \phi_{\alpha }(z,t)+\hc, \\
\vpi(\vecr) &\simeq &\sum_{\alpha q_\so} \me^{iq_\so z} \sqrt{\frac{\hbar \Omega_{\alpha
q_\so }}{2}}\vpi _{\alpha q_\so}(x,y) \int \frac{\dq  }{\sqrt{2\pi 
}}b_{\alpha q}(t)\me^{i(q-q_\so )z}+\hc \\
&= &\sum_{\alpha q_\so} \me^{iq_\so z} \sqrt{\frac{\hbar \Omega_{\alpha q_\so }}{2}}\vpi_{\alpha q_\so}(x,y)\, \phi_{\alpha }(z,t)+\hc.
\end{eqnarray*}
Corrections can be included by expanding the prefactors 
$\sqrt{\hbar \Omega_{\alpha q }/2}\, \vu_{\alpha q }(x,y)$ and 
$\sqrt{\hbar \Omega_{\alpha q }/2}\, \vpi_{\alpha q }(x,y)$ about $q=q_o$, and combining the resulting
powers of $(q-q_o)$ with the $\exp(i(q-q_o)z)$ in the integrals over $q$ to yield
expressions involving derivatives of $\phi_\alpha(z,t)$, but we neglect them here.

Returning to Eq.~\eqref{phononenvelope},
we can derive approximate equations of motion for the $\phi_{\alpha}(z,t)$ 
in the Heisenberg picture by expanding the dispersion relation $\Omalq $ 
of mode $\alpha $ about $ q_\so $:
\begin{align} 
\frac{\partial \phi_\alpha}{\partial t} & = \frac{1}{i \hbar}[\phi_\alpha, \hamilA] \nonumber \\
&= \frac{1}{i\hbar} \int \frac{\dq}{\sqrt{2\pi}} \hbar \Omega_q \me^{i(q-q_\so)z} b_{\alpha q}(t) \nonumber \\
&= -i \int \frac{\dq}{\sqrt{2\pi}}  
\left(\Omega_{\alpha q_0} + v_{\alpha q_\so }(q-q_0) + \frac{1}{2}v_{\alpha q_\so }^\prime (q-q_0)^2 +\ldots\right) \me^{i(q-q_\so)z} b_{\alpha q}(t) \nonumber \\
&\approx -i \Omega_{\alpha q_0} \phi_{\alpha}(z,t) 
-v_{\alpha q_\so }\frac{\partial \phi_{\alpha }(z,t)}{\partial z}+
\frac{1}{2}iv_{\alpha q_\so }^\prime \frac{\partial^{2}\phi_{\alpha }(z,t)
}{\partial t^{2}}+\ldots, \label{commHA}
\end{align}
where 
\begin{subequations}
\begin{eqnarray}
v_{\alpha q_\so } &=&\left( \frac{\dd\Omalq }{\dq  }\right)_{q=q_\so }, \label{eq:valqso}
\\
v_{\alpha q_\so }^\prime  &=&\left( \frac{\dd^{2}\Omalq }{\dq ^{2}}
\right)_{q=q_\so },
\end{eqnarray}
\end{subequations}
etc. An expression for the group velocity $v_{\alpha q}$ 
in terms of the modal fields is worked out in \refsupp{app:acousticvg}.

\subsection{Acoustic powers}
Finally, we establish some expressions concerning 
the power carried by the acoustic modes in the envelope
representation.  
Classically, the acoustic power density at a point in the medium in a direction $\unitn$ is given by 
\begin{equation}
\powdensA_\text{cl} \cdot \unitn = -\frac{\partial u^i (\vecr)}{\partial t}c^{ijkl}(\vecr)S^{kl}(\vecr)n^j , 
\end{equation}
which has the natural interpretation of power being the dot product of an applied force and the
velocity of the point of application~\cite{Auld1990}.
Using~\eqref{eq:eomacoustic}, the power carried by a waveguide mode in the $\unitz$ direction is thus
\begin{equation}
\powA_\text{cl}(z)=-\int \dx\dy\,\frac{\pi^i (\vecr)}{\rho (x,y)}c^{izkl}(x,y)S^{kl}(
\vecr).
\end{equation}
To construct a quantum
operator corresponding to the power density we use the symmetrized form
of the non-commuting operators $\vpi(\vecr)$ and $S^{kl}(\vecr)$, in the usual way, as
shown in \refsupp{app:acousticpower}. This leads to the
result that the operator for the power carried by the acoustic field is
\begin{equation}
\powA(z)=\sum_{\alpha ,\alphap }\int \frac{\dq  \dq ^\prime }{2\pi }
b_{\alphap \qp}^\dagger(t) b_{\alpha q}(t)\me^{i(q-q^\prime )z}\pmodeA_{\alphap \alpha}(\qp,q).  \label{power}
\end{equation}
Here the contribution from the pair of modes $\alpha$ and $\alpha'$ at wavenumbers
$q'$ and $q$ is given by
\begin{eqnarray}
\pmodeA_{\alphap \alpha}(\qp,q)  %
&=&\frac{\hbar }{2}\sqrt{\frac{\Omega_{\alphap \qp}}{
\Omalq }}q\int \dx\dy\,\frac{c^{izkz}(x,y)}{\rho (x,y)}\left(
f_{\alphap \qp}^i (x,y)\right)^*f_{\alpha
q}^k (x,y)  \nonumber \\
&&+\frac{\hbar }{2}\sqrt{\frac{\Omalq }{\Omega_{\alpha^\prime \qp}}}\qp\int \dx\dy\,\frac{c^{izkz}(x,y)}{\rho (x,y)}\left(
f_{\alphap \qp}^k (x,y)\right)^*\left( f_{\alpha
q}^i (x,y)\right)  \nonumber \\
&&-\frac{i\hbar }{2}\sqrt{\frac{\Omega_{\alphap \qp}}{
\Omalq }}\int \dx\dy\,\frac{\left( f_{\alphap q^\prime }^i (x,y)\right)^*}{\sqrt{\rho (x,y)}}c^{izkl}(x,y)\left( \frac{
\partial }{\partial r^l }\left( \frac{\falq^k (x,y)}{\sqrt{\rho
(x,y)}}\right) \right)  \nonumber \\
&&+\frac{i\hbar }{2}\sqrt{\frac{\Omalq }{\Omega_{\alpha^\prime \qp}}}\int \dx\dy\,\frac{\left( \falq^i (x,y)\right) }{\sqrt{
\rho (x,y)}}c^{izkl}(x,y)\left( \frac{\partial }{\partial r^l }\left( \frac{
f_{\alphap \qp}^k (x,y)}{\sqrt{\rho (x,y)}}\right) \right)
^*,  \nonumber
\end{eqnarray}
where we have used \eqref{eq:modamp} to introduce the modified mode functions 
$\vf_{\alpha,q}$.

For a phonon field involving only one mode $\alpha$ and 
assuming we can neglect the $q$ dependence of $p_{\alpha\alpha }(q,\qp)$ 
over the pulse spectrum we obtain the slowly-varying power operator as 
\begin{eqnarray}
\powA_{\text{sv}}(z) &\simeq& \int \frac{\dq  \dqp}{2\pi }b_{\alpha \qp}^\dagger (t)b_{\alpha q}(t)
\me^{i(q-\qp)z}\pmodeA_{\alpha\alpha}(q_\so ,q_\so )
\label{powerwork} \\
&=&\pmodeA_{\alpha\alpha}(q_\so ,q_\so )\phi_{\alpha}^\dagger (z)\phi_{\alpha}(z).  \nonumber
\end{eqnarray}
Finally, it may be shown (see \refsupp{app:acousticpower}), that 
$ \pmodeA_{\alpha\alpha}(q_\so ,q_\so )=\hbar\Omega_{\alpha q_\so} v_{\alpha q_\so}$,
so that the power carried by the acoustic envelope is
\begin{equation} \label{eq:powAsv}
\powA_{\text{sv}}(z)= \hbar \Omega_{\alpha q_\so }\, v_{\alpha q_\so }\, 
\phi_{\alpha}^\dagger (z)\phi_{\alpha}(z),
\end{equation}
and it follows that in this limit, $ \phi_{\alpha}^\dagger (z)\phi_{\alpha}(z)$
has the natural interpretation of a phonon number density operator.

\section{Quantization of the electromagnetic fields } \label{sec:hamilem}

To construct the full opto-acoustic Hamiltonian we will need similar
results for the quantization of the electromagnetic field
in integrated structures. The procedure is well known 
and we simply summarize some essential results~\cite{Sipe2004}.

\subsection{Hamiltonian and modes}
As the fundamental quantum fields we take the electric displacement $\vD\left(\vecr\right) $ 
and magnetic $\mathbf{B}\left( \vecr\right) $ fields
with commutation relations
\begin{eqnarray*}
\left[ D^i (\vecr),D^j (\vecrp)\right] &=&\left[ B^i (
\vecr),B^j (\vecrp)\right] =0, \\
\left[ D^i (\vecr),B^j (\vecrp)\right] &=&i\hbar
\epsilon^{ilj}\frac{\partial }{\partial r^l }\delta (\vecr-\vecrp).
\end{eqnarray*}
This choice, which dates back to Born and Infeld~\cite{Born1934} has the advantage that the transversality 
of the two fields is easily imposed.  
The quantization procedure  has been discussed at length
elsewhere~\cite{Sipe2004}, and we simply  quote the necessary.
The parallels to the acoustic problem in the previous section are very apparent.

The electromagnetic Hamiltonian operator is taken as
\begin{equation}
\hamilEM =\frac{1}{2\muo}\int B^i (\vecr) B^i (\vecr) \, \drrr+\frac{1}{2\epso}\int D^i (\vecr)\beta_\text{ref}(\vecr)D^i (\vecr) \, \drrr ,  
\end{equation}
where
\begin{equation}
\beta_\text{ref}(\vecr)=\frac{1}{\epsilon_\text{ref}(\vecr)} ,
\end{equation}
describes the ``background'' dielectric response of the waveguide structure in terms
of the relative dielectric constant $\epsilon_\text{ref}(\vecr)$, without any acoustic effects. 
Note that it is straightforward to include a tensor response in $\epsilon_\text{ref}(\vecr)$ but
to reduce cluttering the tensor notation, here we treat the material as optically isotropic.
In principle, we could also extend the treatment to include dispersion of the 
dielectric~\cite{Bhat2006} at the expense of considerably more complexity.
For Brillouin processes, the linewidths of the interacting optical waves are usually narrow, 
and we can safely neglect dispersion within each optical field.

As with the acoustic case,
we are interested in optical waveguide modes with translational invariance along $z$.
We find these modes $\{\vB_{\Lambda}(\vecr), \omega_\Lambda \}$ by solution of the vector Helmholtz equation 
\begin{equation}\label{eq:vechelmholtz}
\nabla \times \big[ \beta_\text{ref}(x,y) \nabla \times \vB_\Lambda \big] = \frac{\omega_\Lambda^2}{c^2} \vB,
\end{equation}
together with Ampere's law
\begin{equation}
\vD_\Lambda(\vecr) = \frac{i}{\muo \omega_\Lambda} \nabla \times \vB_\Lambda,
\end{equation}
and then introduce waveguide mode functions $\vd_{\gamma k}$ defined by
\begin{equation}
\vD_{\gamma k}(\vecr) = \frac{\vd_{\gamma k}(x,y)}{\sqrt{2\pi}} \me^{i kz},
\end{equation}
where $\gamma$ indexes the transverse spatial bands of the waveguide.
We choose the normalization 
\begin{eqnarray}
\frac{1}{\epso}\int \dx\dy\,\beta_\text{ref}(\vecr)\, \vd_{\gamma k}^*(x,y)\cdot \vd_{\gamma k}(x,y) &=&1,  \label{norm} 
\end{eqnarray}
and expand the displacement field operator as 
\begin{eqnarray} 
\vD(\vecr) &=&\sum_{\gamma}\int \frac{\dk}{\sqrt{2\pi}} \sqrt{\frac{\hbar \omega_{\gamma k}}{2 }}a_{\gamma k} \vd_{\gamma k}(x,y)\me^{ikz}+\hc.  \label{modeuse} %
\end{eqnarray}
The mode operators $a_{\gamma k}$ satisfy the standard commutation relations
\begin{align}
\left[ a_{\gamma k},a_{\gammap \kp}\right] &=0, \nonumber \\
\left[ a_{\gamma k},a_{\gammap \kp}^\dagger \right] &=\delta _{\gamma\gammap }\delta (k-\kp)\nonumber ,
\end{align}
and neglecting the vacuum energy, the electromagnetic Hamiltonian reduces to 
\begin{align}
\hamilEM =\sum_{\gamma}\int \dk\,\hbar \omega_{\gamma k}\, a_{\gamma k}^\dagger a_{\gamma k}.
\end{align}

\subsection{Envelope operators}
As with the acoustic modes we can introduce envelope function operators 
associated with a mode $\gamma$ and a range of
wavenumbers in the neighborhood of some $k_{j}$: 
\begin{equation}
\label{envelope}
\psi_{\gamma j}(z)=\int \frac{\dk}{\sqrt{2\pi }}a_{\gamma k}(t)\me^{i(k-k_{j})z}.
\end{equation}
The integration is to be taken over the range of wavenumbers $ k $ that we wish to associate with the center value $k_{j}$. 
This allows the introduction of distinct envelope operators for fields that occupy 
the same spatial mode but occupy distinct frequency ranges (such as a pump and
Stokes wave in the same mode).

If the integrals in (\ref{envelope}) were to extend over all $k$ we would obtain 
\begin{equation}
\left[ \psi_{\gamma j}(z,t),\psi_{\gammap j^\prime }^{\dag }(z^\prime ,t)
\right] =\delta_{\gamma\gammap }\delta_{jj^\prime }\delta (z-z^\prime ).
\label{canonical}
\end{equation}
In principle, 
the existence of cutoffs and the possible partitioning of each channel $\gamma$
into separate bands for pump and Stokes waves means that the $k$ integrals
have restricted range, 
but as with the acoustic fields, we assume the excitations are sufficiently narrow
band and away from cutoff that the integrals leading to the Dirac
delta function in~\eqref{canonical} can be safely extended to infinity.
Assuming in fact that only values of $k$ close to $k_{j}$ are important for each mode $\gamma j$, 
as we now label them, we can write
\begin{eqnarray} \label{eq:Denvop}
\vD(\vecr) &=&\sum_{\gamma}\int \frac{\dk}{\sqrt{2\pi}}\,\sqrt{\frac{\hbar \omega_{\gamma k}}{2}}
a_{\gamma k}(t)\vd_{\gamma k}(x,y)\me^{ikz}+\hc  \label{Duse} \\
&=&\sum_{\gamma,j}\me^{ik_{j}z}\int \frac{\dk}{\sqrt{2\pi}}\,\sqrt{\frac{\hbar \omega_{\gamma k}}{2 }}
a_{\gamma k}(t)\vd_{\gamma k}(x,y)\me^{i(k-k_{j})z}+\hc  \nonumber \\
&\approx &\sum_{\gamma,j}\me^{ik_{j}z}\sqrt{\frac{\hbar \omega_{\gamma}^j }{2}}\mathbf{
d}_{\gamma k_{j}}(x,y)\int \frac{\dk}{\sqrt{2\pi }}\;a_{\gamma k}(t)\me^{i(k-k_{j})z}+\hc 
\nonumber \\
&=&\sum_{\gamma,j}\me^{ik_{j}z}\sqrt{\frac{\hbar \omega_{\gamma}^j }{2}}\vd
_{\gamma k_{j}}(x,y)\,\psi_{\gamma j}(z)+\hc,  \nonumber
\end{eqnarray}
where we have put $\omega_{\gamma}^j \equiv \omega_{\gamma k_{j}}$ and neglected
the variation in $\sqrt{\omega_\gamma^j}$ and 
the $k$ dependence of $\vd_{\gamma k}(x,y)$; as in our treatment of acoustic fields,
corrections to these expressions can be easily identified.

Again in analogy with the treatment of the acoustic fields, 
an operator for the slowly-varying part of the power in the waveguide can be
constructed from the Poynting vector (\refsupp{supp:powmodeEM}) which takes the form 
\begin{equation}
\powEM_{\text{sv}}(z)=\sum_{\gamma,\gammap }\int \frac{\dk\dkp }{2\pi }
a_{\gammap \kp}^\dagger(t) a_{\gamma k}(t)\me^{i(k-\kp)z}\pmodeEM_{\gammap \gamma}(\kp,k) .
\end{equation}
For ranges of $k$ and $\kp$ close enough to $k_{j}$, and assuming
that for different $\gamma$ the corresponding $k_{j}$ ranges are distinct, we can
write this as 
\begin{equation} \label{eq:powEM}
\powEM_{\text{sv}}(z)\approx \sum_{\gamma,j}\int \frac{\dk\dkp }{2\pi }
a_{\gamma \kp}^\dagger(t) a_{\gamma k}(t)\me^{i(k-k^\prime )z} \pmodeEM_{\gamma \gamma}(k_{j},k_{j}), 
\end{equation}
and since we find that 
\begin{equation} \label{eq:pmodeEM}
\pmodeEM_{\gamma \gamma}(k_{j},k_{j})=\hbar \omega_{\gamma}^j v_{\gamma}^j , 
\end{equation}
where $v_{\gamma}^j $ is the group velocity of electromagnetic mode type $\gamma$
centered at $k_{j}$, we have 
\begin{equation}\label{eq:PsvEM}
\powEM_{\text{sv}}(z,t)\approx \sum_{\gamma,j}\hbar \omega_{\gamma}^j v_{\gamma}^j \;\psi _{\gamma j}^\dagger (z,t)\psi_{\gamma j}(z,t),
\end{equation}
so that $\psi _{\gamma j}^\dagger (z,t)\psi_{\gamma j}(z,t) $ behaves as a photon 
number density operator; note again the similarity with the treatment of the
acoustic field, for which~\eqref{eq:powAsv} is the corresponding result.
The form in Eqs.~\eqref{eq:pmodeEM} and~\eqref{eq:PsvEM} does
not seem to have been presented earlier,
and is derived in \refsupp{supp:powmodeEM}. 

Finally, in similar fashion to Eqs.~\eqref{commHA}, we find that the envelope operator
obeys the dynamical equation
\begin{eqnarray}\label{commHEM} 
\frac{\partial \psi_{\gamma k_{j}}(z)}{\partial t}
&=&-i\omega_{\gamma}^j \psi_{\gamma j}(z)-v_{\gamma}^j \frac{\partial \psi_{\gamma j}(z)}{
\partial z}+\frac{1}{2}iv_{\gamma}^{j\prime }\frac{\partial^{2}\psi_{\gamma j}(z)}{
\partial z^{2}}+\ldots  
\end{eqnarray}
where
\begin{eqnarray*}
v_{\gamma}^j  &=&\left( \frac{\dd\omega_{\gamma k}}{\dk}\right)_{k=k_{j}}, \\
v_{\gamma}^{j\prime } &=&\left( \frac{\dd^{2}\omega_{\gamma k}}{\dk^{2}}\right)_{k=k_{j}}
\end{eqnarray*}
are respectively the group velocity and group velocity dispersion of mode $\gamma j$ at its reference wavenumber $k_{j}$. 
For the narrow bandwidths involved in SBS physics, higher dispersive terms are unlikely to be
needed.

\section{The complete opto-acoustic Hamiltonian} \label{sec:hamilall}

At last, we can now assemble the complete opto-acoustic Hamiltonian
\begin{eqnarray*}
H &=&\int \frac{\pi^i (\vecr)\pi^i (\vecr)}{2\rho (\vecr)}
\, \drrr+\frac{1}{2}\int S^{ij}(\vecr)c^{ijkl}(\vecr)S^{kl}(
\vecr) \, \drrr 
+\frac{1}{2\muo}\int B^i (\vecr) B^i (\vecr)\,  \drrr+\frac{1}{2\epso}\int D^i (\vecr)\beta
^{ij}(\vecr)D^j (\vecr)  \, \drrr .
\end{eqnarray*}
Being composed of different classes of oscillators, the electromagnetic and acoustic fields commute with each other.
All quantities are taken as position dependent, varying continuously
(if rapidly) across any material boundaries. Only at the end of various calculations
we will allow them to acquire step-wise discontinuities. 

The opto-acoustic coupling is captured by the new quantity $\beta^{ij}(\vecr)$. 
This is the total inverse (relative) dielectric tensor, 
\begin{equation} \label{eq:fullbeta}
\beta^{ij}(\vecr)=\delta^{ij}\beta_\text{ref}(\vecr)+\tilde{\beta}^{ij}(\vecr;\left[ \vu\left( \vecr\right) \right] ), 
\end{equation}
which includes both the purely electromagnetic properties (the background waveguide structure) in $\beta_\text{ref}$,
and the photoelastic and radiation pressure couplings in the correction $\tilde{\beta}^{ij}$.
Naturally, we have $ \beta^{ij}(\vecr)\epsilon^{jk}(\vecr)=\delta^{ik} $,
where $\epsilon^{jk}(\vecr)$ is the complete relative dielectric tensor.
The coupling between sound and light enters because we assume that $\beta^{ij}(\vecr)$ 
depends on the displacement field $\vu\left(\vecr \right) $, 
through both its dependence on the strain in the material and the motion of the interfaces. 
Thus we can write 
\begin{equation}
H=\hamilA + \hamilEM +V,  \label{Htotal}
\end{equation}
where  
the opto-acoustic coupling  is
\begin{equation}
V=\frac{1}{2\epso}\int D^i (\vecr) 
   \tilde{\beta}^{ij}(\vecr ;\left[ \vu\left( \vecr\right) \right] )D^j (\vecr)
\, \drrr .  \label{interaction}
\end{equation}
In principle, an additional coupling  arises from  the dependence of the mechanical density and stiffness 
on the electromagnetic field variables. These effects lead  to
terms quadratic in $\vu(\vecr)$ or $\vpi(\vecr)$ (corresponding to two-phonon-single-photon interactions) 
which are of higher order than we consider here.
They are also of much lower energy and the processes are very unlikely to be phase-matched, so they are safely neglected.

To proceed to a set of coupled mode equations, we seek to expand the interaction 
Hamiltonian (\ref{interaction}) in terms of the mode operators constructed earlier. 
The physics of the opto-acoustic interaction is introduced by writing
\begin{equation} \label{eq:coupling}
\beta^{ij}(\vecr;\left[ \vu(\vecr)\right] )
= p^{ijlm}(x,y)S^{lm}(\vecr) + \delta^{ij}\beta _\text{ref}(\mathbf{r-u(r))}+ \ldots
\end{equation}
keeping only linear terms in the strain in keeping with our neglect of two-photon interactions. The 
photoelastic tensor $ p^{ijlm}(x,y) $ accounts for the conventional 
electrostrictive/photoelastic contribution to SBS. 
The effect of moving boundaries and radiation pressure enters 
through the second term's dependence on the displacement $\vu(\vecr)$. 
Note that while we focus below on the effects associated with material discontinuities,
this expression also accounts for a bulk contribution to 
the radiation pressure in graded index materials for which $\beta_\text{ref}(\vecr)$
varies smoothly in space.
Equation~\eqref{eq:coupling} is a key expression because the symmetric relationship of radiation pressure
and moving boundary effects follow directly from its form.
This identification is what allows us to avoid the rather lengthy thermodynamic arguments which were
required in the previous rigorous derivation of the classical coupled mode equations~\cite{Wolff2015}.

Expanding $\beta_\text{ref}(\mathbf{r-u(r)})$ for small
displacements we take 
\begin{equation}
\beta_\text{ref}(\mathbf{r-u(r))} \approx \beta_\text{ref}(\vecr)-u^l (\vecr)
\frac{\partial \beta_\text{ref}(\vecr)}{\partial r^l },
\end{equation}
and using \eqref{eq:fullbeta} we have  for the opto-acoustic correction 
\begin{equation}
\tilde{\beta}^{ij}(\vecr;\left[ \vu(\vecr)\right] )=
p^{ijlm}(x,y)S^{lm}(\vecr) 
-\delta^{ij}u^l (\vecr)\frac{\partial \beta_\text{ref}(\vecr)}{\partial r^l  } .
\end{equation}
Then interaction (\ref{interaction}) becomes
\begin{eqnarray}
V  \label{Vfull} 
&=&\frac{1}{2\epso}\int D^i (\vecr)D^j \left( \vecr
\right) \left( p^{ijlm}(x,y)S^{lm}(\vecr)-\delta^{ij}\left( \frac{
\partial \beta_\text{ref}(x,y)}{\partial r^l }\right) u^l (\vecr)\right) \drrr.
\end{eqnarray}
Now since the phonon energy is much smaller than the photon energy the only
significant terms in $V$ will involve the creation and annihilation of
photons. 
On substituting~\eqref{modeuse} for $\vD$,
we normal order the photon mode operators that arise and neglect the resulting constant terms corresponding
to vacuum fluctuation corrections to both the photoelastic tensor and the
displacement-induced change in the dielectric properties. We thus find
\begin{eqnarray} \label{eq:Vdef}
V &=&\frac{1}{\epso}\sum_{\gamma,\gammap }\int \dk\dkp\, a_{\gamma k}^\dagger a_{\gammap \kp}\sqrt{\frac{\hbar \omega_{\gamma k}}{
4\pi }}\sqrt{\frac{\hbar \omega_{\gammap \kp}}{4\pi }} \\
&&\times \int \left( d_{\gamma k}^i (x,y)\right)^*d_{\gammap \kp}^j (x,y)\me^{i(\kp-k)z}\left( p^{ijlm}(x,y)S^{lm}(\vecr
)-\delta^{ij}\left( \frac{\partial \beta_\text{ref}(x,y)}{\partial r^l }
\right) u^l (\vecr)\right) \drrr.
\end{eqnarray}
From (\ref{eq:straindef}) and the second of (\ref{eq:uexp}) we can write 
\begin{equation} \label{eq:strainexp}
S^{lm}(\vecr)=\sum_{\alpha }\int \dq  \,\sqrt{\frac{\hbar \Omalq }{2}}b_{\alpha q}\;S_{\alpha q}^{lm}(\vecr)+\hc, 
\end{equation}
where $S_{\alpha q}^{lm}(\vecr)$ is of the form 
\begin{equation}
S_{\alpha q}^{lm}(\mathbf{r})=\frac{s_{\alpha q}^{lm}(x,y)\eiqz}{\sqrt{2\pi }}.
\end{equation}

After some manipulation (see \refsupp{supp:interaction}), 
the interaction can be reduced to the form
\begin{eqnarray}
V  \label{Vwork} 
&=&\sum_{\gamma,\gammap ,\alpha }
\int \frac{\dk\dkp \dq  }{\left( 2\pi \right)^{3/2}}\;a_{\gamma k}^\dagger a_{\gammap \kp}b_{\alpha q}
\int  \Gamma (\gamma k;\gammap \kp;\alpha q)\, \me^{i(\kp-k+q)z} \, \dz \\
&&+\sum_{\gamma,\gammap ,\alpha }
\int \frac{\dk\dkp \dq  }{(2\pi )^{3/2}} \;b_{\alpha q}^\dagger a_{\gammap \kp}^\dagger a_{\gamma k}
\int  \Gamma^*(\gamma k;\gammap \kp;\alpha q)\, \me^{-i(\kp-k+q)z}\,   \dz\nonumber,
\end{eqnarray}
where the coupling parameter is 
\begin{eqnarray}
\Gamma (\gamma k;\gammap \kp;\alpha q)  \label{gammaresult} 
&=&\frac{1}{\epso}\sqrt{\frac{\hbar \omega_{\gamma k}}{2}}\sqrt{\frac{
\hbar \omega_{\gammap \kp}}{2}}\sqrt{\frac{\hbar \Omega_{\alpha
q}}{2}}  \nonumber \\
&&\times \int \dx\dy\,\left( d_{\gamma k}^i (x,y)\right)^*d_{\gammap \kp}^j (x,y)\left( p^{ijlm}(x,y)s_{\alpha q}^{lm}(x,y)-\delta
^{ij}\left( \frac{\partial \beta_\text{ref}(x,y)}{\partial r^l }\right) u_{\alpha q}^l (x,y)\right) .  \nonumber
\end{eqnarray}
Note that if there is a slow 
variation of the nonlinear properties, due to longitudinal 
variation in, say, the composition or waveguide dimensions, then 
$ \Gamma (\gamma k;\gammap \kp;\alpha q) $ will acquire this variation too, 
and the integration over $z$ in~\eqref{Vwork} would capture this effect.

For an infinite homogeneous waveguide  we can do the remaining integral over all $z$ 
in~(\ref{Vwork}) 
to obtain a delta function $\delta (\kp-k+q)$. Using this to eliminate
the $q$ integral in~\eqref{Vwork},
 the total Hamiltonian (\ref{Htotal}) becomes 
\begin{eqnarray*} \label{eq:Htotal}
H &=&\sum_{\gamma}\int \dk\,\hbar \omega_{\gamma k}\, a_{\gamma k}^\dagger a_{\gamma k}+\sum_{\alpha
}\int \dq  \,\hbar \Omalq \, b_{\alpha q}^\dagger b_{\alpha q} \\
&&+\sum_{\gamma,\gammap ,\alpha }\int \frac{\dk\dkp }{\sqrt{ 2\pi }}\;a_{\gamma k}^\dagger a_{\gammap \kp}b_{\alpha
(k-\kp)}\;\Gamma (\gamma k;\gammap \kp;\alpha (k-\kp))
\\
&&+\sum_{\gamma,\gammap ,\alpha }\int \frac{\dk\dkp }{\sqrt{2\pi }}
\;b_{\alpha (k-\kp)}^\dagger a_{\gammap \kp}^{\dagger
}a_{\gamma k}\;\Gamma^*(\gamma k;\gammap \kp;\alpha (k-\kp)).
\end{eqnarray*}
Observe that 
the final two terms explicitly display momentum conservation, and 
are clearly identified as describing anti-Stokes and Stokes processes respectively.
This infinite structure form is a good starting point
for investigating the enhancement and suppression of Brillouin scattering by
adjusting the matrix elements or the density of states of optical or acoustic
modes~\cite{Merklein2015}; a simple Fermi's Golden Rule calculation reveals much of the
underlying physics, as we will show in a subsequent contribution~\cite{SipeFermi2015}.
Technically, the finite phonon lifetime requires the $\delta(k'-k+q)$ function
to be broadened into a linewidth function. In practice the $\delta$ function
is a reasonable approximation, with the impact of loss entering through the 
linear properties as discussed later.

\section{Quantum coupled mode equations}
\label{sec:quantumcme}
\begin{figure}
\centering
\includegraphics[height=6cm]{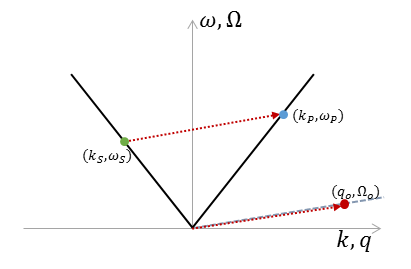} 

\includegraphics[height=6cm]{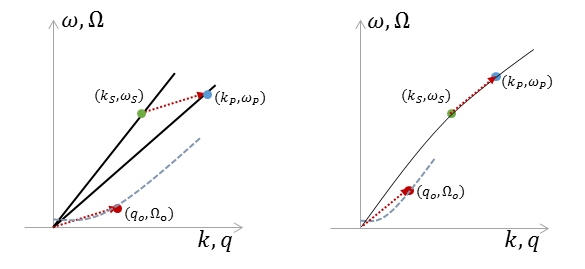} 

\caption{ Phase-matching diagrams for backward SBS (top),  forward intermodal (bottom left)
and forward intramodal SBS (bottom right). Black lines represent optical dispersion relations,
dashed blue lines represent acoustic dispersion relations.}
\label{fig:phasematch}
\end{figure}

To derive coupled mode equations for the envelope function operators,
we assume that the process of interest destroys pump photons with
transverse mode and center wavenumber $(\gamma_\sP,k_\sP)$, creating Stokes
photons with $(\gamma_\sS, k_\sS)$,
and phonons with  mode-wavenumber pair $(\alpha,q)=(\alpha_{\so},q_\so) $. 
Additional processes could be included
to describe cascaded SBS phenomena~\cite{Kang2009,Buttner2014}.
We assume that the center wavenumbers and
frequencies involved satisfy energy conservation and phase matching; that is,
\begin{eqnarray}
\omega_\sP &=&\omega_\sS+\Omega_{\so},  \nonumber \\
k_\sP &=&k_\sS+q_\so ,  \label{centers} 
\end{eqnarray}
where we note the wavevectors can be positive or negative, and we have defined $
\omega_\sP\equiv \omega_{\sP k_\sP}$, $\omega_\sS\equiv \omega_{\sS k_\sS}$,
and $\Omega_{\so}=\Omega_{\alpha_{\so}q_\so }$.  
Figure~\ref{fig:phasematch} indicates the significance of these quantities for the processes of
backward, forward intermodal and backward intermodal SBS. 

Then we can write~\eqref{Vwork} as 
\begin{eqnarray*}
V &=&\int \dz\, \int \frac{\dk\dkp \dq  }{\left( 2\pi \right)^{3/2}}
\;a_{\sP k}^\dagger \me^{-i(k-k_\sP)z}a_{\sS\kp}\me^{i(k^\prime -k_\sS)z}b_{\alpha_{\so}q}\me^{i(q-q_\so )z}\;\Gamma (\gamma_\sP k;\gamma_\sS \kp;\alpha
_{\so}q) \\
&&+\int \dz\, \int \frac{\dk\dkp \dq  }{(2\pi )^{3/2}}\;b_{\alpha_{\so}q}^{\dagger
}\me^{-i(q-q_\so )z}a_{\sS\kp}^\dagger \me^{-i(k^\prime -k_\sS)z}a_{\sP k}\me^{i(k-k_\sP)z}\;\Gamma^*(\gamma_\sP k;\gamma_\sS \kp;\alpha
_{\so}q).
\end{eqnarray*}
Given the narrow bandwidths associated with SBS, 
we treat the coupling strengths $\Gamma(\gamma_\sP k;\gamma_\sS \kp;\alpha_{\so}q)$ as constant over the range of wavenumbers integration:
\begin{equation}
\Gamma (\gamma_\sP k;\gamma_\sS \kp;\alpha_{\so}q)\approx \Gamma (\gamma_\sP k_\sP;\gamma_\sS k_\sS;\alpha
_{\so}q_\so )\equiv \Gamma . 
\end{equation}
Pulling this term out, moving to the Heisenberg picture and using the definitions of the envelope functions in Eqs.~\eqref{phononenvelope} and~\eqref{envelope}, we reach 
\begin{equation}
V=\Gamma \int \dz\,\psi_\sP^\dagger (z,t)\psi_\sS(z,t)\phi (z,t)+\Gamma
^*\int \dz\,\phi^\dagger (z,t)\psi_\sS^\dagger (z,t)\psi
_\sP(z,t), 
\end{equation}
where $\psi_\sS$, $\psi_\sS$, $\phi$ are the pump and Stokes photon fields
and phonon fields respectively.
Using \eqref{commHA} and \eqref{commHEM}, we find
\begin{subequations}
\begin{align}
\frac{\partial \psi_\sP(z,t)}{\partial t} 
&=\frac{1}{i\hbar }\left[ \psi_\sP(z,t),\hamilEM \right] +\frac{1}{i\hbar } \left[ \psi_\sP(z,t),V\right]  \nonumber \\
&=-i\omega_\sP\psi_\sP(z,t)-v_\sP\frac{\partial \psi_\sP(z,t)}{\partial z
}+\frac{1}{2}iv_\sP^\prime \frac{\partial^{2}\psi_\sP(z,t)}{\partial
z^{2}}+\frac{\Gamma }{i\hbar }\psi_\sS(z,t)\phi (z,t),
\intertext{and similarly}
\frac{\partial \psi_\sS(z,t)}{\partial t}
&=-i\omega_\sS\psi_\sS(z,t)-v_\sS\frac{\partial \psi_\sS(z,t)}{\partial z
}+\frac{1}{2}iv_\sS^\prime \frac{\partial^{2}\psi_\sS(z,t)}{\partial
z^{2}}+\frac{\Gamma^*}{i\hbar }\phi^\dagger (z,t)\psi_\sP(z,t), \\
\frac{\partial \phi (z,t)}{\partial t} 
&=-i\Omega_{\so}\phi (z,t)-v_{\so}\frac{\partial \phi (z,t)}{\partial z}+\frac{
1}{2}iv_{\so}^\prime \frac{\partial^{2}\phi (z,t)}{\partial z^{2}}+\frac{
\Gamma^*}{i\hbar }\psi_\sS^\dagger (z,t)\psi_\sP(z,t).
\end{align}
\end{subequations}
 
To remove the fast time-dependence, we define
\begin{subequations}
\begin{align}
\Psi_\sP(z,t) & = \psi_\sP(z,t) \sqrt{\hbar \omega_\sP |v_\sP|} \me^{i\omega_\sP t}, \\
\Psi_\sS(z,t) & = \psi_\sS(z,t) \sqrt{\hbar \omega_\sS |v_\sS|} \me^{i\omega_\sS t}, \\
\Phi(z,t) & = \phi (z,t)    \sqrt{\hbar \Omega_o |v_\so|} \me^{i\Omega_{\so}t},
\end{align}
\end{subequations}
so that
\begin{eqnarray*}
\Psi_\sP^\dagger (z,t)\Psi_\sP(z,t) &=&\hbar \omega_\sP\left\vert v_\sP\right\vert \, \psi_\sP^\dagger (z,t){\psi}_\sP(z,t), \\
\Psi_\sS^\dagger (z,t)\Psi_\sS(z,t) &=&\hbar \omega_\sS\left\vert v_\sS\right\vert \, \psi_\sS^\dagger (z,t){\psi}_\sS(z,t), \\
\Phi^\dagger (z,t)\Phi (z,t) &=&\hbar \Omega_{\so}\left\vert v_{\so}\right\vert \, \phi^\dagger (z,t){\phi}(z,t).
\end{eqnarray*}
Thus $\Psi_\sP^\dagger (z,t)\Psi_\sP(z,t)$, $\Psi_\sS^{\dagger
}(z,t)\Psi_\sS(z,t)$, and $\Phi^\dagger (z,t)\Phi (z,t)$ can be
identified as the power flowing in the waveguide in each mode at
position $z$; this is a positive quantity with the direction along $z$ given
by the sign of the velocity. 

Defining a reduced coupling constant
\begin{equation} \label{eq:gamred}
\bar{\Gamma}(\gamma k; \gammap \kp; \alpha q)  = 
\frac{1}{\sqrt{\left( \hbar \omega_{\gamma k}\right) \left( \hbar \omega_{\gammap \kp}\right) \left( \hbar \Omalq  \right) 
\left\vert v_{\gamma k} v_{\gammap \kp} v_{\alpha q} \right\vert }}\Gamma( \gamma k; \gammap \kp; \alpha q),
\end{equation}
and evaluating it at the center frequencies of our equations:
\begin{equation}
\bar{\Gamma}_0  = \frac{1}{\sqrt{\left( \hbar \omega_\sP\right) \left( \hbar \omega _\sS\right) \left( \hbar \Omega_{\so}\right) \left\vert v_\sP v_\sS v_{\so}\right\vert }}\Gamma( \gamma_\sP k_\sP; \gamma_\sS k_\sS; \alpha_\so q_\so),
\end{equation}
we obtain the Heisenberg evolution equations for the field operators
\begin{subequations}
\label{eq:fullcmes}
\begin{eqnarray}
\frac{\partial \Psi_\sP(z,t)}{\partial t}+v_\sP\frac{\partial \Psi_\sP(z,t)
}{\partial z}-\frac{1}{2}iv_\sP^\prime \frac{\partial^{2}\Psi_\sP(z,t)}{
\partial z^{2}} &=&-i\omega_\sP\left\vert v_\sP\right\vert \bar{\Gamma}_0
\;\Psi_\sS(z,t)\Phi (z,t), \label{eq:fullcmesp}\\
\frac{\partial \Psi_\sS(z,t)}{\partial t}+v_\sS\frac{\partial \Psi_\sS(z,t)
}{\partial z}-\frac{1}{2}iv_\sS^\prime \frac{\partial^{2}\Psi_\sS(z,t)}{
\partial z^{2}} &=&-i\omega_\sS\left\vert v_\sS\right\vert \bar{\Gamma}_0
^*\;\Phi^\dagger (z,t)\Psi_\sP(z,t), \label{eq:fullcmess}\\
\frac{\partial \Phi (z,t)}{\partial t}+v_{\so}\frac{\partial \Phi (z,t)}{
\partial z}-\frac{1}{2}iv_{\so}^\prime \frac{\partial^{2}\Phi (z,t)}{
\partial z^{2}} &=&-i\Omega_{\so}\left\vert v_{\so}\right\vert \bar{\Gamma}_0
^*\;\Psi_\sS^\dagger (z,t)\Psi_\sP(z,t).\label{eq:fullcmesa}
\end{eqnarray}
\end{subequations}

We have thus established a fully quantum form of the opto-acoustic coupled
evolution equations and shown that the correspondence of 
coupling constants for photoelasticity plus moving boundaries and the
reverse processes of electrostriction plus radiation pressure emerge naturally from
the starting point of \eqref{eq:coupling}.
We next evaluate the coupling constants (or ``matrix elements'') 
to connect them to familiar 
forms in the literature.

\subsection{The coupling constants }
We separate  the coupling constants
$\barGam (\gamma k;\gammap \kp;\alpha q)$ in Eq.~\eqref{eq:gamred} into the contributions
\begin{equation}
\barGam (\gamma k;\gammap \kp;\alpha q)=\barGam_\text{bulk}(\gamma k;\gammap \kp;\alpha q)+\Gamma_\text{surf}(\gamma k;\gammap \kp;\alpha
q), 
\end{equation}
where 
\begin{eqnarray}
\barGam_\text{bulk}(\gamma k;\gammap \kp;\alpha q)  \label{gammabulk} 
&=& 
\frac{1}{\epsilon_0}
\frac{1}{2\sqrt{2} \left\vert v_{\gamma k} v_{\gammap \kp} v_{\alpha q} \right\vert}
\int \dx\dy\,\left( d_{\gamma k}^i (x,y)\right)^*d_{\gammap \kp }^j (x,y)p^{ijlm}(x,y)s_{\alpha q}^{lm}(x,y),  
\end{eqnarray}
and 
\begin{eqnarray} \label{gammasurface}
\barGam_\text{surf}(\gamma k;\gammap \kp;\alpha q)  &=&
- \frac{1}{\epsilon_0}
\frac{1}{2\sqrt{2} \left\vert v_{\gamma k} v_{\gammap \kp} v_{\alpha q} \right\vert}
\int \dx\dy\,\left( d_{\gamma k}^i (x,y)\right)^*d_{\gammap \kp }^i (x,y)\left( \frac{\partial \beta_\text{ref}(x,y)}{\partial r^j }\right) u_{\alpha q}^j (x,y).  
\end{eqnarray}
The label ``surface'' for Eq.~\eqref{gammasurface} is perhaps too restrictive since the derivative
$\partial \beta_\text{ref}/\partial r^j$ is non-zero in graded index materials, however
this contribution is typically weak, and 
in current experiments it is the surface contribution that is of most interest.
We seek expressions in terms of the optical modes $\vd_{\gamma k}$ and acoustic modes $\vu_{\alpha q}$ that 
can be obtained from numerical solvers.

\subsubsection{The bulk coupling constant}
The bulk term in~\eqref{gammabulk} describes 
photoelasticity (in~\eqref{eq:fullcmesp} and~\eqref{eq:fullcmess}) and
electrostriction (in \eqref{eq:fullcmesa}).
Its evaluation is straightforward. Even in the limit of
a step discontinuity in material parameters across a waveguide
interface, the $\vd_{\gamma k}(x,y)$ will suffer at most a step
discontinuity, as will $s_{\alpha q}^{lm}(x,y)$ and, by assumption, $p^{ijlm}(x,y)$. 
Hence \eqref{gammabulk} remains well-defined for step discontinuities.

Using the definition of  $s_{\alpha q}^{lm}(x,y)$, one can show that 
for the standard interaction with modes $(\sP, \sS, \so)$, the bulk element 
reduces to the form 
\begin{equation} \label{eq:bargam}
\barGam_\text{bulk} =  \frac{1}{2^{3/2}  \sqrt{|v_\sP v_\sS v_\so|} } \frac{1}{\epsilon_0} 
 \int \dx \dy \,  d_S^i(x,y) (d_\sP^j(x,y))^* p^{ijlm}(x,y)\left[\left(\frac{\partial}{\partial r_l} u_m(x,y)\right) +i q \delta_{mz} (u_l(x,y))^* \right] .
\end{equation}
which is consistent with the results in Wolff~\etal~\cite{Wolff2015} 
for the photoelastic coupling, there termed $\Gamma^{\ePE}$.

\subsubsection{The surface matrix element}
The surface term in~\eqref{gammasurface} describes driving
of the optical fields by moving boundaries (in~\eqref{eq:fullcmesp} and~\eqref{eq:fullcmess}) and radiation pressure (in~\eqref{eq:fullcmesa}).
Its evaluation is more subtle, since in the
limit of a step discontinuity in material parameters $\beta_\text{ref}(x,y)$
will change in a step-like fashion, but its derivative can be Dirac
delta-function-like. 
Consequently, a smoothing operation is required to make sense of this term.
Following Johnson~\etal~\cite{Johnson2002} one can show (see \refsupp{app:surfacematrix})
that for a boundary contour $\vecR_c(s)$ parameterized by arc length $s$ that separates two materials 
with dielectric constants $\epsilon_+$ and $\epsilon_-$ the surface contribution reduces to  
\begin{multline}\label{gammasurfaceuse} 
\barGam_\text{surf}(\gamma k;\gammap \kp;\alpha q)  
= \frac{1}{2^{3/2}\sqrt{\left\vert v_{\gamma k}v_{\gammap \kp}v_{\alpha q}\right\vert }} \\
\times \left[ \frac{1}{\epso}(\frac{1}{\epsilon_{-}}-\frac{1}{\epsilon _{+}})
\int \left[ d_{\gamma k}^{\perp }(\vecR_c (s))\right]^{\ast
}d_{\gammap \kp}^{\perp }(\vecR_c (s))\left[  \unitn(s)\cdot \vu_{\alpha q}(\vecR_c (s))\right]\,  \dRc (s)   \right.  \\
\left.
+\epso\left( \epsilon_{+}-\epsilon_{-}\right) 
\int \left[ \ve_{\gamma k}^{\parallel }(\vecR_c (s))\right]^*\cdot 
\ve_{\gammap \kp}^{\parallel }(\vecR_c (s))\, \left[\unitn (s)\cdot \vu_{\alpha q}(\vecR_c (s))\right] \, \dRc (s)  \right] .
\end{multline}
Here the unit normal $\unitn(s)$ points in the direction from $\epsilon_-$ to $\epsilon_+$.
The full expression for $\Gamma_\text{surf}(\gamma k;\gammap \kp;\alpha
q)$ then involves a sum over all such curves separating distinct dielectrics. 
Note there is no ambiguity in evaluating these terms, 
since $d_{\gamma k}^{\perp }(\vecr)$ is
continuous across a step discontinuity in $\beta_\text{ref}(x,y)$, as is $
\ve_{\gamma k}^{\parallel }(\vecr).$

For the SBS combination of modes $(\sP, \sS, \so)$, 
the surface term can also be written as 
\begin{align}
\bar{\Gamma}^\text{MB} & = 
\left(
\frac{1}{2 \sqrt{2} \sqrt{|v_\sP v_S v_\so|}}  
\int \dRc(s)  \, (\vu^*\cdot \unitn) \left [
\frac{1}{\epsilon_0} \left(\frac{1}{\epsilon_-}-\frac{1}{\epsilon_+}\right) (\vd_S^\perp)^* \vd_\sP^\perp
+\epsilon_0(\epsilon_--\epsilon_+) (\unitn \times \ve_S)^* \cdot (\unitn \times \ve_\sP) 
 \right] 
\right)^*,
\end{align}
which coincides with the expression of Wolff~\etal~\cite{Wolff2015}.

A normalized form of these expressions convenient for working with numerical mode solvers is provided
in \refsupp{app:redmatel}.

\subsection{Recovery of classical coupled mode equations}
\label{sec:classcme}
Finally, the standard classical coupled mode equations~\cite{Wolff2015} can be 
recovered from~\eqref{eq:fullcmes} by dropping the dispersion terms and 
taking mean values for the operators:
\begin{subequations}\label{eq:classcmes}
\begin{eqnarray} 
\frac{\partial \expect{\psi_\sP}}{\partial t}+v_\sP\frac{\partial \expect{\psi_\sP} }{\partial z} 
&=&-i\omega_\sP\left\vert v_\sP\right\vert \bar{\Gamma}_0 \;\expect{\psi_\sS}\expect{\phi} , \\
\frac{\partial \expect{\psi_\sS}}{\partial t}+v_\sS\frac{\partial \expect{\psi_\sS} }{\partial z}
&=&-i\omega_\sS\left\vert v_\sS\right\vert \bar{\Gamma}_0^*\;\expect{\phi}^{*} \expect{\psi_\sP}, \\
\frac{\partial \expect{\phi} }{\partial t}+v_{\so}\frac{\partial \expect{\phi} }{ \partial z} + v_\so \alpha \langle \phi \rangle
&=&-i\Omega_{\so}\left\vert v_{\so}\right\vert \bar{\Gamma}_0^*\;\expect{\psi_\sS}^{*} \expect{\psi_\sP}.
\end{eqnarray}
\end{subequations}
We have introduced the acoustic loss $\alpha$ by hand
following the expression of Wolff~\etal~\cite{Wolff2015}:
\begin{align}
\alpha = \frac{\Omega_o^2}{v_\so} \int \dx\dy\, u_i^* \partial_j (\eta_{ijkl}\partial_k u_l),
\end{align}
where $\eta_{ijkl}$ is the viscosity tensor.

\section{Discussion}
\label{sec:discussion}
Equations~\eqref{eq:fullcmes} represent a full quantum description of guided-wave 
optoacoustic interactions.
Although it requires some preliminary derivations to identify the effective
fields that are involved, our approach provides a direct
derivation of both the photoelastic/electrostrictive and
moving-boundaries/radiation pressure components of the SBS interaction. The
latter has clear contributions from both surface effects at material boundaries
and bulk effects due to smooth variation in dielectric properties.  By avoiding
any discussion of forces and stress tensors, the ambiguities and challenges of
prior treatments do not arise.

The equations of motion have a number of potential applications. In the
classical limit they provide a rigorous confirmation of the earlier
treatment~\cite{Wolff2015}.  In the quantum regime, we have a theory of
opto-acoustic interactions that faithfully represents the photon and number
statistics including any non-classical behavior.  To date, quantum acoustic
effects in guided wave systems have not been observed due to the overwhelming
thermal contribution to the phonon field, though this may change in the near future.
However, we can certainly envisage mixed systems in which a
classical coherent state phonon field interacts with non-classical photon
states in order to transfer quantum information between different channels, and
our treatment is ideal for studies at this quantum-classical boundary.

An obvious and significant extension to our work would be a complete 
treatment of the acoustic dissipation, which we plan to present in the future.  
To incorporate dissipation
into a Hamiltonian picture, a natural approach would be to introduce a thermal
bath of phonon oscillators rather than just the Brillouin excited mode. 
These modes would couple with the coherent modes of interest and with each other
through a three-phonon collision term associated with anharmonicity in the
phonon Hamiltonian.  Tracing over the additional oscillators
would lead to an effective dissipation on the preferred phonon mode.  According
to need or preference, one could derive a dissipative master equation for the
reduced phonon density operator, or perhaps more usefully, a set of Heisenberg
equations with Langevin noise terms associated with the loss.
This kind of treatment would be particularly important in understanding the 
impact of phonon loss on the photon quantum noise.

As mentioned earlier, 
another avenue is the description of  enhancement and  inhibition 
of SBS through density of states engineering~\cite{Merklein2015}.
In the spontaneous regime, this is well handled by a Fermi Golden Rule calculation of phonon generation rates~\cite{SipeFermi2015}, as we will demonstrate in a subsequent work.

\begin{acknowledgments}
This research was supported by the Australian Research Council Centre of Excellence for Ultrahigh bandwidth Devices for Optical Systems (CUDOS) under project number CE110001018.
\end{acknowledgments}

\ifbool{includesupp}{}{%
\end{document}
}

~\newpage

\setcounter{section}{0}
\setcounter{equation}{0}
\setcounter{figure}{0}
\setcounter{table}{0}
\makeatletter
\renewcommand{\thesection}{S.\Roman{section}}
\renewcommand{\thesubsection}{S.\arabic{subsection}}
\renewcommand{\theequation}{S.\arabic{equation}}
\renewcommand{\thefigure}{S.\arabic{figure}}
\renewcommand{\bibnumfmt}[1]{[S#1]}
\renewcommand{\citenumfont}[1]{S#1}

\begin{center}
\textbf{ \large A Hamiltonian treatment of stimulated Brillouin scattering in nanoscale integrated waveguides}

\textbf{ \large --- Supplementary Material }

~ \\

J. E. Sipe$^{1,2}$ and M. J. Steel$^{2,3}$ \\

~ \\
\emph{1. Department of Physics and Institute for Optical Sciences, University of Toronto, Toronto, Ontario M5S 1A7, Canada}

\emph{2. Macquarie University Quantum Science and Technology Centre,
Department of Physics \& Astronomy, Macquarie University, NSW 2109, Australia}

\emph{3.  Centre for Ultrahigh bandwidth Devices for Optical Systems (CUDOS),
MQ Photonics Research Centre,
Department of Physics \& Astronomy, Macquarie University, NSW 2109, Australia}

\end{center}

\newcommand{\vA}{\mathbf{A}}
\newcommand{\vBik}{\vB_{\gamma k}}
\newcommand{\vDik}{\vD_{\gamma k}}
\newcommand{\vEik}{{\cal E}_{Ik}}
\newcommand{\vHik}{{\cal H}_{Ik}}
\newcommand{\vbik}{\mathbf{b}_{Ik}}
\newcommand{\vdik}{\mathbf{d}_{Ik}}
\newcommand{\omik}{\omega_{\gamma k}}
\newcommand{\vik}{v_{Ik}}
\newcommand{\br}{\mathbf{r}}
\newcommand{\nsq}{n^2(x,y)}
\newcommand{\intz}{\int_{-L/2}^{L/2}\mathrm{d}z\,}
\newcommand{\intxy}{\int \mathrm{d}x \mathrm{d}y\,}
\newcommand{\intxyz}{\int_{-L/2}^{L/2} \int_{\cal A} \mathrm{d}\br\, }
\newcommand{\paromk}{\frac{\partial}{\partial \omik}}
\newcommand{\park}{\frac{\partial}{\partial k}}
\newcommand{\calO}{{\cal O}}
\newcommand{\vb}{\mathbf{b}}

\newcommand{\kbar}{\bar{k}}
\newcommand{\kj}{k_j}
\newcommand{\kpj}{k_{j'}}

\section{Heisenberg equations for acoustic field} \label{supp:heisenbergacoustic}

In this section, we derive the equations of motion \eqref{eq:eomacoustic} in \refmain{sec:hamilacoustics}.

Using Eqs.~\eqref{canonical} and \eqref{eq:heisacoustic} we can derive the relations
\begin{eqnarray*}
\frac{\partial u^n(\vecr,t)}{\partial t} &=& \frac{1}{i\hbar} \left[ u^{n}(\vecr),H\right]   \\
 &=&\frac{1}{i\hbar}\int \frac{\left[ u^{n}(\vecr
),\pi ^{i}(\vecrp)\right] \pi ^{i}(\vecrp)}{
2\rho (\vecrp)}\,\drrrp+\frac{1}{i\hbar}\int \frac{\pi ^{i}
(\vecrp)\left[ u^{n}(\vecr),\pi ^{i}(\vecrp)
\right] }{2\rho (\vecrp)}\,\drrrp \\
&=& \delta ^{ni}\int \frac{\delta (\vecr-\vecrp)\pi ^{i}
(\vecrp)}{2\rho (\vecrp)}\,\drrr^{\prime}+ \delta ^{ni}\int \frac{\pi ^{i}(\vecrp)\delta (
\vecr-\vecrp)}{2\rho (\vecrp)}\,\drrrp
\\
&=&\frac{\pi ^{n}(\vecr)}{\rho (\vecr)},
\end{eqnarray*}
and
\begin{eqnarray*}
\frac{\partial \pi^n(\vecr,t)}{\partial t} &=& \frac{1}{i\hbar} \left[ \pi^{n}(\vecr),H\right]   \\
&=& \frac{1}{i\hbar} \frac{1}{2}\int \frac{\partial 
\left[ \pi ^{n}(\vecr),u^{i}(\vecrp)\right] }{
\partial \rp^j}c^{ijkl}(\vecrp)\frac{\partial u^{k}(
\vecrp)}{\partial \rp^l}\,\drrrp +\frac{1}{i \hbar}\frac{1}{2}\int \frac{\partial u^{i}(\vecrp)}{
\partial \rp^j}c^{ijkl}(\vecrp)\frac{\partial \left[
\pi ^{n}(\vecr),u^{k}(\vecrp)\right] }{\partial r^{\prime l}}\,\drrrp \\
&=&-\frac{1 }{2}\delta ^{ni}\int \frac{\partial \delta (\mathbf{r-r}^{\prime })}{\partial \rp^j}c^{ijkl}(\vecrp)\frac{
\partial u^{k}(\vecrp)}{\partial \rp^l}\,\drrr^{\prime } 
-\frac{1 }{2}\delta ^{nk}\int \frac{\partial u^{i}(\vecr^{\prime }
)}{\partial \rp^j}c^{ijkl}(\vecrp)\frac{
\partial \delta (\vecr-\vecrp)}{\partial \rp^l}\,\drrr
^{\prime } \\
&=&\frac{1 }{2}\int \delta (\vecr-\vecrp)\left( \frac{
\partial }{\partial \rp^j}\left( c^{njkl}(\vecrp)\frac{
\partial u^{k}(\vecrp)}{\partial \rp^l}\right) \right) 
\,\drrrp 
+\frac{1 }{2}\int \left( \frac{\partial }{\partial \rp^l}
\left( c^{ijnl}(\vecrp)\frac{\partial u^{i}(\vecr^{\prime})}{\partial \rp^j}\right) \right) \delta (\mathbf{r-r}
^{\prime })\,\drrrp, \\
&=&\frac{1 }{2}\frac{\partial }{
\partial r^{j}}\left( c^{njkl}(\vecr)\frac{\partial u^{k}(\vecr)}{
\partial r^{l}}\right) +\frac{1 }{2}\frac{\partial }{\partial r^{l}}
\left( c^{ijnl}(\vecr)\frac{\partial u^{i}(\mathbf{r)}}{\partial r^{j}}
\right)  \\
&=& \frac{\partial }{\partial r^{j}}\left( c^{njkl}(\vecr)\frac{
\partial u^{k}(\vecr)}{\partial r^{l}}\right)  \\
&=& \frac{\partial }{\partial r^{j}} \big( c^{njkl}(\vecr)S^{kl}(\vecr)\big) ,
\end{eqnarray*}
where in the second last line we have used (\ref{tensorsymmetry}).

\section{Operator $\mathcal{M}^{nk}(\vecr)$ is Hermitian}
\label{app:MHermitian}
Here we show that the operator $\mathcal{M}^{nk}(\vecr)$ of (\ref{Moperator}) is Hermitian.
Consider an integral over an appropriate volume and assume that fields are either
periodic over the volume or vanish at the surface of the
volume. Then for vector functions $\mathbf{C(\vecr)}$ and $\mathbf{D(\vecr)}$, integrating
by parts twice gives
\begin{eqnarray*}
\int \left( D^{n}(\vecr)\right)^*\left( \mathcal{M}^{nk}(
\vecr)C^k (\vecr)\right) \drrr 
&=&-\int \frac{\left( D^{n}(\vecr)\right)^*}{\rho^{1/2}(\mathbf{
r})}\frac{\partial }{\partial r^j }\left( c^{njkl}(\vecr)\frac{
\partial }{\partial r^l }\left( \frac{C^k (\vecr)}{\rho^{1/2}(
\vecr)}\right) \right) \drrr \\
&=&\int \left( \frac{\partial }{\partial r^j }\left( \frac{\left( D^{n}(
\vecr)\right)^*}{\rho^{1/2}(\vecr)}\right) \right) \left(
c^{njkl}(\vecr)\frac{\partial }{\partial r^l }\left( \frac{C^k (
\vecr)}{\rho^{1/2}(\vecr)}\right) \right) \drrr \\
&=&-\int \frac{C^k (\vecr)}{\rho^{1/2}(\vecr)}\frac{\partial }{
\partial r^l }\left( c^{njkl}(\vecr)\left( \frac{\partial }{\partial
r^j }\left( \frac{\left( D^{n}(\vecr)\right)^*}{\rho^{1/2}(
\vecr)}\right) \right) \right) \drrr %
\end{eqnarray*}
Now using (\ref{tensorsymmetry}) we put $c^{njkl}(\vecr)=c^{klnj}(
\vecr)$ and switching the dummy indices $j \leftrightarrow l$, we can write this as
\begin{eqnarray*}
\int \left( D^{n}(\vecr)\right)^*\left( \mathcal{M}^{nk}(
\vecr)C^k (\vecr)\right) \drrr 
&=&-\int \frac{C^k (\vecr)}{\rho^{1/2}(\vecr)}\frac{\partial }{
\partial r^j }\left( c^{kjnl}(\vecr)\left( \frac{\partial }{\partial
r^l }\left( \frac{\left( D^{n}(\vecr)\right)^*}{\rho^{1/2}(
\vecr)}\right) \right) \right) \drrr \\
&=&\left( -\int \frac{\left( C^k (\vecr)\right)^*}{\rho^{1/2}(
\vecr)}\frac{\partial }{\partial r^j }\left( c^{kjnl}(\vecr
)\left( \frac{\partial }{\partial r^l }\left( \frac{\left( D^{n}(\vecr
)\right) }{\rho^{1/2}(\vecr)}\right) \right) \right) \drrr
\right)^* \\
&=&\left( \int \left( C^k (\vecr\right) )^*\left( \mathcal{M}
^{kn}(\vecr)D^{n}(\vecr)\right) \drrr\right)^*,
\end{eqnarray*}
and so the differential operator $\mathcal{M}^{nk}(\vecr)$ is
Hermitian.

\section{Properties of mode functions and partner functions} \label{supp:partners}
Here we establish some useful properties of the mode functions $\vF _\Lambda (\vecr)$
introduced in~\eqref{eq:mcaleig} to~\eqref{complete},
that are required to reduce the acoustic Hamiltonian to canonical harmonic oscillator form 
in \ref{supp:acousticexpansion}.

Note that since $\mathcal{M}^{nk}(\vecr)$ is real, if $\vF _\Lambda (\vecr)$ is an eigenfunction then $\vF _{\Lambda}^{\ast }(\mathbf{r)}$ is also an eigenfunction with the same eigenvalue $\omega_\Lambda$. 
This may happen  simply  because $\vF_\Lambda (\vecr)$ is purely real. 
In fact, as we show in \ref{supp:partnerbuild}, it is always possible to choose the set of
eigenfunctions $\left\{ \vF_\Lambda (\vecr)\right\} $ such
that each of them is purely real. But it is often more
convenient to work with complex eigenfunctions (traveling waves rather than
standing waves, for example). Section~\ref{supp:partnerbuild} establishes that if we include
complex eigenfunctions in the set $\left\{ \vF_\Lambda (\vecr
)\right\} $, the set can be chosen so that each eigenfunction $\vF 
_{\Lambda }(\vecr)$ is either real or, if not, there is another
eigenfunction $\vF _{\bar{\Lambda}}(\vecr)$ in the set such that $
\vF _{\bar{\Lambda}}(\vecr)=\vF_\Lambda ^{\ast }(\mathbf{
r})$. 

Typically the naturally chosen  set of eigenfunctions 
will make this so; for example, if $\vF_\Lambda (\vecr)$ is a
traveling wave to the right, then $\vF _{\bar{\Lambda}}(\vecr)$
is a traveling wave to the left. We refer to $\vF _{\bar{\Lambda}}(
\vecr)$ as the ``partner'' of $\vF_\Lambda (\vecr)$. That
is, each complex eigenfunction in the set has a partner that is also in the
set. If there are purely real eigenfunctions in the set $\left\{ \vF 
_{\Lambda }(\vecr)\right\} $, we take them to be their own partners. \
Then the set of eigenfunctions $\left\{ \vF_\Lambda (\vecr
)\right\} $ is equivalent to the set of $\left\{ \vF _{\bar{\Lambda}}(
\vecr)\right\} $ of partner eigenfunctions, and $\vF _{\bar{
\Lambda}}(\vecr)=\vF_\Lambda ^{\ast }(\vecr)$ for each $
\Lambda $. Since the set of partners is equivalent to the original set, 
then from \eqrefmain{complete} we can also write
\begin{equation}
\int \vF _{\bar{\Lambda}}^{\ast }(\vecr)\cdot \vF _{\bar{
\Lambda}^{\prime }}(\vecr)\, \drrr=\delta_{\bar{\Lambda}\bar{\Lambda}^{\prime }},  \label{vnorm2}
\end{equation}
and
\begin{equation}
\sum_{\bar{\Lambda}}F_{\bar{\Lambda}}^{n}(\vecr)\left( F_{\bar{\Lambda}}^{m}(\vecrp)\right) ^{\ast }=\delta ^{nm}\delta (\mathbf{r-r}
^{\prime }).  \label{complete2}
\end{equation}
Similarly, from \eqrefmain{tildes} we see that we have
\begin{subequations} \label{eq:partnerrules}
\begin{eqnarray} 
\mathbf{\tilde{U}}_{\bar{\Lambda}}(\vecr) &=&\mathbf{\tilde{U}}
_{\Lambda }^{\ast }(\vecr), \\
\mathbf{\tilde{\Pi}}_{\bar{\Lambda}}(\vecr) &=&-\mathbf{\tilde{\Pi} }_{\Lambda}^{\ast }(\vecr).
\end{eqnarray}
\end{subequations}

\section{Acoustic mode expansion} \label{supp:acousticexpansion}
Here we show that through use of the partner functions introduced in \ref{supp:partners},
the acoustic field operators and Hamiltonian
can be expanded in terms of the mode functions as expressed in Eqs.~\eqref{eq:upiexpansion} and~\eqref{eq:hamil}.
Recall that the $\left\{ \vF_\Lambda (\vecr)\right\} $ are the eigenfunctions
of~\eqref{Moperator} and that 
the
$\left\{ \mathbf{\tilde{U}}_{\Lambda }(\vecr)\right\} $
and 
$\left\{ \mathbf{\tilde{\Pi}}_{\Lambda }(\vecr)\right\} $
are defined as in Eq.~\eqref{tildes}.
Both sets of functions are proportional to the $\left\{ \vF_\Lambda (\vecr)\right\} $,
so we can take each to constitute a complete set of
states. We can then expand 
\begin{eqnarray}
\tilde{\vu}(\vecr)  &=&\sum_{\Lambda }\sqrt{\frac{\hbar }{2\Omega
_{\Lambda }}}\mathcal{C}_{\Lambda }^{(1)}\mathbf{\tilde{U}}_{\Lambda }( \vecr),  \label{tildeexpansion} \\
\tilde{\vpi}(\vecr) &=&\sum_{\Lambda }\sqrt{\frac{\hbar }{2\Omega
_{\Lambda }}}\mathcal{C}_{\Lambda }^{(2)}\mathbf{\tilde{\Pi}}_{\Lambda }( \vecr),  \nonumber
\end{eqnarray}
where  $\mathcal{C}_{\Lambda }^{(1)}$ and $\mathcal{C}_{\Lambda }^{(2)}$ are operators
and the factors $ \sqrt{\hbar /2\Omega_{\Lambda }}$ are added for later convenience. 

In the Heisenberg picture,
$\mathcal{C}_{\Lambda }^{(1)}$ and $\mathcal{C}_{\Lambda }^{(2)}$ 
are time-dependent, and therefore so are  $\tilde{\vu}(\vecr)$ and $\tilde{\vpi}(\vecr)$.
However, since $\tilde{\vu}(\vecr)$ and $\tilde{\vpi}(\vecr)$ are 
Hermitian the $\left\{ \mathcal{C}_{\Lambda }^{(1)}\right\} $ are not all
independent, nor are the $\left\{ \mathcal{C}_{\Lambda }^{(2)}\right\} $. 
Using Eqs.~\eqref{eq:partnerrules} we have
\begin{eqnarray*}
\tilde{\vu}^\dagger (\vecr)   &=&\sum_{\Lambda }\sqrt{\frac{
\hbar }{2\Omega_{\Lambda }}}\left( \mathcal{C}_{\Lambda }^{(1)}\right)
^\dagger \mathbf{\tilde{U}}_{\Lambda }^{\ast }(\vecr)=\sum_{\Lambda }
\sqrt{\frac{\hbar }{2\Omega_{\Lambda }}}\left( \mathcal{C}_{\Lambda}^{(1)}\right) ^\dagger \mathbf{\tilde{U}}_{\bar{\Lambda}}(\vecr) \\
\tilde{\vpi}^\dagger (\vecr) &=&\sum_{\Lambda }\sqrt{\frac{
\hbar }{2\Omega_{\Lambda }}}\left( \mathcal{C}_{\Lambda }^{(2)}\right)
^\dagger \mathbf{\tilde{\Pi}}_{\Lambda }^{\ast }(\vecr
)=-\sum_{\Lambda }\sqrt{\frac{\hbar }{2\Omega_{\Lambda }}}\left( \mathcal{C}
_{\Lambda }^{(2)}\right) ^\dagger \mathbf{\tilde{\Pi}}_{\bar{\Lambda}}(
\vecr),
\end{eqnarray*}
so that \eqref{tildeexpansion} may also be written as a sum over partner modes, 
\begin{eqnarray*}
\tilde{\vu}  &=&\sum_{\Lambda }\sqrt{\frac{\hbar }{2\Omega
_{\Lambda }}}\mathcal{C}_{\bar{\Lambda}}^{(1)}\mathbf{\tilde{U}}_{\bar{
\Lambda}}(\vecr), \\
\tilde{\vpi} &=&\sum_{\Lambda }\sqrt{\frac{\hbar }{2\Omega
_{\Lambda }}}\mathcal{C}_{\bar{\Lambda}}^{(2)}\mathbf{\tilde{\Pi}}_{\bar{
\Lambda}}(\vecr),
\end{eqnarray*}
where we have used $\Omega_{\bar{\Lambda}}=\Omega_{\Lambda }$. Then from
the Hermiticity of the canonical fields we see that we require
\begin{eqnarray*}
\left( \mathcal{C}_{\Lambda }^{(1)}\right) ^\dagger  &=&\mathcal{C}_{\bar{
\Lambda}}^{(1)}, \\
-\left( \mathcal{C}_{\Lambda }^{(2)}\right) ^\dagger  &=&\mathcal{C}_{\bar{
\Lambda}}^{(2)},
\end{eqnarray*}
which may be satisfied by setting 
\begin{eqnarray}
\mathcal{C}_{\Lambda }^{(1)} &=&b_{\Lambda }+b_{\bar{\Lambda}}^\dagger ,
\label{Cs} \\
\mathcal{C}_{\Lambda }^{(2)} &=&b_{\Lambda }-b_{\bar{\Lambda}}^\dagger . 
\nonumber
\end{eqnarray}
For real eigenfunctions $\vF_\Lambda (\vecr)$ we have $\bar{
\Lambda}=\Lambda $ and this just says that $\mathcal{C}_{\Lambda }^{(1)}$ is
(proportional to) a coordinate operator, and $\mathcal{C}_{\Lambda }^{(2)}$
is (proportional to) a momentum operator (as we will see). For partners, $
b_{\Lambda }$ and $b_{\bar{\Lambda}}$ are independent operators (or independent
amplitudes in the classical case).  Using (\ref{Cs}) in (\ref{tildeexpansion}) we
have 
\begin{eqnarray} \label{eq:utilexp}
\tilde{\vu}(\vecr) &=&\sum_{\Lambda }\sqrt{\frac{\hbar }{2\Omega_\Lambda }}\left( b_{\Lambda }+b_{\bar{\Lambda}}^\dagger \right) \mathbf{ F}_{\Lambda }(\vecr) \\
&=&\sum_{\Lambda }\sqrt{\frac{\hbar }{2\Omega_{\Lambda }}}b_{\Lambda } \vF_\Lambda (\vecr)+\sum_{\Lambda }\sqrt{\frac{\hbar }{2\Omega_\Lambda }}b_{\bar{\Lambda}}^\dagger \vF _{\bar{\Lambda}}^{\ast }( \vecr) \nonumber \\
&=&\sum_{\Lambda }\sqrt{\frac{\hbar }{2\Omega_{\Lambda }}}b_{\Lambda } \vF_\Lambda (\vecr)+\sum_{\Lambda }\sqrt{\frac{\hbar }{2\Omega_\Lambda }}b_{\Lambda }^\dagger \vF_\Lambda ^{\ast }(\vecr) \nonumber \\
&=&\sum_{\Lambda }\sqrt{\frac{\hbar }{2\Omega_{\Lambda }}}b_{\Lambda } \vF_\Lambda (\vecr)+\hc\nonumber ,
\end{eqnarray}
where in the third line we use the fact that the sum is over the whole set of partner functions. 
Similarly
\begin{eqnarray} \label{eq:pitilexp}
\tilde{\vpi}(\vecr) &=&\sum_{\Lambda }\left( -i\sqrt{\frac{\hbar \Omega_\Lambda }{2}}\right) \left( b_{\Lambda }-b_{\bar{\Lambda}}^\dagger \right) \vF_\Lambda (\vecr) \\
&=&-i\sum_{\Lambda }\sqrt{\frac{\hbar \Omega_{\Lambda }}{2}}b_{\Lambda } \vF_\Lambda (\vecr)+i\sum_{\Lambda }\sqrt{\frac{\hbar \Omega_\Lambda }{2}}b_{\bar{\Lambda}}^\dagger \vF _{\bar{\Lambda}}^{\ast }(\vecr) \nonumber \\
&=&-i\sum_{\Lambda }\sqrt{\frac{\hbar \Omega_{\Lambda }}{2}}b_{\Lambda } \vF_\Lambda (\vecr)+i\sum_{\Lambda }\sqrt{\frac{\hbar \Omega_\Lambda }{2}}b_{\Lambda }^\dagger \vF_\Lambda ^{\ast }(\mathbf{ r}) \nonumber \\
&=&-i\sum_{\Lambda }\sqrt{\frac{\hbar \Omega_{\Lambda }}{2}}b_{\Lambda } \vF_\Lambda (\vecr)+\hc
\end{eqnarray}
in accordance with Eq.~\eqref{eq:upiexpansion}.

Postulating the commutation relations 
\begin{eqnarray}
\left[ b_{\Lambda },b_{\Lambda ^{\prime }}\right] &=&0,  \label{supacomm} \\
\left[ b_{\Lambda },b_{\Lambda ^{\prime }}^\dagger \right] &=&\delta
_{\Lambda \Lambda ^{\prime }},  \nonumber
\end{eqnarray}
we find 
\begin{eqnarray}
\left[ \tilde{u}^{n}(\vecr),\tilde{\pi}^{m}(\vecrp)
\right]  \label{upi} 
&=&i\sum_{\Lambda ,\Lambda ^{\prime }}\sqrt{\frac{\hbar }{2\Omega_{\Lambda }}}\sqrt{\frac{\hbar \Omega_{\Lambda ^{\prime }}}{2}}\left[ b_{\Lambda},b_{\Lambda ^{\prime }}^\dagger \right] F_{\Lambda }^{n}(\vecr
)\left( F_{\Lambda ^{\prime }}^{m}(\vecrp)\right) ^{\ast } 
\nonumber \\
&&-i\sum_{\Lambda ,\Lambda ^{\prime }}\sqrt{\frac{\hbar }{2\Omega_{\Lambda }}}\sqrt{\frac{\hbar \Omega_{\Lambda ^{\prime }}}{2}}\left[ b_{\Lambda}^\dagger ,b_{\Lambda ^{\prime }}\right] \left( F_{\Lambda }^{n}(\vecr
)\right) ^{\ast }F_{\Lambda ^{\prime }}^{m}(\vecrp)  \nonumber
\\
&=&\frac{i\hbar }{2}\sum_{\Lambda }F_{\Lambda }^{n}(\vecr)\left(
F_{\Lambda }^{m}(\vecrp)\right) ^{\ast }+\frac{i\hbar }{2}
\sum_{\Lambda }\left( F_{\Lambda }^{n}(\vecr)\right) ^{\ast }F_{\Lambda}^{m}(\vecrp)  \nonumber \\
&=&\frac{i\hbar }{2}\sum_{\Lambda }F_{\Lambda }^{n}(\vecr)\left(
F_{\Lambda }^{m}(\vecrp)\right) ^{\ast }+\frac{i\hbar }{2}
\sum_{\Lambda }F_{\bar{\Lambda}}^{n}(\vecr)\left( F_{\bar{\Lambda}}^{m}(
\vecrp)\right) ^{\ast }  \nonumber \\
&=&i\hbar \delta ^{nm}\delta (\vecr-\vecrp),  
\end{eqnarray}
where we have used~\eqref{complete} and~\eqref{complete2}. Since 
$ 
[\tilde{\vu}^m(\vecr), \tilde{\vpi}^m(\vecr)] =
[{\vu}^m(\vecr), {\vpi}^m(\vecr)]$ we recover the starting commutation relations~\eqref{canonical2}.
It is possible but more complicated to show that demanding the result (\ref
{upi}) one can find that the $b_{\Lambda }$ and $b_{\Lambda }^\dagger $
must satisfy (\ref{supacomm}). 

Now we look at the Hamiltonian in terms of the $b_{\Lambda }$ and $ b_{\Lambda }^\dagger $.  
From (\ref{Hwork}) we have 
\begin{eqnarray*}
\hamilA= \frac{1}{2}\int \tilde{\pi}^{i}(\vecr)\tilde{\pi}^{i}(\vecr) \, \drrr+
\frac{1}{2}\int \tilde{u}^{i}(\vecr)\mathcal{M}^{ik}(\vecr) \tilde{u}^{k}\, \drrr . 
\end{eqnarray*}
From the above we have 
\begin{eqnarray*}
\tilde{u}^{i}\mathbf{(r}) &=&\sum_{\Lambda }\sqrt{\frac{\hbar }{2\Omega
_{\Lambda }}}b_{\Lambda }F_{\Lambda }^{i}(\vecr )+\sum_{\Lambda }\sqrt{
\frac{\hbar }{2\Omega_\Lambda }}b_{\Lambda }^\dagger \left( F_{\Lambda}^{i}(\vecr )\right) ^{\ast }, \\
\tilde{\pi}^{i}\mathbf{(r}) &=&-i\sum_{\Lambda }\sqrt{\frac{\hbar \Omega
_{\Lambda }}{2}}b_{\Lambda }F_{\Lambda }^{i}(\vecr )+i\sum_{\Lambda }
\sqrt{\frac{\hbar \Omega_\Lambda }{2}}b_{\Lambda }^\dagger \left(
F_{\Lambda }^{i}(\vecr )\right) ^{\ast },
\end{eqnarray*}
so 
\begin{eqnarray*}
\int \tilde{\pi}^{i}\mathbf{(r})\tilde{\pi}^{i}\mathbf{(r})\, \drrr  
&=&-\sum_{\Lambda ,\Lambda ^{\prime }}\frac{\hbar }{2}\sqrt{\Omega _{\Lambda}
\Omega _{\Lambda ^{\prime }}}b_{\Lambda }b_{\Lambda ^{\prime }}\int
F_{\Lambda }^{i}(\vecr )F_{\Lambda ^{\prime }}^{i}(\vecr )\, \drrr 
\\
&&-\sum_{\Lambda ,\Lambda ^{\prime }}\frac{\hbar }{2}\sqrt{\Omega _{\Lambda }\Omega _{\Lambda ^{\prime }}}b_{\Lambda }^\dagger \dagger b_{\Lambda ^{\prime }}^\dagger \left( \int F_{\Lambda }^{i}(\vecr )F_{\Lambda ^{\prime }}^{i}(\vecr )\, \drrr \right) ^{\ast } \\
&&+\sum_{\Lambda ,\Lambda ^{\prime }}\frac{\hbar }{2}\sqrt{\Omega _{\Lambda }\Omega _{\Lambda ^{\prime }}}b_{\Lambda }b_{\Lambda ^{\prime }}^{\dagger }\int F_{\Lambda }^{i}(\vecr )\left( F_{\Lambda ^{\prime }}^{i}(\mathbf{r })\right) ^{\ast }\, \drrr  \\
&&+\sum_{\Lambda ,\Lambda ^{\prime }}\frac{\hbar }{2}\sqrt{\Omega _{\Lambda
}\Omega _{\Lambda ^{\prime }}}b_{\Lambda }^\dagger b_{\Lambda ^{\prime
}}\left( \int F_{\Lambda }^{i}(\vecr )\left( F_{\Lambda ^{\prime }}^{i}( \vecr )\right) ^{\ast }\, \drrr \right) ^{\ast }.
\end{eqnarray*}
In the last two terms, orthogonality gives $\Lambda ^{\prime }=\Lambda $. 
In the first two, we replace the sum over $\Lambda $ by a sum over $\bar{\Lambda }$ and use the fact that $\vF_\Lambda (\vecr )= \vF _{
\bar{\Lambda}}^{\ast }(\vecr )$; then orthogonality demands that $
\Lambda ^{\prime }=\bar{\Lambda}$.  Recalling that $\Omega _{\bar{
\Lambda}}=\Omega_\Lambda $ we then have 
\begin{eqnarray}
\int \tilde{\pi}^{i}\mathbf{(r})\tilde{\pi}^{i}\mathbf{(r})\, \drrr 
\label{suppTwork} 
&=&-\frac{1}{2}\sum_{\Lambda }\hbar \Omega_\Lambda b_{\Lambda }b_{\bar{
\Lambda}}-\frac{1}{2}\sum_{\Lambda }\hbar \Omega_\Lambda b_{\Lambda
}^\dagger b_{\bar{\Lambda}}^\dagger   \nonumber \\
&&+\frac{1}{2}\sum_{\Lambda }\hbar \Omega_\Lambda b_{\Lambda }b_{\Lambda
}^\dagger +\frac{1}{2}\sum_{\Lambda }\hbar \Omega_\Lambda b_{\Lambda
}^\dagger b_{\Lambda } . \nonumber
\end{eqnarray}
Then since 
\begin{eqnarray*}
\mathcal{M}^{ik}(\vecr )\tilde{u}^{k}(\vecr ) 
&=&\sum_{\Lambda }\sqrt{\frac{\hbar }{2\Omega_\Lambda }}\Omega _{\Lambda
}^{2}b_{\Lambda }F_{\Lambda }^{i}(\vecr )+\sum_{\Lambda }\sqrt{\frac{
\hbar }{2\Omega_\Lambda }}\Omega_\Lambda ^{2}b_{\Lambda }^{\dagger
}\left( F_{\Lambda }^{i}(\vecr )\right) ^{\ast },
\end{eqnarray*}
we have 
\begin{eqnarray*}
\int \tilde{u}^{i}(\vecr )\mathcal{M}^{ik}(\vecr )\tilde{u}^{k}(
\vecr )\, \drrr  
&=&\sum_{\Lambda ,\Lambda ^{\prime }}\frac{\hbar }{2\sqrt{\Omega _{\Lambda
}\Omega _{\Lambda ^{\prime }}}}\Omega _{\Lambda ^{\prime }}^{2}b_{\Lambda
}b_{\Lambda ^{\prime }}\int F_{\Lambda }^{i}(\vecr )F_{\Lambda ^{\prime
}}^{i}(\vecr )\, \drrr  \\
&&+\sum_{\Lambda ,\Lambda ^{\prime }}\frac{\hbar }{2\sqrt{\Omega _{\Lambda
}\Omega _{\Lambda ^{\prime }}}}\Omega _{\Lambda ^{\prime }}^{2}b_{\Lambda
}^\dagger b_{\Lambda ^{\prime }}^\dagger \left( \int F_{\Lambda }^{i}(
\vecr )F_{\Lambda ^{\prime }}^{i}(\vecr )\, \drrr \right) ^{\ast }
\\
&&+\sum_{\Lambda ,\Lambda ^{\prime }}\frac{\hbar }{2\sqrt{\Omega _{\Lambda
}\Omega _{\Lambda ^{\prime }}}}\Omega _{\Lambda ^{\prime }}^{2}b_{\Lambda
}b_{\Lambda ^{\prime }}^\dagger \int F_{\Lambda }^{i}(\vecr )\left(
F_{\Lambda ^{\prime }}^{i}(\vecr )\right) ^{\ast }\, \drrr  \\
&&+\sum_{\Lambda ,\Lambda ^{\prime }}\frac{\hbar }{2\sqrt{\Omega _{\Lambda
}\Omega _{\Lambda ^{\prime }}}}\Omega _{\Lambda ^{\prime }}^{2}b_{\Lambda
}^\dagger b_{\Lambda ^{\prime }}\left( \int F_{\Lambda }^{i}(\vecr 
)\left( F_{\Lambda ^{\prime }}^{i}(\vecr )\right) ^{\ast }\, \drrr 
\right) ^{\ast }.
\end{eqnarray*}
Using the same strategy as above this gives \
\begin{eqnarray}
\int \tilde{u}^{i}(\vecr )\mathcal{M}^{ik}(\vecr )\tilde{u}^{k}(
\vecr )\, \drrr   \label{suppVwork} 
&=&\frac{1}{2}\sum_{\Lambda }\hbar \Omega_\Lambda b_{\Lambda }b_{\bar{
\Lambda}}+\frac{1}{2}\sum_{\Lambda }\hbar \Omega_\Lambda b_{\Lambda
}^\dagger b_{\bar{\Lambda}}^\dagger   \nonumber \\
&&+\frac{1}{2}\sum_{\Lambda }\hbar \Omega_\Lambda b_{\Lambda }b_{\Lambda
}^\dagger +\frac{1}{2}\sum_{\Lambda }\hbar \Omega_\Lambda b_{\Lambda
}^\dagger b_{\Lambda }.  \nonumber
\end{eqnarray}
Combining (\ref{suppTwork},\ref{suppVwork}) we have 
\begin{eqnarray*}
H &=&\frac{1}{2}\sum_{\Lambda }\hbar \Omega_\Lambda \left( b_{\Lambda}
b_{\Lambda }^\dagger +b_{\Lambda }^\dagger b_{\Lambda }\right) \\
&=&\sum_{\Lambda }\hbar \Omega_\Lambda \left( b_{\Lambda }^{\dagger}
b_{\Lambda }+\frac{1}{2}\right) .
\end{eqnarray*}

\section{The group velocity}\label{app:acousticvg}
Here we work out the group velocity of the acoustic
modes in terms of the modal field providing an explicit expression for
Eq.~\eqref{eq:valqso}, a result which is needed in the next section.
We take the continuous limit of Eq.~\eqref{modeintro}, writing
\begin{equation}\label{eq:acousticmode}
 \mathcal{F}_{\alpha q}(\vecr)=\sqrt{L} \vF _{\alpha q}(\vecr) = \vF _{\alpha q}(x,y)\eiqz .
\end{equation}
From~\eqref{eq:mcaleig} we have
\begin{equation}\label{eveq}
\Omsqalq \mathcal{F}_{\alpha q}^{n}(\vecr)
  =\calM^{nk}(\vecr)\mathcal{F}_{\alpha q}^k (\vecr)
  =-\frac{1}{\sqrt{\rho(x,y)}}\frac{\partial }{\partial r^j }
     \left( c^{njkl}(x,y)\frac{ \partial }{\partial r^l }\left( \frac{\mathcal{F}_{\alpha q}^k (\vecr) }{\sqrt{\rho(x,y)}}\right) \right) .  
\end{equation}

It is helpful to re-express $\calM^{nk}$ in terms of an operator $\calL_q^{nk}$ operating
on the transverse spatial variables only.
Applying $\calM^{nk}$ to the mode~\eqref{eq:acousticmode} gives
\begin{align}
\left[\calM^{nk} \left[ \falq^k \me^{i q z} \right] \right] \me^{-i q z}
  &= -\left[\frac{1}{\sqrtrhoxy} \parparr{~}{j} \left( \stiffxy{njkl} \parparr{~}{l} \left( \frac{\falq^k \me^{i q z}}{\sqrtrhoxy}\right) \right) \right]  \me^{-i q z}
\\
  &= -\left[\frac{1}{\sqrtrho} \parparr{~}{j} \left( \stiff{njkl}\left( \parparr{~}{l}  \falqonsrho  \right)\eiqz  + i q \stiff{njkz}  \falqonsrho \eiqz \right) \right]  \me^{-i q z} \nonumber \\
  &= -\frac{1}{\sqrtrho} \left[ \left(\parparr{~}{j}\stiff{njkl}\right) \left( \parparr{~}{l}  \falqonsrho \right) 
                                + \stiff{njkl} \left(\frac{\partial^2}{\partial r^j\partial r^l}  \falqonsrho\right)
                                +i q \stiff{nzkl} \left(\parparr{~}{l} \falqonsrho\right)  \right. \nonumber \\
                               &~~~~~~~ 
                          \left. +i q \left(\parparr{~}{j}\stiff{njkz} \right) \falqonsrho 
                                +i q \stiff{njkz} \left(\parparr{~}{j} \falqonsrho\right)
                                -q^2 \stiff{nzkz} \falqonsrho 
                     \right] \nonumber \\
 & \equiv \calL_q^{nk} [\falq^k],  \nonumber
\end{align}
where the last line defines the action of the operator $\calL_q^{nk}$  on $\falq(x,y)$.
It follows from the Hermiticity of $\calM^{nk}$ that $\calL_q^{nk}$ is Hermitian with 
respect to integration over the transverse plane.  We can then write 
\begin{equation}
\Omsqalq\falq^n (x,y)=\calL_q^{nk} \falq^k(x,y), 
\end{equation}
so that the $\falq^n$ are eigenfunctions of $\calL_q^{nk}$ and may be taken as orthogonal.

Taking the inner product with $(\falq^n)^*$ and using the orthogonality of the $\falq$, we have
\begin{align}
\Omsqalq  & =\int \dxdy\,  (\falq^n(x,y))^* \calL_q^{nk} \falq^k(x,y) .
\end{align}
Differentiating with respect to $q$ gives
\begin{align}
2 \Omalq  \frac{\dd \Omalq }{\dq  }  & =\frac{\dd~}{\dq  } \int \dxdy\,  (\falq^n(x,y))^* \calL_q^{nk} \falq^k(x,y) .
\end{align}
Since $\calL_q^{nk}$ is Hermitian, we may invoke the Hellmann-Feynman theorem 
to simplify the right hand side:
\begin{align}
2 \Omalq  \frac{\dd \Omalq }{\dq  }  
   & = \int \dxdy\,  [\falq^n(x,y)]^* \left( \frac{\dd~}{\dq  }\calL_q^{nk} \right) \falq^k(x,y)  \nonumber \\
   & =  \int \dxdy\,  \frac{(\falq^n)^*}{\sqrtrho} 
     \left[  
                                2 q \stiff{nzkz} \falqonsrho 
-i  \stiff{nzkj} \left(\parparr{~}{j} \falqonsrho\right)  
             -i  \left(\parparr{~}{j}\stiff{njkz} \right) \falqonsrho  \right.  %
\left.    -i  \stiff{njkz} \left(\parparr{~}{j} \falqonsrho\right)
\right].
\end{align}
Then the group velocity
\begin{align}
v_{\alpha q} = \left. \frac{\dd \Omega_{\alpha q'}}{\dq  '} \right|_{q'=q},
\end{align}
is given by
\begin{align} \label{eq:vgfinal}
v_{\alpha q}
& = \frac{q}{ \Omalq } \int \dxdy\, \frac{(\falq^n)^*}{\sqrtrho} \stiff{nzkz}  \falqonsrho \nonumber \\
& ~~~~ - \frac{i}{2 \Omalq } \int \dxdy\, \frac{(\falq^n)^*}{\sqrtrho} 
                           \left( \parparr{~}{j} \left ( \stiff{njkz} \falqonsrho\right)
                          +  \stiff{nzkj} \left(\parparr{~}{j} \falqonsrho \right)  \right) \nonumber \\
& =  \frac{q}{ \Omalq } \int \dxdy\, \frac{(\falq^n)^*}{\sqrtrho} \stiff{nzkz}  \falqonsrho  \nonumber \\
& ~~~~ + \frac{i}{2 \Omalq } \int \dxdy\, \left( \parparr{~}{j} \frac{(\falq^n)^*}{\sqrtrho}  \right) \left ( \stiff{njkz} \falqonsrho\right)
       - \frac{i}{2 \Omalq } \int \dxdy\, \frac{(\falq^n)^*}{\sqrtrho}  \stiff{nzkj} \left(\parparr{~}{j} \falqonsrho \right) .
\end{align} 

Swapping the dummy indices $n\leftrightarrow k$ in the second term and using~\eqref{tensorsymmetry} gives
\begin{align} \label{eq:vgfinal2}
v_{\alpha q}
 & =    \frac{q}{ \Omalq }   \int \dxdy\, \frac{(\falq^n)^*}{\sqrtrho} \stiff{nzkz}  \falqonsrho  \nonumber \\
& ~~~~ + \frac{i}{2 \Omalq } \int \dxdy\, \left( \parparr{~}{j} \frac{(\falq^k)^*}{\sqrtrho}  \right)  \stiff{nzkj} \frac{\falq^n}{\sqrtrho} 
       - \frac{i}{2 \Omalq } \int \dxdy\, \frac{(\falq^n)^*}{\sqrtrho}  \stiff{nzkj} \left(\parparr{~}{j} \falqonsrho \right) \nonumber \\
& = \frac{q}{ \Omalq }   \int \dxdy\, \frac{(\falq^n)^*}{\sqrtrho} \stiff{nzkz}  \falqonsrho  
    + \Real\left[\frac{i}{ \Omalq } \int \dxdy\, \left( \parparr{~}{j} \frac{(\falq^k)^*}{\sqrtrho}  \right)  \stiff{nzkj} \frac{\falq^n}{\sqrtrho} \right]\\
& =  q \Omalq  \int \dxdy\, (\ualq^n)^* \stiff{nzkz}  \ualq^k  + \Real\left[i \Omalq \int \dxdy\, \left( \parparr{~}{j} (\ualq^k)^* \right)  \stiff{nzkj} \ualq^n \right] ,
\end{align}
where the final line follows from \eqref{lrelate1}.

\section{Acoustic power flow} \label{app:acousticpower}

Even in the presence of coupling the displacement to the electromagnetic
fields, or other forces, we expect the first of (\ref{solution}) still to
hold, 
\begin{equation}
\frac{\partial }{\partial t}\vu(\vecr,t)=\frac{\vpi(\vecr,t)}{\rho (\vecr)}. 
\end{equation}
Since in general the power density at a point in the medium in a direction $
\unitn$ is classically given by 
\begin{equation}
{\cal P}_{\unitn}= -\frac{\partial u^i (\vecr)}{\partial t}c^{ijlm}(\vecr)S^{lm}( \vecr)n^j , 
\end{equation}
the power in the waveguide in the $\unitz $ direction, integrated
over the $xy$ plane, is
\begin{align}
P_\text{cl}(z) & =-\int \dx\dy\,\;\frac{\pi^i (\vecr)}{\rho (x,y)}c^{izlm}(x,y)S^{lm}(
\vecr)  \\
 & =-\int \dx\dy\,\;\frac{\pi^i (\vecr)}{\rho (x,y)}c^{izlm}(x,y)\frac{
\partial u^l (\vecr)}{\partial r^{m}}, 
\end{align}
where the second line follows from the symmetry properties of the stiffness tensor.

We form the operator corresponding to the classical $P_\text{cl}(z)$ by a usual procedure.
Since $P_\text{cl}(z)$ involves the product of the classical fields $\pi^i(\vecr)$ and 
$\partial u^l(\vecr)/\partial r^m$, we obtain the operator $P(z)$ by using the symmetrized
version of the operators corresponding to $\pi^i(\vecr)$ and $\partial u^l(\vecr)/\partial r^m$:
\begin{eqnarray*}
P(z) &=&-\frac{1}{2}\int \dx\dy\,\frac{c^{izlm}(x,y)}{\rho (x,y)}\left( \pi^i (
\vecr)\frac{\partial u^l (\vecr)}{\partial r^{m}}+\frac{\partial
u^l (\vecr)}{\partial r^{m}}\pi^i (\vecr)\right) \\
&=&-\frac{1}{2}\int \dx\dy\,\frac{c^{izlm}(x,y)}{\rho (x,y)}K^{ilm}(\vecr),
\end{eqnarray*}
where we put 
\begin{equation}
K^{ilm}(\vecr)\equiv \pi^i (\vecr)\frac{\partial u^l (\vecr
)}{\partial r^{m}}+\frac{\partial u^l (\vecr)}{\partial r^{m}}\pi^i (
\vecr). 
\end{equation}
Using (\ref{uandpi}), we see that $K^{ilm}(\vecr)$ has the form 
\begin{equation}
K^{ilm}(\vecr)=\frac{\hbar }{4\pi }\sum_{\alpha ,\alphap }\int
\dq  \dq ^\prime \sqrt{\Omalq \Omega_{\alphap \qp}}
\kappa_{\alpha \alphap }^{ilm}(q,\qp),  \label{Kikl}
\end{equation}
where 
\begin{equation}
\kappa_{\alpha \alphap }^{ilm}(q,\qp)=\tilde{\kappa}_{\alpha \alphap }^{ilm}(q,\qp)
 +\bar{\kappa}_{\alpha \alpha^\prime }^{ilm}(q,\qp). 
\end{equation}
The first term contains parts rapidly-varying in space and time:
\begin{eqnarray*}
\tilde{\kappa}_{\alpha \alphap }^{ilm}(q,\qp) 
&=&b_{\alphap \qp}b_{\alpha q} \\
&&\times \left[ \left( \pi_{\alphap q^\prime }^i (x,y)\me^{i\qp z}\right) \left( \frac{\partial }{\partial r^{m}}
\left( u_{\alpha q}^l (x,y)\eiqz\right) \right) +\left( \frac{\partial }{
\partial r^{m}}\left( u_{\alphap \qp}^l (x,y)\me^{iq^\prime z}\right) \right) \left( \pi_{\alpha q}^i (x,y)\eiqz\right) \right] \\
&&+b_{\alphap \qp}^\dagger b_{\alpha q}^\dagger  \\
&&\times \left[ \left( \pi_{\alphap q^\prime }^i (x,y)\me^{i\qp z}\right)^*\left( \frac{\partial }{\partial
r^{m}}\left( u_{\alpha q}^l (x,y)\eiqz\right) \right)^*+\left( 
\frac{\partial }{\partial r^{m}}\left( u_{\alphap q^\prime }^l (x,y)\me^{i\qp z}\right) \right)^*\left( \pi_{\alpha
q}^i (x,y)\eiqz\right)^*\right] ,
\end{eqnarray*}
and $\bar{\kappa}_{\alpha \alphap }^{ilm}(q,\qp)$ contains the slowly-varying terms, 
\begin{eqnarray}
\bar{\kappa}_{\alpha \alphap }^{ilm}(q,\qp)
\label{kappabar} 
&=&b_{\alphap \qp}^\dagger b_{\alpha q}  \nonumber \\
&&\times \left[ \left( \pi_{\alphap q^\prime }^i (x,y)\me^{i\qp z}\right)^*\left( \frac{\partial }{\partial
r^{m}}\left( u_{\alpha q}^l (x,y)\eiqz\right) \right) +\left( \frac{
\partial }{\partial r^{m}}\left( u_{\alphap q^\prime }^l (x,y)\me^{i\qp z}\right) \right)^*\left( \pi_{\alpha
q}^i (x,y)\eiqz\right) \right]  \nonumber \\
&&+b_{\alphap \qp}b_{\alpha q}^\dagger   \nonumber \\
&&\times \left[ \left( \pi_{\alphap q^\prime }^i (x,y)\me^{i\qp z}\right) \left( \frac{\partial }{\partial r^{m}}
\left( u_{\alpha q}^l (x,y)\eiqz\right) \right)^*+\left( \frac{
\partial }{\partial r^{m}}\left( u_{\alphap q^\prime }^l (x,y)\me^{i\qp z}\right) \right) \left( \pi_{\alpha
q}^i (x,y)\eiqz\right)^*\right] .  \nonumber
\end{eqnarray}
We write 
\begin{equation}
P(z)=P_{\text{rv}}(z)+P_{\text{sv}}(z), 
\end{equation}
where $P_{\text{rv}}(z)$ contains the contributions from $\tilde{\kappa}_{\alpha
\alphap }^{ilm}(q,\qp)$ and $P_{\text{sv}}(z)$ those from $\bar{
\kappa}_{\alpha \alphap }^{ilm}(q,\qp)$. Our interest is in the latter.
Since the sums and integrals in (\ref{Kikl}) are over all $\alpha ,\alphap ,q,$ and $ \qp$,
we may switch the dummy indices in the second term on the right-hand-side of \eqref{kappabar}: 
\begin{eqnarray*}
\bar{\kappa}_{\alpha \alphap }^{ilm}(q,\qp) 
&\rightarrow &b_{\alphap \qp}^\dagger b_{\alpha q}\left[
\left( \pi_{\alphap \qp}^i (x,y)\me^{i\qp z}\right)
^*\left( \frac{\partial }{\partial r^{m}}\left( u_{\alpha q}^l (x,y)\eiqz\right) \right) +
\left( \pi_{\alpha q}^i (x,y)\eiqz\right) \left( \frac{\partial }{\partial r^{m}}
\left( u_{\alphap \qp}^l (x,y)\me^{i\qp z} \right) \right)^*
\right] \\
&&+b_{\alpha q}b_{\alphap \qp}^\dagger \left[ \left( \pi
_{\alpha q}^i (x,y)\eiqz\right) \left( \frac{\partial }{\partial r^{m}}
\left( u_{\alphap \qp}^l (x,y)\me^{i\qp z}\right)
\right)^*+
\left( \pi_{\alphap \qp}^i (x,y)\me^{i\qp z}\right)^*
\left( \frac{\partial }{\partial r^{m}}\left( u_{\alpha q}^l (x,y)\eiqz\right) \right) 
\right] .
\end{eqnarray*}
Moving to normal-ordering with 
\begin{equation}
b_{\alpha q}b_{\alphap \qp}^\dagger =b_{\alpha^\prime \qp}^\dagger b_{\alpha q}+\delta_{\alpha \alpha^\prime }\delta (q-\qp), 
\end{equation}
we have
\begin{eqnarray} \label{kappabaruse} 
\bar{\kappa}_{\alpha \alphap }^{ilm}(q,\qp)
&=&2 L^{ilm}_{\alpha'\alpha}(q',q;x,y) b_{\alphap \qp}^\dagger b_{\alpha q}
    +T_{\alpha }^{ilm}(q)\delta_{\alpha a^\prime }\delta (\qp-q), 
\end{eqnarray}
where
\begin{align} \label{eq:Ldef}
L_{\alphap \alpha }^{ilm}(\qp,q;x,y) &=
\left( \pi_{\alphap \qp}^i (x,y)\me^{i\qp z}\right)
^*\left( \frac{\partial }{\partial r^{m}}\left( u_{\alpha q}^l (x,y)\eiqz\right) \right)
+
\left( \pi_{\alpha q}^i (x,y)\eiqz\right) \left( \frac{\partial }{\partial r^{m}}
\left( u_{\alphap \qp}^l (x,y)\me^{i\qp z} \right) \right)^*,
\\
T_{\alpha }^{ilm}(q)&= 
\left( \pi_{\alpha q}^i (x,y)\eiqz\right)^* \left( \frac{\partial }{\partial r^{m}}\left( u_{\alpha q}^l (x,y)\eiqz\right) \right) 
+ 
\left( \pi_{\alpha q}^i (x,y)\eiqz\right) \left( \frac{\partial }{\partial r^{m}}\left( u_{\alpha q}^l (x,y)\eiqz\right) \right)^*.
\end{align}

The term involving $ T_{\alpha }^{ilm}(q) $ in Eq.~\eqref{kappabaruse} represents vacuum 
zero-point contributions and should give
no net contribution to $P_\text{sv}(z)$, which is a directed quantity.
Indeed, using Eqs.~\eqref{lrelate1} and the 
property $f_{\alpha q}^k(x,y)=(f_{\alpha (-q)}^k(x,y))^*$, which follows
from the Hermiticity of $\calM^{nk}$ (see \ref{supp:partners}), it
can be shown that its contribution to Eq.~\eqref{Kikl} vanishes.

The remaining contribution to $P_\text{sv}$ can be written
\begin{equation}
P_{\text{sv}}(z)=\sum_{\alpha ,\alphap }\int \frac{\dq  \dq ^\prime }{2\pi }
b_{\alphap \qp}^\dagger b_{\alpha q}\me^{i(q-q^\prime )z}\pmodeA_{\alphap a}(\qp,q),  \label{powersv}
\end{equation}
where the pairwise term
\begin{eqnarray*}
\pmodeA_{\alphap a}(\qp,q) 
&=&-\frac{\hbar }{2}\sqrt{\Omalq \Omega_{\alpha^\prime \qp}}\int \dx\dy\,\frac{\stiff{izkl}}{\rho }L_{\alpha^\prime \alpha }^{ikl}(\qp,q;x,y).
\end{eqnarray*}
Evaluating the derivatives in Eq.~\eqref{eq:Ldef},
\begin{eqnarray*}
L_{\alphap \alpha }^{ikl}(\qp,q;x,y) 
&=&\delta_{lz}\left( iq\left( \pi_{\alphap q^\prime }^i \right)^*u_{\alpha q}^k -i\qp\left(
u_{\alphap \qp}^k \right)^*\left( \pi_{\alpha
q}^i \right) \right) \\
&&+\left( \pi_{\alphap \qp}^i \right)^*\left( 
\frac{\partial }{\partial r^l } u_{\alpha q}^k  \right)
+\left( \frac{\partial }{\partial r^l } u_{\alphap q^\prime }^k  \right)^*\left( \pi_{\alpha q}^i \right) .
\end{eqnarray*}
Using \eqref{lrelate1} we have 
\begin{eqnarray*}
L_{\alphap \alpha }^{ikl}(\qp,q;x,y) 
&=&\delta_{lz}\left[ -\frac{q}{\Omalq }\left( f_{\alpha^\prime \qp}^i \right)^*\falq^k -\frac{\qp}
{\Omega_{\alphap \qp}}\left( f_{\alpha^\prime \qp}^k \right)^*\left( \falq^i \right)
\right] \\
&&+i\sqrt{\rho }\left[ \left( f_{\alphap q^\prime }^i \right)^*\left( \frac{\partial }{\partial r^l }\left( 
\frac{\falq^k }{\Omalq \sqrt{\rho }}\right)
\right) -\left( \frac{\partial }{\partial r^l }\left( \frac{f_{\alpha
^\prime \qp}^k }{\Omega_{\alphap \qp}\sqrt{
\rho }}\right) \right)^*\left( \falq^i \right)
\right] .
\end{eqnarray*}
Consequently,
\begin{eqnarray}
\pmodeA_{\alphap \alpha}(\qp,q)  
&=&\frac{\hbar }{2}\sqrt{\frac{\Omega_{\alphap \qp}}{ \Omalq }}q\int \dx\dy\,\frac{\stiff{izkz}}{\rho }\left( f_{\alphap \qp}^i \right)^*f_{\alpha q}^k   
+\frac{\hbar }{2}\sqrt{\frac{\Omalq }{\Omega_{\alpha^\prime \qp}}}\qp\int \dx\dy\,\frac{\stiff{izkz}}{\rho }\left( f_{\alphap \qp}^k \right)^*\left( f_{\alpha q}^i \right)  \nonumber \\
&&-\frac{i\hbar }{2}\sqrt{\frac{\Omega_{\alphap \qp}}{ \Omalq }}\int \dx\dy\,\frac{\left( f_{\alphap q^\prime }^i \right)^*}{\sqrt{\rho }}\stiff{izkl}\left( \frac{ \partial }{\partial r^l }\left( \frac{\falq^k }{\sqrt{\rho }}\right) \right)  
+\frac{i\hbar }{2}\sqrt{\frac{\Omalq }{\Omega_{\alpha^\prime \qp}}}\int \dx\dy\,\frac{\left( \falq^i \right) }{\sqrt{ \rho }}\stiff{izkl}\left( \frac{\partial }{\partial r^l }\left( \frac{ f_{\alphap \qp}^k }{\sqrt{\rho }}\right) \right)
^*,  \nonumber
\end{eqnarray}
so that
\begin{eqnarray*}
\pmodeA_{\alpha \alpha }(q,q) 
&=&\frac{\hbar }{2}q\int \dx\dy\,
\frac{ \left( f_{\alpha q}^i \right)^* }{\sqrt{\rho}} c^{izkz} \frac{ f_{\alpha q}^k  }{\sqrt{\rho}}
\; +\; \frac{\hbar }{2}q\int \dx\dy\, 
  \frac{ \left( f_{\alpha q}^k \right)^*  }{\sqrt{\rho}} c^{izkz} \frac{ f_{\alpha q}^i  }{\sqrt{\rho}}
\\
&&-\frac{i\hbar }{2}\int \dx\dy\,\frac{\left( f_{\alpha
q}^i \right)^*}{\sqrt{\rho }}c^{izkl}\left( 
\frac{\partial }{\partial r^l }\left( \frac{f_{\alpha q}^k }{
\sqrt{\rho }}\right) \right) 
\,
+\, \frac{i\hbar }{2}\int \dx\dy\,\frac{\left( f_{\alpha
q}^i \right) }{\sqrt{\rho }}c^{izkl}\left( \frac{
\partial }{\partial r^l }\left( \frac{f_{\alpha q}^k }{\sqrt{
\rho }}\right) \right)^*.
\end{eqnarray*}
In the second term we may exchange $i$ and $k$ because the other elements of
the stiffness tensor are both the same to obtain
\begin{align}
\frac{\pmodeA_{\alpha \alpha }(q,q)}{\hbar \Omalq }
 & = \frac{q}{\Omalq }\int \dx\dy\,
\frac{ \falq^i  }{\sqrt{\rho}}
  c^{izkz} \frac{\left( f_{\alpha q}^k \right)^* }{\sqrt{\rho}} \nonumber \\
& ~~~+\frac{i}{2\Omalq }\int \dx\dy\,
 \frac{ f_{\alpha q}^i  }{\sqrt{\rho }}c^{izkl}\left( \frac{
\partial }{\partial r^l }\left( \frac{f_{\alpha q}^k }{\sqrt{
\rho }}\right) \right)^* 
-\frac{i}{2\Omalq }\int \dx\dy\,\frac{\left( f_{\alpha
q}^i \right)^*}{\sqrt{\rho }}c^{izkl}\left( 
\frac{\partial }{\partial r^l }\left( \frac{f_{\alpha q}^k }{
\sqrt{\rho }}\right) \right)   \nonumber \\
& =  q \Omalq  \int \dxdy\, (\ualq^k)^* \stiff{izkz}  \ualq^i  + \Real\left[i \Omalq \int \dxdy\, \left( \parparr{~}{l} \ualq^k \right)^*  \stiff{izkl} \ualq^i \right] ,
\end{align}
which by Eq.~\eqref{eq:vgfinal} is simply the group velocity of the acoustic mode.
The desired result~\eqref{eq:powAsv} then follows from~\eqref{powerwork}.

\section{Electromagnetic power flow} \label{supp:powmodeEM}

Here we justify the relations~\eqref{eq:powEM} to~\eqref{eq:PsvEM} in the main paper
for the optical power transport in terms of the optical field envelope operators. 

The operator for the power carried by the field is given by the Poynting vector which we write in the symmetrized form
\begin{equation}\label{eq:poynting}
\vS(\vecr, t) = \frac{1}{2} \big[ \vE(\vecr,t) \times \vH(\vecr,t) -\vH(\vecr,t) \times \vE(\vecr,t) \big].
\end{equation}
Following \eqref{eq:Denvop}, the $\vE$ and $\vH$ field operators are given by 
\begin{align}
\vE(\vecr,t) &= \sum_{\gamma,j}\me^{ik_{j}z} \int \frac{\dk}{\sqrt{2\pi}} \sqrt{\frac{\hbar \omega_{\gamma k} }{2}}\ve_{\gamma k}(x,y)  \,
a_{\gamma k} \, \me^{i(k-k_j)z} + \hc \\
\vH(\vecr,t) & =\sum_{\gamma,j}\me^{ik_{j}z} \int \frac{\dk}{\sqrt{2\pi}} \sqrt{\frac{\hbar \omega_{\gamma k} }{2}}\vh_{\gamma k}(x,y) \,
a_{\gamma k} \, \me^{i(k-k_j)z} + \hc,
\end{align}
where the mode functions satisfy
\begin{align} 
\label{eq:Eexpan}
\ve_{\gamma k}(x,y) & = \frac{\vd_{\gamma k}(x,y)}{\epsilon_0 \epsilon(x,y)} \\
\vh_{\gamma k}(x,y) & = \frac{1}{i\omega_{\gamma k}\mu_0} \left[ \nabla \times (\ve_{\gamma k}(x,y)\me^{ikz})\right] \me^{-ikz} .
\label{eq:Hexpan}
\end{align}
It also follows from Maxwell's equations that in lossless systems, for each mode $\gamma k$, there is a partner mode $\gamma \kbar$ 
with $\kbar=-k$, $\omega_{\gamma \kbar}=\omega_{\gamma k}$
and
\begin{align}\label{eq:ehpartners}
\ve_{\gamma  \kbar}(x,y) & = \ve^*_{\gamma k}(x,y) \\
\vh_{\gamma  \kbar}(x,y) & = -\vh^*_{\gamma k}(x,y) .
\end{align}

Using~\eqref{eq:Eexpan} and~\eqref{eq:Hexpan} in~\eqref{eq:poynting},
the operator describing the total power flow in the waveguide is 
\begin{align}
\powEM(z) & =   \int\dx\dy\,  \vS(\vecr, t) \cdot \unitz \nonumber \\
& = \frac{1}{2} \int\dx\dy\, \sum_{\gamma, \gamma',j,j'} \int \frac{\dk \dkp}{2\pi}    
\sqrt{\frac{\hbar \omega_{\gamma k} }{2}}  \sqrt{\frac{\hbar \omega_{\gammap \kp} }{2}} \nonumber \\
& ~~~~~ \unitz \cdot \Big \{ \left[ \ve_{\gamma k} a_{\gamma k} \, \me^{i(k-\kj)z} + (\ve_{\gamma k})^* a_{\gamma k}^\dagger \, \me^{-i(k-\kj)z} \right] 
 \times 
\left[ \vh_{\gammap \kp} a_{\gammap \kp} \, \me^{i (\kp-\kpj) z} +(\vh_{\gammap \kp})^* a_{\gammap \kp}^\dagger \, \me^{-i(\kp-\kpj) z} \right]  \nonumber \\
& ~~~~- \left[ \vh_{\gammap \kp} a_{\gammap \kp} \, \me^{i (\kp-\kpj) z} +(\vh_{\gammap \kp})^* a_{\gammap \kp}^\dagger \, \me^{-i(\kp-\kpj) z} \right]  
  \times \left[ \ve_{\gamma k} a_{\gamma k} \, \me^{i(k-\kj)z} + (\ve_{\gamma k})^* a_{\gamma k}^\dagger \, \me^{-i(k-\kj)z} \right] \Big \}
\end{align}
The temporally slowly-varying part of this expression is 
\begin{align} \label{eq:powsvem}
\powEM_\text{sv} 
 & = \frac{1}{2} \sum_{\gamma, \gamma'} \int \frac{\dk \dkp}{2\pi}  
   \sqrt{\frac{\hbar \omega_{\gamma k} }{2}}  \sqrt{\frac{\hbar \omega_{\gammap \kp} }{2}}   \int\dx\dy\, \nonumber \\
 &~~~~~~  \unitz \cdot \Big[
 \ve_{\gamma k}  \times (\vh_{\gammap \kp})^*  a_{\gamma k} a_{\gammap \kp}^\dagger \, \me^{i[(k-\kp)-(\kj-\kpj)]z}  
   + (\ve_{\gamma k})^*  \times \vh_{\gammap \kp} a_{\gamma k}^\dagger a_{\gammap \kp} \, \me^{-i[(k-\kp)-(\kj-\kpj)]z} \nonumber \\
 &~~~~~~~~~ -
    \vh_{\gammap \kp} \times (\ve_{\gamma k})^*  a_{\gammap \kp} a_{\gamma k}^\dagger \, \me^{-i[(k-\kp)-(\kj-\kpj)]z}  
   - ( \vh_{\gammap \kp} )^* \times \ve_{\gamma k} a_{\gammap \kp}^\dagger a_{\gamma k} \, \me^{i[(k-\kp)-(\kj-\kpj)]z} \Big]  \nonumber \\
 & = \frac{1}{2} \sum_{\gamma, \gamma'} \int \frac{\dk \dkp}{2\pi}  
   \sqrt{\frac{\hbar \omega_{\gamma k} }{2}}  \sqrt{\frac{\hbar \omega_{\gammap \kp} }{2}}   \int\dx\dy\, \nonumber \\
 &~~~~~~  \unitz \cdot \Big[
 \ve_{\gamma k}  \times (\vh_{\gammap \kp})^*  
  \left( a_{\gamma k} a_{\gammap \kp}^\dagger  + a_{\gammap \kp}^\dagger a_{\gamma k} \right)\, \me^{i[(k-\kp)-(\kj-\kpj)]z} \nonumber \\
 & ~~~~~~~~~
   + (\ve_{\gamma k})^*  \times \vh_{\gammap \kp} \left( a_{\gamma k}^\dagger a_{\gammap \kp} 
                                 +a_{\gammap \kp} a_{\gamma k}^\dagger \right) \me^{-i[(k-\kp)-(\kj-\kpj)]z}   \Big] \nonumber\\
 & =  \sum_{\gamma, \gamma'} \int \frac{\dk \dkp}{2\pi}  
   \sqrt{\frac{\hbar \omega_{\gamma k} }{2}}  \sqrt{\frac{\hbar \omega_{\gammap \kp} }{2}}   \int\dx\dy\, \nonumber \\
 &~~~~~~  \unitz \cdot \Big[ \ve_{\gamma k}  \times (\vh_{\gammap \kp})^*  a_{\gammap \kp}^\dagger a_{\gamma k} \, \me^{i[(k-\kp)-(\kj-\kpj)]z} 
   + (\ve_{\gamma k})^*  \times \vh_{\gammap \kp}   a_{\gamma k}^\dagger a_{\gammap \kp}  \,  \me^{-i[(k-\kp)-(\kj-\kpj)]z} \Big] \nonumber \\
 & ~~~~+ \frac{1}{2} \sum_{\gamma, \gamma'} \int \frac{\dk \dkp}{2\pi}  
   \sqrt{\frac{\hbar \omega_{\gamma k} }{2}}  \sqrt{\frac{\hbar \omega_{\gammap \kp} }{2}}   \int\dx\dy\, 
\delta_{\gamma\gammap}\delta(k-\kp)  
\nonumber \\
 &~~~~~~~~~~  \unitz \cdot \Big[ \ve_{\gamma k}  \times (\vh_{\gammap \kp})^*  \, \me^{i[(k-\kp)-(\kj-\kpj)]z}  
             + (\ve_{\gamma k})^*  \times \vh_{\gammap \kp}  \me^{-i[(k-\kp)-(\kj-\kpj)]z}  \Big] \nonumber \\
 & =  \sum_{\gamma, \gamma'} \int \frac{\dk \dkp}{2\pi}  
   \sqrt{\frac{\hbar \omega_{\gamma k} }{2}}  \sqrt{\frac{\hbar \omega_{\gammap \kp} }{2}}   \int\dx\dy\, \nonumber \\
 &~~~~~~  \unitz \cdot \Big[ \ve_{\gamma k}  \times (\vh_{\gammap \kp})^*  a_{\gammap \kp}^\dagger a_{\gamma k} \, \me^{i[(k-\kp)-(\kj-\kpj)]z} 
   + (\ve_{\gamma k})^*  \times \vh_{\gammap \kp}   a_{\gamma k}^\dagger a_{\gammap \kp}  \,  \me^{-i[(k-\kp)-(\kj-\kpj)]z} 
\Big] \nonumber \\
 & ~~~~+ \frac{1}{2} \sum_{\gamma} \int \frac{\dk }{2\pi}  \frac{\hbar \omega_{\gamma k} }{2}  \dx\dy\, 
\unitz \cdot \Big( \ve_{\gamma k}  \times (\vh_{\gamma k})^*   + (\ve_{\gamma k})^*  \times \vh_{\gamma k}  \Big) .
\end{align}
Since the $k$ integral is over all wavenumbers including all partner modes, it is easy to show using~\eqref{eq:ehpartners} that the second term 
in this expression, associated with vacuum contributions, vanishes, as we would expect for a signed quantity.

For the remaining non-vacuum contribution, since the sums and integrals are over all values
we may swap the indices $\gamma, \gammap$ and $k,\kp$ in the second term in square brackets to give  
\begin{align} \label{eq:powsvem2}
\powEM_\text{sv} 
 & = 
 \sum_{\gamma, \gamma'} \int \frac{\dk \dkp}{2\pi}  
   \sqrt{\frac{\hbar \omega_{\gamma k} }{2}}  \sqrt{\frac{\hbar \omega_{\gammap \kp} }{2}}   \int\dx\dy\, 
\unitz \cdot \Big[ \ve_{\gamma k}  \times (\vh_{\gammap \kp})^*  + (\ve_{\gammap \kp})^*  \times \vh_{\gamma k}\Big]
a_{\gammap \kp}^\dagger a_{\gamma k} \, \me^{i[(k-\kp)-(\kj-\kpj)]z} \nonumber \\
& = 
 \sum_{\gamma, \gamma'} \int \frac{\dk \dkp}{2\pi}  
\pmodeEM_{\gammap,\gamma}(\kp,k)
a_{\gammap \kp}^\dagger a_{\gamma k} \, \me^{i[(k-\kp)-(\kj-\kpj)]z} ,
\end{align} 
where we have introduced the quantity
\begin{align} \label{eq:pmodeb}
\pmodeEM_{\gammap,\gamma}(\kp,k)
= \sqrt{\frac{\hbar \omega_{\gamma k} }{2}}  \sqrt{\frac{\hbar \omega_{\gammap \kp} }{2}}   \int\dx\dy\, 
\unitz \cdot \left( \ve_{\gamma k}  \times (\vh_{\gammap \kp})^* + (\ve_{\gammap \kp})^*  \times \vh_{\gamma k} \right).
\end{align}
Finally, if the different modes $\gamma$ have very different center wavenumbers $k_j$, then only the $\gamma=\gammap$ terms will
contribute significantly to \eqref{eq:powsvem2} and we may approximate 
\begin{align} \label{eq:poweemsv}
\powEM_\text{sv}  \approx 
 \sum_{\gamma} \int \frac{\dk \dkp}{2\pi}  a_{\gamma \kp}^\dagger a_{\gamma k} \, \me^{i(k-\kp)z}  
\pmodeEM_{\gamma\gamma}(k,k),
\end{align} 
with $\pmodeEM_{\gamma\gamma}(k,k)$ the power carried by the normalized mode functions $\gamma$ at center wavenumber $k$.

\subsection{Interpretation as the photon number density operator}
To convert the result in~\eqref{eq:poweemsv} to a simple expression involving the photon envelope operators we require the group
velocity in terms of the fields.

Noting that the basis functions $\vB_{\gamma k}(\vecr)=\vb_{\gamma k}(x,y)\me^{ikz}$ are eigenmodes of the 
vector Helmholtz equation~\eqref{eq:vechelmholtz},
the transverse mode functions $\vb_{\gamma k}$ are eigenfunctions of the equation
\begin{align}\label{eq:Okuse}
\calO_k \vb_{\gamma k} = \frac{\omega^2_{\gamma k}}{c^2} \vb_{\gamma k},
\end{align}
where the $k$-dependent operator $\calO_k$ operates on a vector function $\vf$ as 
\begin{align}\label{eq:Okdef}
\calO_k  \vf= \nabla_t \times \left(\frac{1}{n^2} \nabla_t \times \vf\right) - \frac{k^2}{n^2} \unitz \times \unitz\times\vf + 
  ik \left[ \unitz \times \frac{1}{n^2} \nabla_t \times \vf + \nabla_t \times\left(\frac{1}{n^2} \unitz \times \vf\right) \right],
\end{align}
and where $\nabla_t=[\partial_x, \partial_y, 0]$.
It can be shown that $\calO_k$ is Hermitian such that
\begin{align}
\intxy \vf_{1}^* \cdot (\calO_k \vf_{2}) = \left(\intxy \vf_{2} \cdot (\calO_k \vf_{1}) \right)^* .
\end{align}
From Ampere's law, we also have that
\begin{align} \label{eq:amp}
\nabla_t \times \vb_{\gamma k}= -i\muo \omega_{\gamma k} \vd_{\gamma k}-i k \unitz \times \vb_{\gamma k}.
\end{align}

We now take the inner product with $\vb_{\gamma k}^*$ in~\eqref{eq:Okuse} and differentiate both sides with respect to $k$:
\begin{align}\label{eq:bomlhs}
\frac{\partial}{\partial k} \intxy \vb_{\gamma k}^* \cdot \calO_k \vb_{\gamma k} 
      & = \frac{\partial}{\partial k} \left( \frac{\omega^2_{\gamma k}}{c^2} \intxy \vb_{\gamma k} ^* \cdot\vb_{\gamma k} \right) \nonumber \\
      & = \muo\frac{\partial}{\partial k} \frac{\omega^2_{\gamma k}}{c^2}  \nonumber \\
      & = \frac{ 2 \muo \omega_{\gamma k} }{c^2} \frac{\partial \omega_{\gamma k}}{\partial k} ,
\end{align}
where we used the normalization $\intxy \vb_{\gamma k} ^* \cdot\vb_{\gamma k}/\muo=1$ which follows from~\eqref{norm}
and Maxwell's equations.

By the Hermiticity of $\calO_k$, we can invoke the Hellmann-Feynman theorem to write the left hand side as
\begin{align}\label{eq:bomrhs}
\intxy \vb_{\gamma k}^* \cdot \left( \frac{\partial}{\partial k} \calO_k \right) \vb_{\gamma k}  
& = - 2k \intxy \vb_{\gamma k}^* \cdot \unitz \times (\unitz\times \vb_{\gamma k}) \frac{1}{n^2} \nonumber \\
  & ~~~~~+ i \intxy \left [ \vb_{\gamma k}^* \cdot \unitz \times \frac{1}{n^2} \nabla_t \times \vb_{\gamma k}
  + \vb_{\gamma k}^* \cdot \nabla_t \times\left(\frac{1}{n^2} \unitz \times \vb_{\gamma k}\right) \right]\nonumber \\
& = - 2k \intxy \vb_{\gamma k}^* \cdot \unitz \times (\unitz\times \vb_{\gamma k})  \frac{1}{n^2}\nonumber \\
& ~~~~~ + i \intxy \left[ \left(\vb_{\gamma k}^* \cdot \unitz \times \frac{1}{n^2} \left(-i\muo \omega_{\gamma k}\vd_{\gamma k} -i k \unitz \times \vb_{\gamma k} \right) \right)
  + \left(\nabla_t \times \vb_{\gamma k}^*\right) \cdot \left(\frac{1}{n^2} \unitz \times \vb_{\gamma k}\right) \right] \nonumber \\
& = - 2k \intxy \vb_{\gamma k}^* \cdot \unitz \times (\unitz\times \vb_{\gamma k})  \frac{1}{n^2}\nonumber \\
& ~~~~~ + i \intxy \vb_{\gamma k}^* \cdot \unitz \times \left[\frac{1}{n^2} \left(-i\muo \omega_{\gamma k}\vd_{\gamma k} -i k \unitz \times \vb_{\gamma k} \right) \right] \nonumber \\
& ~~~~~ + i \intxy \left(i \muo \omega_{\gamma k} \vd_{\gamma k}^*+ik \unitz \times \vb_{\gamma k}^*\right) \cdot \left(\frac{1}{n^2} \unitz \times \vb_{\gamma k}\right) \nonumber \\
& =  \frac{\omega_{\gamma k}}{c^2}  \intxy \left[ \vb_{\gamma k}^* \cdot \unitz \times \ve_{\gamma k} - \ve_{\gamma k}^* \cdot \unitz \times \vb_{\gamma k} \right] \nonumber \\
& =  \frac{\muo \omega_{\gamma k}}{c^2} \unitz \cdot \intxy \ve_{\gamma k} \times \vh_{\gamma k}^* + \ve_{\gamma k}^* \times \vh_{\gamma k}
\end{align}

Comparing~\eqref{eq:bomlhs} and~\eqref{eq:bomrhs} yields
\begin{align}
 \frac{\partial \omega_{\gamma k}}{\partial k}  
 & = \frac{1}{2} \unitz \cdot \intxy \ve_{\gamma k} \times \vh_{\gamma k}^* + \ve_{\gamma k}^* \times \vh_{\gamma k} \nonumber \\
 & = \frac{1}{\hbar \omega_{\gamma k}} \pmodeEM_{\gamma\gamma}(k,k) ,
\end{align}

Finally, from~\eqref{eq:pmodeb} we then have $\pmodeEM_{\gamma\gamma}(k,k) = \hbar \omega^j_\gamma v_\gamma^j $, and from~\eqref{eq:poweemsv} with~\eqref{envelope}
we obtain~\eqref{eq:PsvEM}
\begin{equation}
\powEM_{\text{sv}}(z)\approx \sum_{\gamma,j}\hbar \omega_{\gamma}^j v_{\gamma}^j \;\psi _{\gamma j}^\dagger (z)\psi_{\gamma j}(z),
\end{equation}
in exact analogy with the acoustic result in \eqref{eq:powAsv} but allowing for the sum
over electromagnetic modes.

\section{Simplification of the opto-acoustic interaction term}\label{supp:interaction}
Here we show how the interaction~\eqref{eq:Vdef} may be reduced to the form shown 
in~\eqref{Vwork}.
Inserting the expansion of the strain tensor~\eqref{eq:strainexp} into~\eqref{eq:Vdef}
gives
\begin{eqnarray*}
V &=&\frac{1}{\epso}\sum_{\gamma,\gammap ,\alpha }\int \dk\dkp  \dq   \, a_{\gamma k}^\dagger a_{\gammap \kp}b_{\alpha q}\sqrt{\frac{\hbar
\omega_{\gamma k}}{4\pi }}\sqrt{\frac{\hbar \omega_{\gammap \kp}}{
4\pi }}\sqrt{\frac{\hbar \Omalq }{4\pi }}\int \me^{i(k^\prime -k+q)z}\dz \\
&&\times \int \left( d_{\gamma k}^i (x,y)\right)^*d_{\gammap \kp }^j (x,y)\left( p^{ijlm}(x,y)s_{\alpha q}^{lm}(x,y)-\delta^{ij}\left( 
\frac{\partial \beta_\text{ref}(x,y)}{\partial r^l }\right) u_{\alpha q}^l (x,y)\right) \dx\dy\, \\
&&+\frac{1}{\epso}\sum_{\gamma,\gammap ,\alpha }\int \dk\dkp  \dq   \, a_{\gamma k}^\dagger a_{\gammap \kp}b_{\alpha q}^\dagger \sqrt{
\frac{\hbar \omega_{\gamma k}}{4\pi }}\sqrt{\frac{\hbar \omega_{\gammap \kp}}{4\pi }}\sqrt{\frac{\hbar \Omalq \Omalq }{4\pi }}\int
\me^{i(\kp-k-q)z}\dz \\
&&\times \int \left( d_{\gamma k}^i (x,y)\right)^*d_{\gammap \kp }^j (x,y)\left( p^{ijlm}(x,y)\left( s_{\alpha q}^{lm}(x,y)\right)^{\ast
}-\delta^{ij}\left( \frac{\partial \beta_\text{ref}(x,y)}{\partial r^l } \right) \left( u_{\alpha q}^l (x,y)\right)^*\right) \dx\dy\,.
\end{eqnarray*}
Since the inverse dielectric tensor is symmetric even under strain, we
have $p^{ijlm}(x,y)=p^{jilm}(x,y)$, and swapping the dummy indices $k,\kp$ in the second term gives
\begin{eqnarray*}
V 
&=&\frac{1}{\epso}\sum_{\gamma,\gammap ,\alpha }\int \dk\dkp  \dq   \, a_{\gamma k}^\dagger a_{\gammap \kp}b_{\alpha q}\sqrt{\frac{\hbar
\omega_{\gamma k}}{4\pi }}\sqrt{\frac{\hbar \omega_{\gammap \kp}}{
4\pi }}\sqrt{\frac{\hbar \Omalq }{4\pi }}\int \me^{i(k^\prime -k+q)z}\dz \\
&&\times \int \left( d_{\gamma k}^i (x,y)\right)^*d_{\gammap \kp }^j (x,y)\left( p^{ijlm}(x,y)s_{\alpha q}^{lm}(x,y)-\delta^{ij}\left( 
\frac{\partial \beta_\text{ref}(x,y)}{\partial r^l }\right) u_{\alpha q}^l (x,y)\right) \dx\dy\, \\
&&+\frac{1}{\epso}\sum_{\gamma,\gammap ,\alpha }\int \dk\dkp  \dq   \, a_{\gammap \kp}^\dagger a_{\gamma k}b_{\alpha q}^\dagger \sqrt{
\frac{\hbar \omega_{\gamma k}}{4\pi }}\sqrt{\frac{\hbar \omega_{\gammap \kp}}{4\pi }}\sqrt{\frac{\hbar \Omalq }{4\pi }}\int
\me^{i(k-\kp-q)z}dz \\
&&\times \int \left( d_{\gammap \kp}^j (x,y)\right)^{\ast}
d_{\gamma k}^i (x,y)\left( p^{ijlm}(x,y)\left( s_{\alpha q}^{lm}(x,y)\right)
^*-\delta^{ij}\left( \frac{\partial \beta_\text{ref}(x,y)}{\partial r^l }
\right) \left( u_{\alpha q}^l (x,y)\right)^*\right) \dx\dy.
\end{eqnarray*}

We can now write this as
\begin{eqnarray}
V  \label{eq:suppVwork} 
&=&\sum_{\gamma,\gammap ,\alpha }
\int \frac{\dk\dkp \dq  }{\left( 2\pi \right)^{3/2}}\;a_{\gamma k}^\dagger a_{\gammap \kp}b_{\alpha q}
\int  \Gamma (\gamma k;\gammap \kp;\alpha q)\, \me^{i(\kp-k+q)z} \, \dz \\
&&+\sum_{\gamma,\gammap ,\alpha }
\int \frac{\dk\dkp \dq  }{(2\pi )^{3/2}} \;b_{\alpha q}^\dagger a_{\gammap \kp}^\dagger a_{\gamma k}
\int  \Gamma^*(\gamma k;\gammap \kp;\alpha q)\, \me^{-i(\kp-k+q)z}\,   \dz\nonumber,
\end{eqnarray}
where the coupling is characterized by the slowly-varying coefficients
\begin{eqnarray}
\Gamma (\gamma k;\gammap \kp;\alpha q)  \label{supp:gammaresult} 
&=&\frac{1}{\epso}\sqrt{\frac{\hbar \omega_{\gamma k}}{2}}\sqrt{\frac{
\hbar \omega_{\gammap \kp}}{2}}\sqrt{\frac{\hbar \Omega_{\alpha
q}}{2}}  \nonumber \\
&&\times \int \dx\dy\,\left( d_{\gamma k}^i (x,y)\right)^*d_{\gammap \kp}^j (x,y)\left( p^{ijlm}(x,y)s_{\alpha q}^{lm}(x,y)-\delta
^{ij}\left( \frac{\partial \beta_\text{ref}(x,y)}{\partial r^l }\right) u_{\alpha q}^l (x,y)\right) .  \nonumber
\end{eqnarray}

\section{Smoothing the surface matrix element}\label{app:surfacematrix}
In the $(x,y)$ plane we can in general identify a number of curves $C$ that
indicate where $\beta_\text{ref}(x,y)$ will change discontinuously from one
value to another. These may be straight or curved lines.
We only contemplate discontinuous changes in $\beta_\text{ref}(x,y)$, 
adding up the neighborhoods of all such curves identifies the regions
where $\beta_\text{ref}(x,y)$ is assumed to vary in the $xy$ plane. We
write $\vecR=(x,y)$, and parameterize such a curve by 
$\vecR_c(s)$ $=(x_c(s),y_c(s))$, and for a given curve let $s$ range from $0$
to $1$. 
As $s$ increases along the curve we have 
\begin{eqnarray*}
\vecdR_c (s) &=& \unitx \frac{\dx_c (s)}{\ds}\ds+\unity \frac{\dy_c (s)}{\ds}\ds \\
&=&\mathbf{\hat{u}}(s)\, \dR_c ,
\end{eqnarray*}
where the length 
\begin{equation}
\dR_c =\ds\sqrt{\left( \frac{\dx_c (s)}{\ds}\right)^{2}+\left( \frac{\dy_c (s)
}{\ds}\right)^{2}}, 
\end{equation}
and the unit vector 
\begin{equation}
\unitu (s)=\frac{\unitx \frac{\dx_c (s)}{\ds}+\unity \frac{\dy_c (s)}{\ds}}{\sqrt{\left( \frac{\dx_c (s)}{\ds}\right)^{2}+\left( 
\frac{\dy_c (s)}{\ds}\right)^{2}}}. 
\end{equation}

\begin{figure}
\includegraphics[width=8cm]{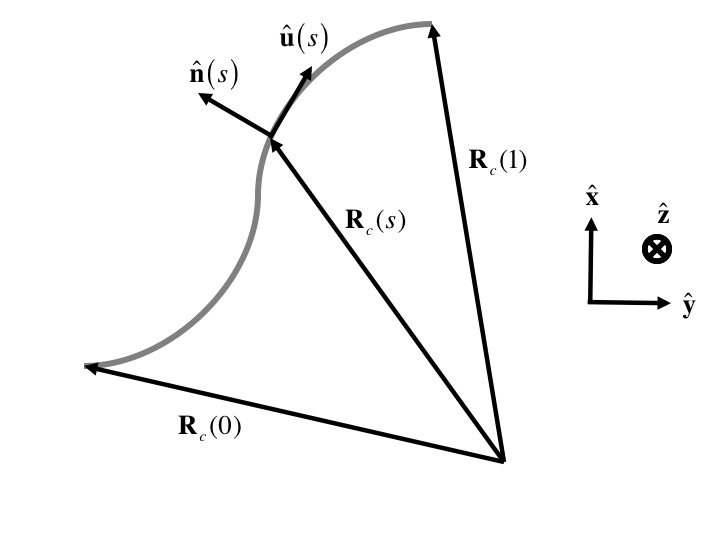}
\caption{ Geometry for smoothing of discontinuous fields at waveguide interfaces. \label{fig:smooth}}
\end{figure}

We introduce a normal to the curve as 
\begin{eqnarray*}
\unitn(s) &\equiv &\unitz \times \unitu (s) \\
&=&\frac{- \unity \frac{\dx_c (s)}{\ds}+\unitx \frac{ \dy_c (s)}{\ds}}{\sqrt{\left( \frac{\dx_c (s)}{\ds}\right)^{2}
  +\left( \frac{ \dy_c (s)}{\ds}\right)^{2}}},
\end{eqnarray*}
and in the neighborhood of the curve we can specify the points in the $xy$
plane by $(s,\zeta )$, where 
\begin{equation}
\mathbf{R=R}_c (s)+\zeta \unitn(s), 
\end{equation}
or 
\begin{eqnarray*}
x &=&x_c (s)+\zeta \left( \mathbf{\hat{x}\cdot \hat{n}}(s)\right) , \\
y &=&y_c (s)+\zeta \left( \mathbf{\hat{y}\cdot \hat{n}}(s)\right) .
\end{eqnarray*}
For fixed $s$, $\beta_\text{ref}$ changes as $\zeta $ passes from $<0$ to $>0$;
that is, it is only a function of $\zeta $.   We assume now that the
change in $\beta_\text{ref}(x,y)$ occurs only in such a small region about $
\mathbf{R=R}_c (s)$ (we will eventually take that change to be a Dirac
delta function there) that the mapping from $(\zeta ,s)$ to $(x,y)$ is
one-to-one. Then we can write 
\begin{equation}
\left( \frac{\partial \beta_\text{ref}(x,y)}{\partial r^k }\right) u_{\alpha
q}^k (x,y)=\left( \frac{\dd\beta_\text{ref}(\zeta )}{\dd\zeta }\right) 
\unitn(s)\cdot \vu_{\alpha q}(x(s,\zeta ),y(s,\zeta )), 
\end{equation}
and we have 
\begin{multline}
-\int \dx\dy\,\left( d_{\gamma k}^i (x,y)\right)^*d_{\gammap \kp }^i (x,y)\left( \frac{\partial \beta_\text{ref}(x,y)}{\partial r^k }\right)
u_{\alpha q}^k (x,y)  \label{workmatrix} \\
=-\int \left\vert J(s,\zeta )\right\vert \ds\dd\zeta \left(
d_{\gamma k}^i (x,y)\right)^*d_{\gammap \kp}^i (x,y)\left( 
\frac{d\beta_\text{ref}(\zeta )}{d\zeta }\right) \unitn(s)\cdot 
\vu_{\alpha q}(x,y),  %
\end{multline}
where in the second line we understand $x=x(s,\zeta )$ and $y=y(s,\zeta )$;
that is, we have switched integration variables from $x$ and $y$ to $s$ and $
\zeta $. The Jacobian 
\begin{equation}
J(s,\zeta )=\left\vert 
\begin{array}{cc}
\frac{\partial x}{\partial s} & \frac{\partial y}{\partial s} \\ 
\frac{\partial x}{\partial \zeta } & \frac{\partial y}{\partial \zeta }
\end{array}
\right\vert =\left( \left( \frac{\dx_c (s)}{\ds}+\zeta \unitx \cdot 
\frac{\partial \unitn(s)}{\mathbf{\partial s}}\right) \unity\cdot\unitn(s)-\left( \frac{\dy_c (s)}{\ds}+\zeta \unity \cdot 
\frac{\partial \unitn(s)}{\mathbf{\partial s}}\right) \unitx \cdot \unitn (s)\right) . 
\end{equation}
Now at each point $(s,\zeta )$ we can use the local reference frame to identify
\begin{equation}
\left( d_{\gamma k}^i (x,y)\right)^*d_{\gammap \kp }^i (x,y)=\left( d_{\gamma k}^{\perp }(x,y)\right)^*d_{\gamma^\prime \kp}^{\perp }(x,y)+\left( \vd_{\gamma k}^{\parallel }(x,y)\right)
^*\cdot \vd_{\gammap \kp}^{\parallel }(x,y), 
\end{equation}
where 
\begin{eqnarray*}
d_{\gammap \kp}^{\perp }(x,y)&=&\unitn(s)\cdot \mathbf{d}
_{\gammap \kp}(x,y), \\
\vd_{\gammap \kp}^{\parallel }(x,y)&=&\mathbf{\hat{z}\hat{z
}}\cdot \vd_{\gammap \kp}(x,y)+\mathbf{\hat{u}}\left(
s\right) \mathbf{\hat{u}}\left( s\right) \cdot \vd_{\gamma^\prime \kp}(x,y) \\
&=&\vd_{\gammap \kp}(x,y)-\unitn\left( s\right) 
\unitn\left( s\right) \cdot \vd_{\gammap \kp }(x,y),
\end{eqnarray*}
again recalling that $x=x(s,\zeta )$ and $y=y(s,\zeta )$. Then we can
write (\ref{workmatrix}) as 
\begin{multline}
-\int \dx\dy\,\left( d_{\gamma k}^i (x,y)\right)^*d_{\gammap \kp }^i (x,y)\left( \frac{\partial \beta_\text{ref}(x,y)}{\partial r^k }\right)
u_{\alpha q}^k (x,y)  \label{workmatrix2} \\
=-\int \left\vert J(s,\zeta )\right\vert \ds\dd\zeta \left( d_{\gamma k}^{\perp
}(x,y)\right)^*d_{\gammap \kp}^{\perp }(x,y)\left( \frac{ \dd\beta_\text{ref}(\zeta )}{\dd\zeta }\right) \unitn(s)\cdot \vu
_{\alpha q}(x,y)  \\
-\int \left\vert J(s,\zeta )\right\vert \ds\dd\zeta \left( \vd
_{\gamma k}^{\parallel }(x,y)\right)^*\cdot \vd_{\gamma^\prime \kp}^{\parallel }(x,y)\left( \frac{\dd\beta_\text{ref}(\zeta )}{\dd\zeta }
\right) \unitn(s)\cdot \vu_{\alpha q}(x,y).  
\end{multline}
From the relation
\begin{equation} \label{epdef}
\varepsilon_\text{ref}(\zeta )=\frac{1}{\beta_\text{ref}(\zeta )}
\end{equation}
follows
\begin{equation}
\frac{\dd\beta_\text{ref}(\zeta )}{\dd\zeta }=-\frac{1}{\varepsilon_\text{ref}^{2}(\zeta
)}\frac{\dd\varepsilon_\text{ref}(\zeta )}{\dd\zeta } . \label{relatederivs}
\end{equation}

Now the simplest characterization of the variation of $\beta_\text{ref}(\zeta )$
would be to write 
\begin{equation}
\beta_\text{ref}(\zeta )=\beta_{-}+(\beta_{+}-\beta_{-})\theta (\zeta ),
\end{equation}
where $\theta(\zeta)$ is the step function, 
$\beta_{-}$ is the value of $\beta_\text{ref}(\zeta )$ for negative $
\zeta $, and $\beta_{+}$ is the value of $\beta_\text{ref}(\zeta )$ for
positive $\zeta $. Similarly, from~\eqref{epdef}
we can write 
\begin{equation}
\varepsilon_\text{ref}(\zeta )=\frac{1}{\beta_{-}}+\left( \frac{1}{\beta_{+}}-
\frac{1}{\beta_{-}}\right) \theta (\zeta ).
\end{equation}
To differentiate with respect to $\zeta $ and then integrate in (\ref
{workmatrix2}) we smooth these functions. We introduce a smoothing
function $g_{l}(\zeta )$ which is non-negative, peaked at $\zeta =0$,
satisfies 
\begin{equation}
\int g_{l}(\zeta )d\zeta =1  \label{gnorm},
\end{equation}
for each $l$, and approaches a Dirac delta function as $l\rightarrow 0$. \
Then for finite $l$ we have smoothed functions 
\begin{eqnarray*}
\bar{\beta}_\text{ref}(\zeta ) &=&\int g_{l}(\zeta -\zeta^\prime )\beta
_\text{ref}(\zeta^\prime )\, \dd\zeta^\prime , \\
\bar{\varepsilon}_\text{ref}(\zeta ) &=&\int g_{l}(\zeta -\zeta^\prime )\varepsilon_\text{ref}(\zeta^\prime )\, \dd\zeta^\prime .
\end{eqnarray*}
One strategy for evaluating $\partial \beta_\text{ref}(\zeta )/\partial \zeta $
is to take 
\begin{equation}
\frac{\dd\beta_\text{ref}(\zeta )}{\dd\zeta }\rightarrow \frac{\dd\bar{\beta}
_\text{ref}(\zeta )}{\dd\zeta }=(\beta_{+}-\beta_{-})g_{l}(\zeta ).
\label{firstderiv}
\end{equation}
Alternately, using (\ref{relatederivs}), we could take 
\begin{eqnarray}
\frac{\dd\beta_\text{ref}(\zeta )}{\dd\zeta } &\rightarrow &-\frac{1}{\bar{
\varepsilon}_\text{ref}^{2}(\zeta )}\frac{\dd\bar{\varepsilon}_\text{ref}(\zeta )}{
d\zeta }  \label{secondderiv} \\
&=&-\frac{1}{\bar{\varepsilon}_\text{ref}^{2}(\zeta )}\left( \frac{1}{\beta_{+}}-
\frac{1}{\beta_{-}}\right) g_{l}(\zeta ).  \nonumber
\end{eqnarray}
Using (\ref{firstderiv}) and (\ref{secondderiv}) 
in the two right-hand 
expressions of (\ref{workmatrix2}) respectively, gives
\begin{multline}
-\int \dx\dy\,\left( d_{\gamma k}^i (x,y)\right)^*d_{\gammap \kp }^i (x,y)\left( \frac{\partial \beta_\text{ref}(x,y)}{\partial r^k }\right)
u_{\alpha q}^k (x,y) \\
=-(\beta_{+}-\beta_{-})\int \left\vert J(s,\zeta )\right\vert \ds\dd\zeta \, 
\left( d_{\gamma k}^{\perp }(x,y)\right)^*d_{\gammap \kp}^{\perp
}(x,y)g_{l}(\zeta )\unitn(s)\cdot \vu_{\alpha q}(x,y) \\
+\left( \frac{1}{\beta_{+}}-\frac{1}{\beta_{-}}\right) \int \left\vert
J(s,\zeta )\right\vert \ds\dd\zeta \, \frac{\left( \vd_{\gamma k}^{\parallel
}(x,y)\right)^*\cdot \vd_{\gammap \kp}^{\parallel
}(x,y)}{\bar{\varepsilon}_\text{ref}^{2}(\zeta )}g_{l}(\zeta )\unitn
(s)\cdot \vu_{\alpha q}(x,y),
\end{multline}
where we still understand $x=x(s,\zeta )$ and $y=y(s,\zeta )$. Now we  can
let $l\rightarrow 0$ in both terms, because the rest of the
integrands are continuous about $\zeta =0$. Recalling (\ref{gnorm}), we
have
\begin{multline}
-\int \dx\dy\,\left( d_{\gamma k}^i (x,y)\right)^*d_{\gammap \kp }^i (x,y)\left( \frac{\partial \beta_\text{ref}(x,y)}{\partial r^k }\right)
u_{\alpha q}^k (x,y) \\
\rightarrow  
-(\beta_{+}-\beta_{-})\int \left\vert J(s,0)\right\vert \left(
d_{\gamma k}^{\perp }(\vecR_c (s))\right)^*d_{\gammap \kp }^{\perp }(\vecR_c (s))\;\unitn(s)\cdot \vu_{\alpha
q}(\vecR_c (s))\, \ds \\
+\epso^{2}\left( \frac{1}{\beta_{+}}-\frac{1}{\beta_{-}}\right)
\int \left\vert J(s,0)\right\vert \left( \ve_{\gamma k}^{\parallel }(
\vecR_c (s))\right)^*\cdot \ve_{\gammap \kp }^{\parallel }(\vecR_c (s))\;\unitn(s)\cdot \vu
_{\alpha q}(\vecR_c (s))\, \ds.
\end{multline}
where we have used the fact that $x(s,\zeta )\to x_c (s)$, and 
$ y(s,\zeta ) \to y_c (s)$, as $\zeta \rightarrow 0$. Finally, we have 
\begin{eqnarray*}
\left\vert J(s,0)\right\vert \ds &=&\left\vert \left( \frac{\dx_c (s)}{\ds}
\right) 
\unity \cdot \unitn(s)
-\left( \frac{\dy_c (s)}{\ds}\right) 
\unitx \cdot \unitn(s)
\right\vert \ds \\
&=&\sqrt{\left( \frac{\dx_c (s)}{\ds}\right)^{2}+\left( \frac{\dy_c (s)}{\ds}
\right)^{2}}\ds=\dR_c (s),
\end{eqnarray*}
the element of length along the curve. So we can write 
\begin{align}
 -\int \dx\dy\,
&\left[ d_{\gamma k}^i (x,y)\right]^*d_{\gammap \kp }^i (x,y)\left( \frac{\partial \beta_\text{ref}(x,y)}{\partial r^k }\right)
u_{\alpha q}^k (x,y) \nonumber \\
 ~~~~~~~~~&
\rightarrow  -(\beta_{+}-\beta_{-})\int \left[ d_{\gamma k}^{\perp }(\vecR _c (s))\right]^* 
                                   d_{\gammap \kp}^{\perp }(\vecR _c (s))
    \left[ \unitn(s)\cdot \vu_{\alpha q}(\vecR_c (s))\right] \, \dR_c (s)  \nonumber \\
& ~~~~~~+\epso^{2}\left( \frac{1}{\beta_{+}}-\frac{1}{\beta_{-}}\right)
\int \left( \ve_{\gamma k}^{\parallel }(\vecR_c (s))\right)^{\ast }\cdot \ve_{\gammap \kp}^{\parallel }(\vecR_c (s))
\left[ \unitn(s)\cdot \vu_{\alpha q}(\vecR_c (s))\right] \, \dR_c (s)  \nonumber \\
& =(\frac{1}{\varepsilon_{-}}-\frac{1}{\varepsilon_{+}})\int \left(
d_{\gamma k}^{\perp }(\vecR_c (s))\right)^*d_{\gammap \kp }^{\perp }(\vecR_c (s))\left[ \unitn(s)\cdot \vu
_{\alpha q}(\vecR_c (s))\right] \, \dR_c (s)  \nonumber \\
& ~~~~~~+\epso^{2}\left( \varepsilon_{+}-\varepsilon_{-}\right) \int
\left( \ve_{\gamma k}^{\parallel }(\vecR_c (s))\right)^*\cdot 
\ve_{\gammap \kp}^{\parallel }(\vecR_c (s))
 \left[ \unitn(s)\cdot \vu_{\alpha q}(\vecR_c (s)) \right] \, \dR_c (s),
\end{align}
from whence~\eqref{eq:bargam} follows.
The full expression for $\bar{\Gamma}_\text{surf}(\gamma k;\gammap \kp;\alpha
q)$ then involves a sum over all such curves where a transition from one
dielectric constant to another occurs. Note there is no ambiguity in
evaluating these terms, since $\vd_{\gamma k}^{\perp }(\vecr)$ is
continuous across a step discontinuity in $\beta_\text{ref}(x,y)$, as is $
\ve_{\gammap \kp}^{\parallel }(\vecr).$

\section{Reduced matrix elements}\label{app:redmatel}

Using the normalization conditions \eqref{eq:unorm} and \eqref{norm} 
we can write the matrix elements in the form
\begin{eqnarray}
\bar{\Gamma}(\gamma k;\gammap \kp;\alpha q) \label{gresult} 
&=&
\frac{1}{2^{3/2}\Omalq \sqrt{\left\vert v_{\gamma k}v_{\gammap \kp}v_{\alpha q}\right\vert }}
\\
&&
\times \left[
\frac{\int \dx\dy\,\left( d_{\gamma k}^i (x,y)\right)^*d_{\gamma^\prime \kp}^j (x,y)p^{ijlm}(x,y)S_{\alpha q}^{lm}(x,y)}{\left[ \int
\dx\dy\,\beta_\text{ref}(\vecr)\vd_{\gamma k}^*(x,y)\cdot \vd
_{\gamma k}(x,y)\right] \left[ \int \dx\dy\,\rho (x,y)\vu_{\alpha q}^{\ast
}(x,y)\cdot \vu_{\alpha q}(x,y)\right]^{1/2}}  \right. \nonumber \\
&&+(\frac{1}{\varepsilon_{-}}-\frac{1}{\varepsilon_{+}})\frac{\int \left(
d_{\gamma k}^{\perp }(\vecr_c (s))\right)^*d_{\gammap \kp }^{\perp }(\vecr_c (s))\left( \unitn(s)\cdot \vu
_{\alpha q}(\vecr_c (s)\right) dR_c (s)}{\left[ \int \dx\dy\,\beta_\text{ref}(
\vecr)\vd_{\gamma k}^*(x,y)\cdot \vd_{\gamma k}(x,y)\right] 
\left[ \int \dx\dy\,\rho (x,y)\vu_{\alpha q}^*(x,y)\cdot \vu_{\alpha q}(x,y)\right]^{1/2}}  \nonumber \\
&& \left . +\left( \varepsilon_{+}-\varepsilon_{-}\right) \frac{\epsilon
_{0}^{2}\int \left( \ve_{\gamma k}^{\parallel }(\vecr_c (s))\right)
^*\cdot \ve_{\gammap \kp}^{\parallel }(\vecr
_c (s))\unitn(s)\cdot \vu_{\alpha q}(\vecr
_c (s))dR_c (s)}{\left[ \int \dx\dy\,\beta_\text{ref}(\vecr)\vd
_{\gamma k}^*(x,y)\cdot \vd_{\gamma k}(x,y)\right] \left[ \int \dx\dy\,\rho
(x,y)\vu_{\alpha q}^*(x,y)\cdot \vu_{\alpha q}(x,y)
\right]^{1/2}} \right] .  \nonumber
\end{eqnarray}
The advantage of this form is that it can now be used regardless of how the
mode fields are normalized.  Again, the last two lines should be summed
over all curves that contribute.

\section{The organization of eigenfunctions} \label{supp:partnerbuild} 
This section establishes the basic properties of partner eigenfunctions for a Hermitian
operator that are invoked in section~\ref{supp:partners}.

Consider a Hermitian operator, schematically $H(x,\frac{\partial }{\partial x
},\ldots)$; the eigenvalue equation is 
\begin{equation}
H(x,\frac{\partial }{\partial x},\ldots)f(x)=\lambda f(x). 
\end{equation}
Hermiticity guarantees real eigenvalues and the fact that eigenfunctions of
different eigenvalues are orthogonal. The inner product of two such
functions vanishes, where the inner product of $g(x)$ with $f(x)$ is 
\begin{equation}
\int g^*(x)f(x)dx 
\end{equation}
We consider eigenfunctions that are normalized, so 
\begin{equation}
\int f^*(x)f(x)dx=1. 
\end{equation}
\qquad

We want to consider first a number of degenerate eigenfunctions, all with
the same eigenvalue. Suppose now that besides being Hermitian, $H$ is also
real. Then if $f(x)$ is an eigenfunction, $f^*(x)$ will also be an
eigenfunction with the same eigenvalue. 
\begin{equation}
H(x,\frac{\partial }{\partial x},\ldots)f^*(x)=\lambda f^*(x) 
\end{equation}
For a given $f(x)$, of course one possibility is that $f^*(x)$ is
just a constant phase factor times $f(x)$. Then $f(x)$ could be readjusted
to be purely real (or purely imaginary), for example.

Suppose this is not the case. Then $f(x)$ and $f^*(x)$ are linearly
independent, and they span a two-dimensional space. Of course, they need
not be orthogonal. That is, there is no guarantee that the inner product
of $f^*(x)$ with $f(x)$, 
\begin{equation}
\int f(x)f(x)dx  \label{ff}
\end{equation}
vanishes. If it does, we call $f(x)$ and $f^*(x)$ ``partner''
eigenfunctions. Suppose now that (\ref{ff}) does not vanish. We can
construct partner eigenfunctions from $f(x)$ and $f^*(x)$ in the
following way. 

First find 
\begin{equation}
c(x)=N(f(x)+f^*(x)), 
\end{equation}
where $N$ is a real normalization constant; $c(x)$ does not vanish, because
by assumption $f^*(x)$ is not just a multiple of $f(x)$. If we choose $
N$ to be real, then $c(x)$ is also purely real. Now take out from $f(x)$
the amount proportional to $c(x)$, 
\begin{equation}
\bar{f}(x)=f(x)-c(x)\int c(x^\prime )f(x^\prime )dx^\prime , 
\end{equation}
where we do not need $c^*(x^\prime )$ in the integral because $
c(x^\prime )$ is real. Of course $\bar{f}(x)$ cannot vanish everywhere
because otherwise $f(x)$ would just be proportional to $c(x)$ and then $f(x)$
would just be a phase factor times a real function. Now by construction $
c(x)$ is orthogonal to $\bar{f}(x),$
\begin{equation}
\int c(x)\bar{f}(x)dx=0.  \label{orthog}
\end{equation}
Perhaps $\bar{f}(x)$ is purely real; if so, normalize it and call the result 
$s(x)$. Perhaps $\bar{f}(x)$ is purely imaginary; if so, divide by $i$,
normalize it and call the result $s(x)$. If $\bar{f}(x)$ is neither, note
that from (\ref{orthog}) we have 
\begin{equation}
\int c(x)\bar{f}^*(x)dx=0, 
\end{equation}
since $c(x)$ is purely real. Then 
\begin{equation}
\bar{f}(x)+\bar{f}^*(x) 
\end{equation}
is a real function that is orthogonal to $c(x)$; it cannot vanish everywhere
because we have assumed that $\bar{f}(x)$ is not purely imaginary. Now
normalize this function and call it $s(x)$.

Whatever route we have taken to get $s(x)$, we now have two real functions $
c(x)$ and $s(x)$ that are orthogonal to each other and normalized, 
\begin{eqnarray*}
\int c^{2}(x)dx &=&1, \\
\int s^{2}(x)dx &=&1, \\
\int c(x)s(x)dx &=&0.
\end{eqnarray*}
They span the space spanned by $f(x)$ and $f^*(x)$. We can then form
partner functions for this subspace, 
\begin{eqnarray*}
f_{1}(x) &=&\frac{1}{\sqrt{2}}(c(x)+is(x)), \\
f_{1}^*(x) &=&\frac{1}{\sqrt{2}}(c(x)-is(x)).
\end{eqnarray*}
These functions are normalized, 
\begin{eqnarray*}
\int \left[ f_{1}(x)\right]^*f_{1}(x)dx &=&1, \\
\int \left[ f_{1}^*(x)\right]^*f_{1}^*(x)dx &=&1,
\end{eqnarray*}
and orthogonal,
\begin{eqnarray*}
\int \left[ f_{1}^*(x)\right]^*f_{1}(x)dx &=&0, \\
\int \left[ f_{1}(x)\right]^*f_{1}^*(x)dx &=&0.
\end{eqnarray*}

So we have constructed partner eigenfunctions $f_{1}(x)$ and $f_{1}^{\ast
}(x)$ that span the subspace spanned by $f(x)$ and $f^*(x)$. Suppose
now there are more eigenfunctions with the same eigenvalue, which are
orthogonal to $f(x)$ and $f^*(x)$. Call one of them $g(x)$. Then $
g(x)$ must be orthogonal to $f_{1}(x)$ and $f_{1}^*(x)$ since they
span the same subspace as $f(x)$ and $f^*(x),$
\begin{eqnarray}
\int g^*(x)f_{1}(x)dx &=&0,  \label{cross} \\
\int g^*(x)f_{1}^*(x)dx &=&0.  \nonumber
\end{eqnarray}
Now if $g(x)$ is an eigenfunction of $H(x)$, then $g^*(x)$ is an
eigenfunction of $H(x)$ with the same eigenvalue. Suppose $g^*(x)$
is not just a constant phase factor times $g(x)$; then $g^*(x)$ and $
g(x)$ span a two dimensional subspace that, since from (\ref{cross}) we have
immediately 
\begin{eqnarray*}
\int g(x)f_{1}(x)dx &=&0, \\
\int g(x)f_{1}^*(x)dx &=&0,
\end{eqnarray*}
has no overlap with the subspace spanned by $f_{1}(x)$ and $f_{1}^*(x)$
. So from $g(x)$ and $g^*(x)$ we can form two partner wave functions 
$f_{2}(x)$ and $f_{2}^*(x)$ that are orthogonal to each other and each
orthogonal to each of $f_{1}(x)$ and $f_{1}^*(x)$. 

Thus we can proceed and organize our eigenfunctions. As we investigate all
the eigenfunctions of a particular eigenvalue we will sometimes find it is
possible to immediately make an eigenfunction real (as we could have, for
example, if $g^*(x)$ had simply been proportional to $g(x)$ with a
constant phase factor), or otherwise we can establish partners. So we can
imagine listing all our wave functions grouped in the following manner, 
\begin{eqnarray}
&&f_{1}(x)\;\;f_{1}^*(x)  \label{organization} \\
&&f_{2}(x)\;\;f_{2}^*(x)  \nonumber \\
&&f_{3}(x)\;\;f_{3}^*(x)  \nonumber \\
&&\vdots \ \ \ \ \ \ \ \ \vdots  \nonumber \\
&&f_{N}(x)\;\;f_{N}^*(x)  \nonumber \\
&&\;\;\;\;\;f_{I}(x)  \nonumber \\
&&\;\;\;\;\;f_{II}(x)  \nonumber \\
&&\;\;\;\;\;\;\vdots  \nonumber
\end{eqnarray}
Here the Roman numerals indicate real wave functions that ``don't have
partners''; we take them to be purely real. Of course, if we have an even
number of real wave functions without partners we can start combining them
into partners. For example, in the list above we could replace $f_{I}(x)$
and $f_{II}(x)$ by the partners
\begin{eqnarray*}
f_{N+1}(x) &=&\frac{1}{\sqrt{2}}\left( f_{I}(x)+if_{II}(x)\right) , \\
f_{N+1}^*(x) &=&\frac{1}{\sqrt{2}}\left( f_{I}(x)-if_{II}(x)\right) .
\end{eqnarray*}
If we have an even number of eigenfunctions of a particular eigenvalue, then
we could pair them all up in partnerships. If we have an odd number then
there must be at least one ``unpartnered'' wave function. It is also
possible to ``divorce'' some partners; for example, in place of $f_{3}(x)$ and 
$f_{3}^*(x)$ we could choose the real functions 
\begin{eqnarray*}
c_{3}(x) &=&\frac{1}{\sqrt{2}}\left( f_{3}(x)+f_{3}^*(x)\right) , \\
s_{3}(x) &=&-\frac{i}{\sqrt{2}}\left( f_{3}(x)-f_{3}^*(x)\right) .
\end{eqnarray*}
But it is often convenient and natural to have wave functions in
partnerships. In any case, we assume that we have eigenfunctions organized
according to (\ref{organization}). However, we henceforth write $
f_{1}^*(x)$ as $f_{\bar{1}}(x)$, and so on, so the list (\ref
{organization}) can be given as 
\begin{eqnarray}
&&f_{1}(x)\;\;f_{\bar{1}}(x)  \label{organizationnew} \\
&&f_{2}(x)\;\;f_{\bar{2}}(x)  \nonumber \\
&&f_{3}(x)\;\;f_{\bar{3}}(x)  \nonumber \\
&&\vdots \ \ \ \ \ \ \ \ \vdots  \nonumber \\
&&f_{N}(x)\;\;f_{\bar{N}}(x)  \nonumber \\
&&\;\;\;\;\;f_{I}(x)  \nonumber \\
&&\;\;\;\;\;f_{II}(x)  \nonumber \\
&&\;\;\;\;\;\;\vdots  \nonumber
\end{eqnarray}
Then if we denote a general eigenfunction by $f_{\alpha }(x)$, the list of
possible $Js$ is 
\begin{equation}
1,\bar{1},2,\bar{2},3,\bar{3},\ldots N,\bar{N},I,II,III \ldots 
\end{equation}
These eigenfunctions are all orthogonal, 
\begin{equation}
\int f_{\alpha }^*(x)f_{\alphap }(x)dx=\delta_{\alpha \alpha
^\prime }. 
\end{equation}
as $\alpha $ and $\alphap $ range over this list. Associated with
a list of $\alpha s$ we introduce a list of $\bar{\alpha}s,$
\begin{equation}
\bar{1},1,\bar{2},2,\bar{3},3,\ldots\bar{N},N,I,II,III\ldots 
\end{equation}
That is, if $\alpha $ is one of a partnership, $\bar{\alpha}$ is the other
partner; if $\alpha $ is a real wave function, $\bar{\alpha}$ is that wave
function itself. Clearly 
\begin{equation}
\sum_{\bar{\alpha}}=\sum_{\alpha }, 
\end{equation}
and 
\begin{equation}
f_{\bar{\alpha}}^*(x)=f_{\alpha }(x), 
\end{equation}
either because $\bar{\alpha}$ identifies the partner of $\alpha $, or
because $f_{\alpha }(x)$ is real, in which case $f_{\alpha }(x)$ can be
considered its own partner.

Now if we consider the eigenfunctions of a whole range of eigenvalues $
\lambda $ we can do the same sort of organization within the subspace of
each eigenvalue. Then we can let $\alpha $ range over the whole set of
labels of all eigenfunctions of all eigenvalues. For a given $\alpha $ we
identify the eigenvalue by $\lambda_{\alpha }$. Then over this whole
range of $\alpha s$ we have 
\begin{equation}
\int f_{\alpha }^*(x)f_{\alphap }(x)dx=\delta_{\alpha \alpha
^\prime }, 
\end{equation}
where between eigenfunctions associated with different eigenvalues the
orthogonality holds because of Hermiticity of the operator, while between
eigenfunctions associated with the same eigenvalue the orthogonality holds
because of the construction we have adopted. We still have generally 
\begin{equation}
f_{\bar{\alpha}}^*(x)=f_{\alpha }(x), 
\end{equation}
and of course
\begin{equation}
\lambda_{\bar{\alpha}}=\lambda_{\alpha }. 
\end{equation}

\end{document}